\documentclass[8pt]{article}
\usepackage{a4wide}
\usepackage{amssymb}
\usepackage[english]{babel}
\usepackage{epsfig}
\usepackage{graphicx}
\usepackage{color}
\usepackage{amsmath}
\usepackage{boxedminipage}
\begin{document}
\newcommand{\lsla}{\mbox{$\not{\! l}$}}
\newcommand{\psla}{\mbox{$\not{\! p}$}}
\newcommand{\qsla}{\mbox{$\not{\! q}$}}
\newcommand{\drsla}{\mbox{$\not{\! \partial}$}}
\def\vectrl #1{\buildrel\leftrightarrow \over #1}
\def\partrl{\vectrl{\partial}}
\parskip5pt
\numberwithin{equation}{section}
\renewcommand{\thefootnote}{\fnsymbol{footnote}}
\def\db{\delta_{\rm BRS}}

%\input{../common/susydef}
%\input{../common/shortcuts}
%\input{commandes}

% \documentclass[12pt]{article}
% \usepackage{graphicx}
% \usepackage{cite}
% \usepackage{amsmath}
% \usepackage{hyperref}
% \usepackage{graphicx}
% \usepackage{amssymb}
% \usepackage{amsmath}
% \usepackage{epsfig}
%
% \textheight 24cm
% \topmargin     -0.5in
% \textwidth     16.cm
% \parskip .2cm
% \oddsidemargin  0in
% \evensidemargin 0in

%\begin{document}

%\input{shortcuts}
%%%%%%%%%Fawzi's Short CUT: GENERAL%%%%%%%%%%%%%%%%%

\newcommand{\be}{\begin{equation}}
\newcommand{\beq}{\begin{equation}}
\newcommand{\eeq}{\end{equation}}
\newcommand{\ee}{\end{equation}}

\newcommand{\refeq}[1]{Eq.\ref{eq:#1}}
\newcommand{\refig}[1]{Fig.\ref{fig:#1}}
\newcommand{\refsec}[1]{Sec.\ref{sec:#1}}

\newcommand{\beqn}{\begin{eqnarray}}
\newcommand{\eeqn}{\end{eqnarray}}
\newcommand{\bea}{\begin{eqnarray}}
\newcommand{\ena}{\end{eqnarray}}
\newcommand{\ra}{\rightarrow}
\newcommand{\susy}{{{\cal SUSY}$\;$}}
\newcommand{\su}{$ SU(2) \times U(1)\,$}

\newcommand{\gag}{$\gamma \gamma$ }
\newcommand{\gagt}{\gamma \gamma }
\newcommand{\gam}{\gamma \gamma }
\def\W{{\mbox{\boldmath $W$}}}
\def\B{{\mbox{\boldmath $B$}}}
\def\V{{\mbox{\boldmath $V$}}}
\newcommand{\np}{Nucl.\,Phys.\,}
\newcommand{\pl}{Phys.\,Lett.\,}
\newcommand{\pr}{Phys.\,Rev.\,}
\newcommand{\prl}{Phys.\,Rev.\,Lett.\,}
\newcommand{\prep}{Phys.\,Rep.\,}
\newcommand{\zp}{Z.\,Phys.\,}
\newcommand{\sovjnp}{{\em Sov.\ J.\ Nucl.\ Phys.\ }}
\newcommand{\nuclinst}{{\em Nucl.\ Instrum.\ Meth.\ }}
\newcommand{\annp}{{\em Ann.\ Phys.\ }}
\newcommand{\intjmp}{{\em Int.\ J.\ of Mod.\  Phys.\ }}

%GENERAL
\newcommand{\eps}{\epsilon}
\newcommand{\mw}{M_{W}}
\newcommand{\mww}{M_{W}^{2}}
\newcommand{\mwmw}{M_{W}^{2}}
\newcommand{\mhmh}{M_{H}^2}
\newcommand{\mz}{M_{Z}}
\newcommand{\mzz}{M_{Z}^{2}}

\newcommand{\cw}{c_W}
\newcommand{\sw}{s_W}
\newcommand{\tw}{\tan\theta_W}
\def\tww{\tan^2\theta_W}
\def\stw{s_{2w}}

\newcommand{\smw}{s_M^2}
\newcommand{\cmw}{c_M^2}
\newcommand{\seff}{s_{{\rm eff}}^2}
\newcommand{\ceff}{c_{{\rm eff}}^2}
\newcommand{\seffl}{s_{{\rm eff\;,l}}^{2}}
\newcommand{\sww}{s_W^2}
\newcommand{\cww}{c_W^2}
\newcommand{\swo}{s_W}
\newcommand{\cwo}{c_W}

\newcommand{\epm}{$e^{+} e^{-}\;$}
\newcommand{\epemt}{$e^{+} e^{-}\;$}
\newcommand{\epem}{e^{+} e^{-}\;}
\newcommand{\ememt}{$e^{-} e^{-}\;$}
\newcommand{\emem}{e^{-} e^{-}\;}

\newcommand{\ord}{{\cal O}}

\newcommand{\lra}{\leftrightarrow}
\newcommand{\tr}{{\rm Tr}}
\def\ls1{{\not l}_1}
\newcommand{\cms}{centre-of-mass\hspace*{.1cm}}

%W Physics and anomalous

\newcommand{\dkg}{\Delta \kappa_{\gamma}}
\newcommand{\dkz}{\Delta \kappa_{Z}}
\newcommand{\dz}{\delta_{Z}}
\newcommand{\dgz}{\Delta g^{1}_{Z}}
\newcommand{\dgzt}{$\Delta g^{1}_{Z}\;$}
\newcommand{\la}{\lambda}
\newcommand{\lag}{\lambda_{\gamma}}
\newcommand{\lambdae}{\lambda_{e}}
\newcommand{\laz}{\lambda_{Z}}
\newcommand{\lnl}{L_{9L}}
\newcommand{\lnr}{L_{9R}}
\newcommand{\lt}{L_{10}}
\newcommand{\lu}{L_{1}}
\newcommand{\ld}{L_{2}}
\newcommand{\eeww}{e^{+} e^{-} \ra W^+ W^- \;}
\newcommand{\eewwt}{$e^{+} e^{-} \ra W^+ W^- \;$}
\newcommand{\epemww}{e^{+} e^{-} \ra W^+ W^- }
\newcommand{\epemwwt}{$e^{+} e^{-} \ra W^+ W^- \;$}
\newcommand{\eennhht}{$e^{+} e^{-} \ra \nu_e \bar \nu_e HH\;$}
\newcommand{\eennhh}{e^{+} e^{-} \ra \nu_e \bar \nu_e HH\;}
\newcommand{\eennht}{$e^{+} e^{-} \ra \nu_e \bar \nu_e H\;$}
\newcommand{\eennh}{e^{+} e^{-} \ra \nu_e \bar \nu_e H\;}
\newcommand{\ppwg}{p p \ra W \gamma}
\newcommand{\wwhh}{W^+ W^- \ra HH\;}
\newcommand{\wwhht}{$W^+ W^- \ra HH\;$}
\newcommand{\ppwz}{pp \ra W Z}
\newcommand{\ppwgt}{$p p \ra W \gamma \;$}
\newcommand{\ppwzt}{$pp \ra W Z \;$}
\newcommand{\gamgamt}{$\gamma \gamma \;$}
\newcommand{\gamgam}{\gamma \gamma \;}
\newcommand{\egamt}{$e \gamma \;$}
\newcommand{\egam}{e \gamma \;}
\newcommand{\gamgamwwt}{$\gamma \gamma \ra W^+ W^- \;$}
\newcommand{\gamgamwwht}{$\gamma \gamma \ra W^+ W^- H \;$}
\newcommand{\gamgamwwh}{\gamma \gamma \ra W^+ W^- H \;}
\newcommand{\gamgamwwhht}{$\gamma \gamma \ra W^+ W^- H H\;$}
\newcommand{\gamgamwwhh}{\gamma \gamma \ra W^+ W^- H H\;}
\newcommand{\ggww}{\gamma \gamma \ra W^+ W^-}
\newcommand{\ggwwt}{$\gamma \gamma \ra W^+ W^- \;$}
\newcommand{\ggwwht}{$\gamma \gamma \ra W^+ W^- H \;$}
\newcommand{\ggwwh}{\gamma \gamma \ra W^+ W^- H \;}
\newcommand{\ggwwhht}{$\gamma \gamma \ra W^+ W^- H H\;$}
\newcommand{\ggwwhh}{\gamma \gamma \ra W^+ W^- H H\;}
\newcommand{\ggwwz}{\gamma \gamma \ra W^+ W^- Z\;}
\newcommand{\ggwwzt}{$\gamma \gamma \ra W^+ W^- Z\;$}

\newcommand{\veps}{\varepsilon}

\newcommand{\ptu}{p_{1\bot}}
\newcommand{\vecptu}{\vec{p}_{1\bot}}
\newcommand{\ptd}{p_{2\bot}}
\newcommand{\vecptd}{\vec{p}_{2\bot}}
\newcommand{\ie}{{\em i.e.}}
\newcommand{\cm}{{{\cal M}}}
\newcommand{\cl}{{{\cal L}}}
\newcommand{\cd}{{{\cal D}}}
\newcommand{\cv}{{{\cal V}}}
\def\slashc{c\kern -.400em {/}}
\def\slashp{p\kern -.400em {/}}
\def\slashq{q\kern -.450em {/}}
\def\slashL{L\kern -.450em {/}}
\def\slashcl{\cl\kern -.600em {/}}
\def\slashr{r\kern -.450em {/}}
\def\slashk{k\kern -.500em {/}}
\def\Ww{{\mbox{\boldmath $W$}}}
\def\B{{\mbox{\boldmath $B$}}}
\def\noi{\noindent}
\def\nn{\noindent}
\def\sm{SM }
\def\smn{SM}
\def\smp{SM}
%%%
\def\nph{${\cal{N}} {\cal{P}}\;$}
\def\sb{$ {\cal{S}}  {\cal{B}}\;$}
\def\ssb{${\cal{S}} {\cal{S}}  {\cal{B}}\;$}
\def\ssbe{{\cal{S}} {\cal{S}}  {\cal{B}}}
\def\cviol{${\cal{C}}\;$}
\def\pviol{${\cal{P}}\;$}
\def\cpviol{${\cal{C}} {\cal{P}}\;$}

\newcommand{\lgg}{\lambda_1\lambda_2}
\newcommand{\lww}{\lambda_3\lambda_4}
\newcommand{\ppin}{ P^+_{12}}
\newcommand{\pmin}{ P^-_{12}}
\newcommand{\ppout}{ P^+_{34}}
\newcommand{\pmout}{ P^-_{34}}
\newcommand{\sinsq}{\sin^2\theta}
\newcommand{\cossq}{\cos^2\theta}
\newcommand{\yt}{y_\theta}
\newcommand{\hppll}{++;00}
\newcommand{\hpmll}{+-;00}
\newcommand{\hpplt}{++;\lambda_30}
\newcommand{\hpmlt}{+-;\lambda_30}
\newcommand{\hpptt}{++;\lambda_3\lambda_4}
\newcommand{\hpmtt}{+-;\lambda_3\lambda_4}
\newcommand{\dk}{\Delta\kappa}
\newcommand{\klam}{\Delta\kappa \lambda_\gamma }
\newcommand{\kac}{\Delta\kappa^2 }
\newcommand{\lac}{\lambda_\gamma^2 }
\def\gamgamtzz{$\gamma \gamma \ra ZZ \;$}
\def\gamgamtww{$\gamma \gamma \ra W^+ W^-\;$}
\def\gamgamtwwe{\gamma \gamma \ra W^+ W^-}

%%%LOOPS%%%%%%%%%
\def\intfd{ \int \frac{d^4 r}{(2\pi)^4} }
\def\intnd{ \int \frac{d^n r}{(2\pi)^n} }
\def\intnmu{ \mu^{4-n} \int \frac{d^n r}{(2\pi)^n} }
\newcommand{\Dkm}{[(r+k)^2-m_2^2]}
\newcommand{\Dkom}{[(r+k_1)^2-m_2^2]}
\newcommand{\Dkotm}{[(r+k_1+k_2)^2-m_3^2]}
\def\piggt{$\Pi_{\gamma \gamma}\;$}
\def\pigg{\Pi_{\gamma \gamma}}
\newcommand{\mn}{{\mu \nu}}
\newcommand{\mzb}{M_{Z,0}}
\newcommand{\mzbs}{M_{Z,0}^2}
\newcommand{\mwb}{M_{W,0}}
\newcommand{\mwbs}{M_{W,0}^2}
\newcommand{\dgg}{\frac{\delta g^2}{g^2}}
\newcommand{\dee}{\frac{\delta e^2}{e^2}}
\newcommand{\dss}{\frac{\delta s^2}{s^2}}
\newcommand{\dmw}{\frac{\delta \mww}{\mww}}
\newcommand{\dmz}{\frac{\delta \mzz}{\mzz}}
\def\pigz{\Pi_{\gamma Z}}
\def\pizz{\Pi_{Z Z}}
\def\piww{\Pi_{WW}}
\def\pioo{\Pi_{11}}
\def\pitt{\Pi_{33}}
\def\pitq{\Pi_{3Q}}
\def\piqq{\Pi_{QQ}}
\def\delr{\Delta r}
\def\calm{{\cal {M}}}
\def\gww{G_{WW}}
\def\gzz{G_{ZZ}}
\def\goo{G_{11}}
\def\gtt{G_{33}}
\def\szz{s_Z^2}
\def\estk{e_\star^2(k^2)}
\def\sstk{s_\star^2(k^2)}
\def\cstk{c_\star^2(k^2)}
\def\sstz{s_\star^2(\mzz)}
\def\mzst{{M_Z^{\star}}(k^2)^2}
\def\mwst{{M_W^{\star}}(k^2)^2}
\def\epo{\varepsilon_1}
\def\epd{\varepsilon_2}
\def\ept{\varepsilon_3}
\def\dro{\Delta \rho}
\def\gmu{G_\mu}
\def\alpz{\alpha_Z}
\def\danpmz{\Delta\alpha_{{\rm NP}}(\mzz)}
\def\danpk{\Delta\alpha_{{\rm NP}}(k^2)}
\def\calt{{\cal {T}}}
\def\piggh{\pigg^h(s)}
\def\cuv{C_{UV}}
\def\pilr{G_{LR}}
\def\pill{G_{LL}}
\def\dak{\Delta \alpha(k^2)}
\def\damz{\Delta \alpha(\mzz)}
\def\dahmz{\Delta \alpha^{(5)}_{{\rm had}}(\mzz)}
\def\sth{s_{\theta}^2}
\def\cth{c_{\theta}^2}
%%%%% KATO-san shortcuts start here%%%%%%%%%%%%%%
% Macros---------------------------------------
\newcommand{\siki}[1]{Eq.\ref{eq:#1}}
\newcommand{\zu}[1]{Fig.\ref{fig:#1}}
\newcommand{\setu}[1]{Sec.\ref{sec:#1}}
% Macros:NLG-----------------------------------
\newcommand{\anlg}{\tilde\alpha}
\newcommand{\bnlg}{\tilde\beta}
\newcommand{\dnlg}{\tilde\delta}
\newcommand{\enlg}{\tilde\varepsilon}
\newcommand{\knlg}{\tilde\kappa}
\newcommand{\xiw}{\xi_W}
\newcommand{\xiz}{\xi_Z}
\newcommand{\dbr}{\delta_B}
\newcommand{\bothd}{{ \leftrightarrow \atop{\partial^{\mu}} } }

\newcommand{\BARE}[1]{\underline{#1}}
\newcommand{\ZF}[1]{\sqrt{Z}_{#1}}
\newcommand{\ZFT}[1]{\tilde{Z}_{#1}}
\newcommand{\ZH}[1]{\delta Z_{#1}^{1/2}}
\newcommand{\ZHb}[1]{\delta Z_{#1}^{1/2\,*}}
\newcommand{\DM}[1]{\delta M^2_{#1}}
\newcommand{\DMS}[1]{\delta M_{#1}}
\newcommand{\Dm}[1]{\delta m_{#1}}
\newcommand{\tree}[1]{\langle {#1}\rangle}

\newcommand{\Cuv}{C_{UV}}
\newcommand{\logw}{\log M_W^2}
\newcommand{\logz}{\log M_Z^2}
\newcommand{\logh}{\log M_H^2}
\newcommand{\swt}{s_W^2}
\newcommand{\cwt}{c_W^2}
\newcommand{\swf}{s_W^4}
\newcommand{\cwf}{c_W^4}
\newcommand{\MWt}{M_W^2}
\newcommand{\MZt}{M_Z^2}
\newcommand{\MHt}{M_H^2}

\newcommand{\VECsl}[1]{\not{#1}}

\newcommand{\Bphi}{\mbox{\boldmath$\phi$}}
%%%%% KATO-san shortcuts end here%%%%%%%%%%%%%%
\newcommand{\eetth}{$e^+ e^-\ra t \bar{t} H$}
\newcommand{\eettht}{$e^+ e^-\ra t \bar{t} H\;$}
\newcommand{\nnhet}{$\epem \ra \nu_e \bar{\nu}_e H \;$}
\newcommand{\nnhe}{$\epem \ra \nu_e \bar{\nu}_e H$}
\newcommand{\eezh}{$\epem \ra Z H$}
\newcommand{\eezht}{$\epem \ra Z H \;$}
\newcommand{\eezhh}{$\epem \ra Z  H H$}
\newcommand{\eezhht}{$\epem \ra Z H H\;$}
\newcommand{\eeeeht}{$\epem \ra e^+ e^- H \;$}
\newcommand{\eeeeh}{$\epem \ra e^+ e^-  H$}
\newcommand{\eenngt}{$\epem \ra e^+ e^- \gamma \;$}
\newcommand{\eenng}{$\epem \ra e^+ e^-  \gamma$}

%%KANEKO"S shortcut%%%%%%
\def\al{\alpha}
\def\bt{\beta}
\def\gm{\gamma}
\def\Gm{\Gamma}
\def\et{\eta}
\def\del{\delta}
\def\Del{\Delta}
\def\kp{\kappa}
\def\lm{\lambda}
\def\Lm{\Lambda}
\def\th{\theta}
\def\zt{\zeta}
\def\ro{\rho}
\def\sig{\sigma}
\def\Sig{\Sigma}
\def\eps{\epsilon}
\def\vare{\varepsilon}
\def\vphi{\varphi}
\def\om{\omega}
\def\Om{\Omega}
 %---------------------  general
\def\bar{\overline}
\def\d{{\rm d}}
\def\pdf{\partial}
\def\Int{\int\nolimits}
\def\det{{\rm det}}
\def\non{\nonumber}
%------------------------ \begin -- \end
\def\eqn{\begin{equation}}
\def\eqne{\end{equation}}
\def\eqa{\begin{eqnarray}}
\def\eqae{\end{eqnarray}}
\def\ary{\begin{array}}
\def\arye{\end{array}}
\def\dsc{\begin{description}}
\def\dsce{\end{description}}
\def\itm{\begin{itemize}}
\def\itme{\end{itemize}}
\def\enu{\begin{enumerate}}
\def\enue{\end{enumerate}}
\def\ct{\begin{center}}
\def\cte{\end{center}}
%-------------------------  specific
\def\D{{\cal D}}
\def\bfD{{\bf D}}

\newcommand{\cha}{{tt CHANEL}}

\def\sinb{\sin\beta}
\def\cosb{\cos\beta}
\def\sinbb{s_ {2\beta}}
\def\cosbb{c_{2 \beta}}
\def\tgb{\tan \beta}
\def\tgbt{$\tan \beta\;\;$}
\def\tgbsq{\tan^2 \beta}
\def\tgbsqt{$\tan^2 \beta\;$}
\def\sinal{\sin\alpha}
\def\cosal{\cos\alpha}
%%%short notation of angles%%%%%
\def\sb{s_\beta}
\def\cb{c_\beta}
\def\tb{t_\beta}
\def\ttb{t_{2 \beta}}
\def\sa{s_\alpha}
\def\ca{c_\alpha}
\def\ta{t_\alpha}
\def\stb{s_{2\beta}}
\def\ctb{c_{2\beta}}
\def\sbb{s_ {2\beta}}
\def\cbb{c_{2 \beta}}
\def\sta{s_{2\alpha}}
\def\cta{c_{2\alpha}}
\def\sbma{s_{\beta-\alpha}}
\def\cbma{c_{\beta-\alpha}}
\def\sbpa{s_{\beta+\alpha}}
\def\cbpa{c_{\beta+\alpha}}
\def\lone{\lambda_1}
\def\ltwo{\lambda_2}
\def\lthree{\lambda_3}
\def\lfour{\lambda_4}
\def\lfive{\lambda_5}
\def\lsix{\lambda_6}
\def\lseven{\lambda_7}
%%%%%%%%%%%%%%%%%%%%%%%%5
\def\stop{\tilde{t}}
\def\sto{\tilde{t}_1}
\def\stt{\tilde{t}_2}
\def\stl{\tilde{t}_L}
\def\str{\tilde{t}_R}
\def\msto{m_{\sto}}
\def\mstosq{m_{\sto}^2}
\def\mstt{m_{\stt}}
\def\msttsq{m_{\stt}^2}
\def\mt{m_t}
\def\mtsq{m_t^2}
\def\sint{\sin\theta_{\stop}}
\def\sintt{\sin 2\theta_{\stop}}
\def\cost{\cos\theta_{\stop}}
\def\sintsq{\sin^2\theta_{\stop}}
\def\costsq{\cos^2\theta_{\stop}}
\def\mqtt{\M_{\tilde{Q}_3}^2}
\def\mutt{\M_{\tilde{U}_{3R}}^2}
%%%%%%%%%%%%%%%%%%%%%
\def\sbottom{\tilde{b}}
\def\sbo{\tilde{b}_1}
\def\sbt{\tilde{b}_2}
\def\sbl{\tilde{b}_L}
\def\sbr{\tilde{b}_R}
\def\msbo{m_{\sbo}}
\def\msbosq{m_{\sbo}^2}
\def\msbt{m_{\sbt}}
\def\msbtsq{m_{\sbt}^2}
\def\mt{m_t}
\def\mtsq{m_t^2}
%%%%%%%%%%%%%%%%%%%%%
\def\selectron{\tilde{e}}
\def\seo{\tilde{e}_1}
\def\set{\tilde{e}_2}
\def\sel{\tilde{e}_L}
\def\ser{\tilde{e}_R}
\def\mseo{m_{\seo}}
\def\mseosq{m_{\seo}^2}
\def\mset{m_{\set}}
\def\msetsq{m_{\set}^2}
\def\msel{m_{\sel}}
\def\mser{m_{\ser}}
\def\me{m_e}
\def\mesq{m_e^2}
%%%%%%%%%%%%%%%%%%%%%
\def\snu{\tilde{\nu}}
\def\snue{\tilde{\nu_e}}
\def\set{\tilde{e}_2}
\def\snul{\tilde{\nu}_L}
\def\msnue{m_{\snue}}
\def\msnuesq{m_{\snue}^2}
%%%%%%%%%%%%%%%%%%%%%
\def\smuon{\tilde{\mu}}
\def\smul{\tilde{\mu}_L}
\def\smur{\tilde{\mu}_R}
\def\msmul{m_{\smul}}
\def\msmulsq{m_{\smul}^2}
\def\msmur{m_{\smur}}
\def\msmursq{m_{\smur}^2}
%%%%%%%%%%%%%%%%%%%%%%%%%%
\def\stau{\tilde{\tau}}
\def\stauo{\tilde{\tau}_1}
\def\staut{\tilde{\tau}_2}
\def\staul{\tilde{\tau}_L}
\def\staur{\tilde{\tau}_R}
\def\mstauo{m_{\stauo}}
\def\mstauosq{m_{\stauo}^2}
\def\mstaut{m_{\staut}}
\def\mstautsq{m_{\staut}^2}
\def\mtau{m_\tau}
\def\mtausq{m_\tau^2}
%%%%%%%%%%%%%%%%%%%%%%%%%%%%%%
\def\gluino{\tilde{g}}
\def\mgluino{m_{\tilde{g}}}
\def\mchi{m_\chi^+}
\def\neuto{\tilde{\chi}_1^0}
\def\mneuto{m_{\tilde{\chi}_1^0}}
\def\neutt{\tilde{\chi}_2^0}
\def\mneutt{m_{\tilde{\chi}_2^0}}
\def\neutth{\tilde{\chi}_3^0}
\def\mneutth{m_{\tilde{\chi}_3^0}}
\def\neutf{\tilde{\chi}_4^0}
\def\mneutf{m_{\tilde{\chi}_4^0}}
\def\chargop{\tilde{\chi}_1^+}
\def\mchargo{m_{\tilde{\chi}_1^+}}
\def\chargtp{\tilde{\chi}_2^+}
\def\mchargt{m_{\tilde{\chi}_2^+}}
\def\chargom{\tilde{\chi}_1^-}
\def\chargtm{\tilde{\chi}_2^-}
\def\bino{\tilde{b}}
\def\wino{\tilde{w}}
\def\photino{\tilde{\gamma}}
\def\zino{tilde{z}}
%%%%%%%%%%%%%%%%%%%%%%%%%%%%%%%%%
\def\sdowno{\tilde{d}_1}
\def\sdownt{\tilde{d}_2}
\def\sdownl{\tilde{d}_L}
\def\sdownr{\tilde{d}_R}
\def\supo{\tilde{u}_1}
\def\supt{\tilde{u}_2}
\def\supl{\tilde{u}_L}
\def\supr{\tilde{u}_R}
%%%%%%%%%%Higgses masses%%%%%%%%%%%%
\def\mh{m_h}
\def\mht{m_h^2}
\def\MH{M_H}
\def\MHt{M_H^2}
\def\MA{M_A}
\def\MAt{M_A^2}
\def\MHp{M_H^+}
\def\MHm{M_H^-}
%%%%%%%%%%NEEDED FOR rge%%%%%%%%%%%%
\def\mqt{\M_{\tilde{Q}_3}}
\def\mut{\M_{\tilde{U}_{3R}}}
\def\mqtz{\M_{\tilde{Q}_3(0)}}
\def\mutz{\M_{\tilde{U}_{3R}(0)}}
\def\mqtzt{\M_{\tilde{Q}_3^2(0)}}
\def\mutzt{\M_{\tilde{U}_{3R}^2(0)}}

\def\mhf{M_{1/2}}

\renewcommand{\topfraction}{0.85}
\renewcommand{\textfraction}{0.1}
\renewcommand{\floatpagefraction}{0.75}
\newcommand{\drbar}{{\overline{\rm DR}}}

\def\dtb{\delta t_\beta}

\def\xb{s_{\beta}}

\begin{titlepage}

\vspace*{0.1cm}\rightline{LAPTH-1250/08}

%\today

\vspace{1mm}
\begin{center}

{\Large {\bf Automatised full one-loop renormalisation of the
MSSM\\
I: The Higgs sector, the issue of $\tgb$ and gauge invariance}}

\vspace{.5cm}

{\large N.~Baro${}^{1)}$, F.~Boudjema${}^{1)}$ and A.~Semenov${}^{2)}$ }\\

\vspace{4mm}

%\vspace{4mm}

{\it 1) LAPTH, Universit\'e de Savoie, CNRS, \\
BP 110, F-74941 Annecy-le-Vieux Cedex, France}
\\ {\it
2) Joint Institute of Nuclear Research, JINR, 141980 Dubna, Russia }\\

\vspace{10mm}

\abstract{\noi We give an extensive description of the renormalisation
of the Higgs sector of the minimal supersymmetric model in {\tt
SloopS}. {\tt SloopS} is an automatised code for the computation
of one-loop processes in the MSSM. In this paper, the first in a
series, we study in detail the non gauge invariance of some
definitions of $\tgb$. We rely on a general non-linear gauge
fixing constraint to make the gauge parameter dependence of
different schemes for $\tgb$ at one-loop explicit. In so doing, we
update, within these general gauges, an important
Ward-Slavnov-Taylor identity on the mixing between the
pseudo-scalar Higgs, $A^0$, and the $Z^0$. We then compare the
$\tgb$ scheme dependence of a few observables. We find that the
best $\tgb$ scheme is the one based on the decay $A^0 \ra \tau^+
\tau^-$ because of its gauge invariance, being unambiguously
defined from a physical observable, and because it is numerically
stable. The oft used $\bar{{\rm DR}}$ scheme performs almost as
well on the last count, but is usually defined from non-gauge
invariant quantities in the Higgs sector. The use of the heavier
scalar Higgs mass in {\it lieu} of $\tgb$ though related to a
physical parameter induces too large radiative corrections in
many instances and is therefore not recommended. }

\end{center}

%\vspace*{\fill} {\bf \Large \today}

\normalsize

\end{titlepage}

\section{Introduction}
Were it not for the radiative corrections to the lightest Higgs
mass \cite{RCmh1st}, the minimal supersymmetric model or MSSM would
have been a forgotten elegant model a long time ago. Indeed, at
tree-level the mass of the lightest Higgs is predicted to be less
than the mass of the $Z^{0}$ boson, $M_{Z^0}$. That would have been a
real pity from a model whose most appealing and foremost motivation
was to solve the hierarchy problem and make the Higgs more natural,
beside providing a very good Dark Matter candidate. The
renormalisation of the Higgs sector of the MSSM is therefore
important. It is also important because it provides a link to the
other parameters of the Standard Model, namely all the masses of the
particles. It also encodes another parameter that can be seen to
describe the relative scale of the two vacuum expectation values
needed for each Higgs doublet of the \smp, often referred to as
$\tgb$ and which permeates all the other sectors of the MSSM: the
gaugino/higgsino sector and the sfermion sector. Many
renormalisation schemes or definitions of this parameter are
unsatisfactory, as we will see, mainly because they lack a direct
physical interpretation or do not correspond to a physical and gauge
independent parameter.\\

\noi The aim of this paper is to give an extensive description of the
renormalisation of the Higgs sector in {\tt SloopS} at one-loop.
{\tt SloopS} is a fully automated code for the one-loop calculation
of any cross section or decay in the MSSM at one-loop. Although
there have been a few studies of the renormalisation of the Higgs
sector, (see \cite{Djouadi_Susy_Higgs,MartinReview} for a recent
review of the Higgs in supersymmetry) some performed even beyond the
one-loop approximation especially as concerns the mass of the
lightest CP-even Higgs \cite{mh-2loop,mh-3loop}, looking at the
problem afresh while keeping the issue of gauge invariance in mind,
will prove rewarding. Moreover our motivation in developing {\tt
SloopS} was also to have a {\em full} one-loop renormalisation of
all the sectors of the MSSM in a coherent way and therefore the
study of the Higgs sector is a first step. We will point at the non
gauge invariance of some definitions of $\tgb$. Although this has
been known, see for example \cite{stockinger02}, and pointed out at
two-loop in the usual linear gauge \cite{Yamada-2looptb}, most
practitioners have kept the usage of some non-gauge invariant
definitions of $\tgb$ because of their simplicity at the technical
level being based on definitions involving two-point function
self-energies. With the automatisation of the loop calculations,
considerations and definitions of $\tgb$ based on three-point
functions (decays) are hardly more involved than those based solely
on two-point functions describing self-energies, including
transitions.

\noi In the approach adopted within {\tt SloopS}, we strive for an
on-shell, OS, renormalisation scheme in particular for $\tgb$. We
rely on a general non-linear gauge fixing constraint to make the
gauge parameter dependence of different schemes for $\tgb$ at
one-loop explicit. In so doing we rederive and update the
Ward-Slavnov-Taylor identity on the $A^0 Z^0/H^\pm W^\pm$ mixing in
the non-linear gauge. We then compare qualitatively and
quantitatively the $\tgb$ scheme dependence of a few observables.
$A^0$ is the CP-odd Higgs scalar and $H^\pm$ are the charged
Higgses. We find that the best $\tgb$ scheme is the one based on the
decay $A^0 \ra \tau^+ \tau^-$ because of its gauge invariance, being
unambiguously defined from a physical observable, and because it is
numerically stable. The oft used $\bar{{\rm DR}}$ scheme performs
almost as well on the last count, but is usually defined from
non-gauge invariant quantities in the Higgs sector. The use of the
heavier CP-even scalar Higgs mass in {\it lieu} of $\tgb$ though
related to a physical parameter induces too large radiative
corrections in many instances and is therefore not acceptable. It
has been argued that the definitions within the Higgs sector may be
considered universal compared to a definition involving a particular
Higgs decay for example. However, as stressed
in \cite{DavidsonHaber}, staying within the confines of the Higgs
sector and the Higgs potential, one faces the issue that many
 definitions may be basis
dependent, as we will see this will translate at one-loop into
issues about gauge invariance for these definitions. As concerns the
application to the corrections to the lightest Higgs mass our
one-loop treatment is certainly not up-to-date, however our
motivation is to stress the gauge dependence issues and compare the
impact of the scheme dependence for $\tgb$ for many observables
starting for those directly related to the properties of the Higgses
of the MSSM, before reviewing in our forthcoming
studies \cite{Sloops-allren} the impact of $\tgb$ on observables
in the charginos/neutralinos as well as the sfermion sectors. We
feel that this issue is of importance as is a consistent one-loop OS
implementation.

\noi The present paper is organised as follows. In Section~2 we review
the Higgs sector of the MSSM at tree-level. This may, by now, be
considered trivial but it is a necessary step before we embark on
the renormalisation procedure. We also detail this part in order to
show what might qualify as a physical basis independent observable.
Section~3 presents the non-linear gauge fixing condition that we
use. This includes $8$ gauge fixing parameters which are crucial in
studying many issues related to gauge invariance that are not easily
uncovered when one works within the usual linear gauge. Section~4
constitutes the theoretical core of our analyses and deals with
renormalisation, introducing counterterms for the Lagrangian
parameters and the field renormalisation constants. We expose our
renormalisation conditions and update the Slavnov-Taylor identities
involving the $A^0-Z^0$ and $H^{\pm}-W^{\pm}$ transitions. Section~5 is
devoted to defining $\tgb$. We consider a few schemes. Before
turning to applications and numerical results we briefly describe
how our automatic code is set-up in Section~6. In Section~7 a
numerical investigation of the scheme dependence and gauge
dependence of these schemes is studied taking as examples loop
corrections to Higgs masses, decays of the Higgses to fermions and
to gauge and Higgs bosons. Section~8 gives our conclusions. The
paper contains two appendices. Appendix~A details the derivation of
Slavnov-Taylor identity for the $A^0-Z^0$ transition. Field
renormalisation may be introduced at the level of the {\em
unphysical} fields before rotation to the physical fields is
performed, Appendix~B relates these field renormalisation constants
on the Higgs fields to the one we introduce directly after the
physical fields are defined. This may
help in comparing different approaches in the literature. \\

\noi To avoid clutter we use some abbreviations for the trigonometric
functions. For example for an angle $\theta$, $\cos \theta$ will be
abbreviated as $c_\theta$, {\it etc}... so that we will from now on
use $\tb$ for $\tan \beta$.

\section{The Higgs sector of the MSSM at tree-level}

\subsection{The Higgs Potential}
As known, see for instance \cite{MartinReview}, the MSSM requires two
Higgs doublets $H_1$ and $H_2$ with opposite hypercharge. The Higgs
potential in the MSSM is given by:
\begin{eqnarray}
V&=&m_{1}^{2}|H_{1}|^2+m_{2}^{2} |H_{2}|^2+m_{12}^{2}(H_{1}\wedge H_{2}+h.c.)\\
&+&\frac{1}{8}(g^{2}+g'^{2})(|H_{1}|^2-|H_{2}|^2)^{2}
+\frac{g^{2}}{2}|H_{1}^{\dagger}H_{2}|^{2}\nonumber\\
&\quad&\textrm{with}\quad H_{1}\wedge
H_{2}=H_{1}^{a}H_{2}^{b}\epsilon_{ab}\quad
(\epsilon_{12}=-\epsilon_{21}=1, \epsilon_{ii}=0)\, .\nonumber
\end{eqnarray}

\noindent The mass terms are all soft masses even if both $m_1^2$
and $m_2^2$ contain the SUSY preserving $|\mu|^2$ term which
originates from the F-terms. $g,g^\prime$ are, respectiveley, the
$SU(2)_W$ and $U(1)_Y$ gauge couplings. Decomposing each Higgs
doublet field $H_{1,2}$ in terms of its components,
\begin{eqnarray}
H_{1}&=&\left(\begin{array}{c}
H_{1}^{0}\\H_{1}^{-}\end{array}\right)=\left(\begin{array}{c}
(v_{1}+\phi_{1}^{0}-i\varphi_{1}^{0})/\sqrt{2} \\
-\phi_{1}^{-}\end{array}\right) \, ,\\
%%%%%%%%%%%%%%
H_{2}&=&\left(\begin{array}{c}
H_{2}^{+}\\H_{2}^{0}\end{array}\right)=\left(\begin{array}{c}
\phi_{2}^{+} \\ (v_{2}+\phi_{2}^{0}+i\varphi_{2}^{0})/\sqrt{2}
\end{array}\right)\, ,
\label{doublet_H1H2}
\end{eqnarray}
\noi the tree-level Higgs potential writes as
\begin{eqnarray}
V=V_{const}+V_{linear}+V_{mass}+V_{cubic}+V_{quartic} \, ,
\end{eqnarray}
where,
\begin{eqnarray}
V_{linear}&=&\left(m_{1}^{2}v_{1}+m_{12}^{2}v_{2}+\frac{g^{2}+g'^{2}}{8}(v_{1}^{2}-v_{2}^{2})v_{1}\right)\phi_{1}^{0}\nonumber\\
&+&\left(m_{2}^{2}v_{2}+m_{12}^{2}v_{1}-\frac{g^{2}+g'^{2}}{8}(v_{1}^{2}-v_{2}^{2})v_{2}\right)\phi_{2}^{0}\nonumber\\
&\equiv&T_{\phi_{1}^{0}}\phi_{1}^{0}+T_{\phi_{2}^{0}}\phi_{2}^{0} \, ,
\end{eqnarray}
\noi and
%%%%%%%%%%%%%%
\begin{eqnarray}
V_{mass}&=&\frac{1}{2}\left(\begin{array}{cc} \varphi_{1}^{0}&
\varphi_{2}^{0}
\end{array}\right)\underbrace{\left(\begin{array}{cc}
m_{1}^{2}+\frac{g^{2}+g'^{2}}{8}(v_{1}^{2}-v_{2}^{2})&m_{12}^{2}\\
m_{12}^{2}&m_{2}^{2}-\frac{g^{2}+g'^{2}}{8}(v_{1}^{2}-v_{2}^{2})\end{array}\right)}_{M_{\varphi^{0}}^{2}}\left(\begin{array}{c}
\varphi_{1}^{0}\\ \varphi_{2}^{0}\end{array}\right)\nonumber\\
&+&\frac{1}{2}\left(\begin{array}{cc}\phi_{1}^{0}&\phi_{2}^{0}\end{array}\right)\underbrace{\Bigg(M_{\varphi^{0}}^{2}+
\frac{g^{2}+g'^{2}}{4}\left(\begin{array}{cc}v_{1}^{2}&-v_{1}v_{2}\\-v_{1}v_{2}&v_{2}^{2}\end{array}\right)\Bigg)}
_{M_{\phi^{0}}^{2}}
\left(\begin{array}{c}\phi_{1}^{0}\\
\phi_{2}^{0} \end{array}\right)
\nonumber\\
&+&\left(\begin{array}{cc}
\phi_{1}^{-}&\phi_{2}^{-}\end{array}\right)\underbrace{\Bigg(M_{\varphi^{0}}^{2}+\frac{g^{2}}{4}\left(\begin{array}{cc}v_{2}^{2}&-v_{1}v_{2}\\
-v_{1}v_{2}&v_{1}^{2}\end{array}\right)\Bigg)}_{M_{\phi^{\pm}}^{2}}\left(\begin{array}{c}\phi_{1}^{+}\\
\phi_{2}^{+} \end{array}\right) \, .\label{Vmass}
\end{eqnarray}
It is illuminating to express the mass matrices in terms of the
tadpoles especially for the pseudoscalar states
\begin{eqnarray}
M_{\varphi^{0}}^{2}&=&\left(\begin{array}{cc}
\frac{T_{\phi_{1}^{0}}}{v_1}& 0 \\
0 & \frac{T_{\phi_{2}^{0}}}{v_2}\end{array}\right)\;-\;
\frac{m_{12}^{2}}{v_1 v_2} N_{GP} \quad {\rm with} \quad
N_{GP}=\left(\begin{array}{cc}
v_2^2& -v_1 v_2 \\
-v_1 v_2 & v_1^2 \end{array}\right) \, , \nonumber \\
M_{\phi^{c}}^{2}&=&\left(\begin{array}{cc}
\frac{T_{\phi_{1}^{0}}}{v_1}& 0 \\
0 & \frac{T_{\phi_{2}^{0}}}{v_2}\end{array}\right)\;-\;
\left(\frac{m_{12}^{2}}{v_1 v_2}-\frac{g^2}{4} \right) N_{GP} \, .
\end{eqnarray}

\noi The requirement that $v_1$ and $v_2$ correspond to the true
vacua is a requirement on the vanishing of the tadpoles. The
tadpoles, by the way, are also a trade-off for $m_1^2$ and $m_2^2$.
Indeed note that expressing everything in terms of
$T_{\phi_{1,2}^{0}}$, all explicit dependence on $m_1^2$ and $m_2^2$
has disappeared, even in the scalar (CP-even) sector. Note that once
the tadpole condition has been imposed
\beqn
T_{\phi_{1,2}^{0}}=0 \, ,
\eeqn
we immediately find that in both the charged sector and
pseudo-scalar sector, there is a Goldstone boson ({\it i.e.} a zero
mass eigenvalue). This is immediate from the fact that
\beqn
\det(N_{GP})=0 \, .
\eeqn
\noi The masses of the physical charged Higgs, $M_{H^\pm}$ and the
pseudoscalar Higgs, $M_{A^0}$, are then just set from {\em the
invariant} obtained from
\beqn \tr (N_{GP})=v_1^2 + v_2^2=v^2 \, ,
\eeqn
which is another way of seeing that the combination $v$ is a proper
``observable". Indeed after gauging we will find that the masses of
the weak gauge bosons are
\beqn
M_{W^{\pm}}^2&=&\frac{1}{4} g^2 v^2 \, , \nonumber \\
M_{Z^{0}}^2&=& \frac{1}{4} (g^2+g'^2) v^2 \, .
\eeqn
Then
\begin{eqnarray}
\label{ma-m12} M_{A^{0}}^2&=& \tr \left( M_{\varphi^{0}}^{2} \right)= -
m_{12}^2 \frac{v^2}{v_1 v_2}=m_1^2+m_2^2 \, ,\\
\label{mhp-tree} M_{H^{\pm}}^2&=&M_{A^0}^2 + M_{W^{\pm}}^2 \, .
\end{eqnarray}
\noi In Eq.~(\ref{ma-m12}), the first equality does show an implicit
dependence on the ratio of vev ($t_\beta$), but not through
$m_1^2+m_2^2$. The latter must be basis independent, as is the {\em
combination} $m_{12}^2/v_1 v_2$. This is to keep in mind.

\noi It is also interesting to note that for the scalar Higgses,
there is a simple sum rule that does not involve any ratio of vev's.
Indeed, taking the trace of $M_{\phi^0}^2$ and call the two physical
CP-even Higgses $h^{0}$, with mass $M_{h^0}$, and $H^{0}$, with mass
$M_{H^0}$, that would be obtained after rotation, we get the sum rule
\begin{eqnarray}
M_{h^0}^2 + M_{H^0}^2=M_{A^0}^2+M_{Z^0}^2 \, . \label{trace-mh}
\end{eqnarray}
\noi $h^{0}$ will denote the lightest CP-even Higgs. Let us as a
\underline{book-keeping device} introduce the angle $\beta$. At the
moment this is just to help have easy notations:
\begin{eqnarray}
c_\beta=\frac{v_1}{v}\, , \quad s_\beta=\frac{v_2}{v} \quad \textrm{with}\; v=\sqrt{v_{1}^2+v_{2}^2}\, .
\label{def_beta_vs}
\end{eqnarray}
The determinant of the scalar Higgses on the other hand gives
\begin{eqnarray}
M_{h^0}^2 \; M_{H^0}^2=M_{A^0}^2 \; M_{Z^0}^2 \; c_{2\beta}^2 \, . \label{det-mh}
\end{eqnarray}
This shows that if we take $M_{H^0}, M_{A^0}, M_{Z^0}$ as input parameters, we
first derive $M_{h^0}$ from Eq.~(\ref{trace-mh}), then $c_{2\beta}^2$
from Eq.~(\ref{det-mh}). In general with a set of input parameters
$M_{H^0},M_{A^0},M_{Z^0}$, $c_{2\beta}^{2} \leq 1$ is not guaranteed though. We
could of course fix $c_{2\beta}^2$ ($t_{\beta}$) and derive
\underline{$M_{H^0}$ and
$M_{h^0}$} which is what is usually done. \\
The soft SUSY breaking mass parameters $m_{1,2,12}^{2}$ can be
expressed in terms of the physical quantities, $M_{A^0}, M_{Z^0}$
and $c_\beta$ (as for example derived from
Eqs.~(\ref{trace-mh}-\ref{det-mh})):
\begin{eqnarray}
m_{1}^{2}&=&s_{\beta}^{2}M_{A^0}^{2}-\frac{1}{2}c_{2\beta} M_{Z^0}^{2} \, ,\\
m_{12}^{2}&=&-\frac{1}{2}s_{2\beta} M_{A^0}^{2} \, ,\\
m_{2}^{2}&=&c_{\beta}^{2}M_{A^0}^{2}+\frac{1}{2}c_{2\beta} M_{Z^0}^{2} \, .
\end{eqnarray}

\subsection{Basis and rotations}
So far the properties of the physical fields like their masses
have been derived without reverting to a specific basis. The angle
$\beta$ defined in Eq.~(\ref{def_beta_vs}) was just a book-keeping
device. Still, to go from the fields at the Lagrangian level to
the physical fields one needs to perform a rotation. This should
have no effect on physical observables. This naive observation is
important especially when we move to one-loop. The rotations we
will perform will get rid of field mixing. With the tadpole
condition set to zero, it is clear that the pseudoscalar and
charged scalars eigenstates are diagonalised through the same
unitary matrix. At tree-level this is defined precisely through
the same angle $\beta$ as in Eq.~(\ref{def_beta_vs}),
\begin{eqnarray}
N_{GP}=U(-\beta) \left(\begin{array}{cc}
0 &0\\
0 & 1\end{array}\right) U(\beta)\, , \quad
U(\beta)=\left(\begin{array}{cc}
c_\beta &s_\beta \\
-s_\beta & c_\beta\end{array}\right)\, , \quad
U^\dagger(\beta)=U(-\beta) \, .
\end{eqnarray}
Call ${\cal T}_v$, the tadpole matrix defined as
\begin{eqnarray}
{\cal T}_v=\left(\begin{array}{cc}
\frac{T_{\phi_{1}^{0}}}{v_1}& 0 \\
0 & \frac{T_{\phi_{2}^{0}}}{v_2}\end{array}\right) \, .
\end{eqnarray}
The tadpole is, of course, set to zero. But we will leave this zero
there in the notation as we will need this when we go to the
one-loop counterterms. Then the mass matrices for the CP-even,
CP-odd and charged scalars write
\begin{eqnarray}
M_{\varphi^{0}}^{2}&=&{\cal T}_v+M_{A^0}^2 N_{GP} \, ,\label{HiggsMassMatrix1}\\
M_{\phi^{\pm}}^{2}&=&{\cal T}_v+(M_{A^0}^2+M_{W^{\pm}}^2) N_{GP} \, ,\label{HiggsMassMatrix2}\\
M_{\phi^{0}}^{2}&=&{\cal T}_v+M_{A^0}^2 N_{GP} +M_{Z^0}^2 U(\beta)
\left(\begin{array}{cc}
1& 0 \\
0& 0 \end{array}\right) U(-\beta) \, . \label{HiggsMassMatrix3}
\end{eqnarray}
The neutral Higgs is diagonalised through a rotation $\alpha$ such
that
\begin{eqnarray}
U(\alpha) M_{\phi^{0}}^{2} U(-\alpha)&=&\left(\begin{array}{cc}
M_{H^0}^2& 0 \\
0& M_{h^0}^2 \end{array}\right)= U(\alpha) {\cal T}_v U(-\alpha) + \\
& &M_{A^0}^2 U(\alpha-\beta) \left(\begin{array}{cc}
0& 0 \\
0& 1 \end{array}\right) U(\beta-\alpha) +M_{Z^0}^2
U(\alpha+\beta) \left(\begin{array}{cc}
1& 0 \\
0& 0 \end{array}\right) U(-(\alpha+\beta)) \, . \nonumber
\end{eqnarray}
The diagonalisation procedure also produces other, sometimes
useful, constraints and relations:
\begin{eqnarray}
M_{H^0}^2&=&M_{A^0}^2 s_{\alpha-\beta}^2 +M_{Z^0}^2 c_{\alpha+\beta}^2 \, ,\\
M_{h^0}^2&=&M_{A^0}^2 c_{\alpha-\beta}^2 +M_{Z^0}^2 s_{\alpha+\beta}^2 \, ,\\
M_{A^0}^2 s_{2(\alpha-\beta)} &=& M_{Z^0}^2 c_{2(\alpha+\beta)} \, ,
 \\
t_{2\alpha}&=&t_{2\beta} \frac{M_{A^0}^2+M_{Z^0}^2}{M_{A^0}^2-M_{Z^0}^2} \, .
\end{eqnarray}
Note that in the decoupling limit, $M_{A^0} \gg M_{Z^0}$, one has in
effect decoupled one of the Higgs doublet, the other has the
properties of the SM Higgs doublet. The decoupling parameter is
also measured with the parameter $\cbma \ra M_{Z^0}^2/M_{A^0}^2$ for $M_{A^0}
\gg M_{Z^0}$.\\
Therefore, the mass eigenstates in the Higgs sector are given by
\begin{eqnarray}
\left(\begin{array}{c} G^{0}\\A^{0}\end{array}\right)
&=&U(\beta) \left(\begin{array}{c} \varphi_{1}^{0}\\
\varphi_{2}^{0}\end{array}\right)=\left(\begin{array}{cc} c_{\beta} & s_{\beta}\\
-s_{\beta} &
c_{\beta}\end{array}\right) \left(\begin{array}{c} \varphi_{1}^{0}\\
\varphi_{2}^{0}\end{array}\right) \, ,\nonumber\\
%%%%%%%%%%%%%%%%%%%%%%%%%%%%%%%%%%
\left(\begin{array}{c} G^{\pm}\\H^{\pm}\end{array}\right)
&=&U(\beta) \left(\begin{array}{c} \phi_{1}^{\pm}\\
\phi_{2}^{\pm}\end{array}\right)=\left(\begin{array}{cc} c_{\beta} & s_{\beta}\\
-s_{\beta} &
c_{\beta}\end{array}\right) \left(\begin{array}{c} \phi_{1}^{\pm}\\
\phi_{2}^{\pm}\end{array}\right) \, , \nonumber\\
%%%%%%%%%%%%%%%%%%%%%%%%%%%%%%%%%%
\left(\begin{array}{c} H^{0}\\h^{0}\end{array}\right)
&=&U(\alpha) \left(\begin{array}{c} \phi_{1}^{0}\\
\phi_{2}^{0}\end{array}\right)=\left(\begin{array}{cc} c_{\alpha} & s_{\alpha}\\
-s_{\alpha} &
c_{\alpha}\end{array}\right) \left(\begin{array}{c} \phi_{1}^{0}\\
\phi_{2}^{0}\end{array}\right) \, .\label{diagonal_beta_alpha}
\end{eqnarray}

\subsection{Counting parameters}
Before we embark on the technicalities of renormalisation and the
choice of judicious input parameters, it is best to review how we
are going to proceed in general and how to make contact with the
renormalisation of the SM. This will help clarify what are the
fundamental parameters and which are the physical parameters that
can be used for a legitimate renormalisation scheme. Moreover since
some observables {\em belong} to the SM, like the $W^{\pm}$, $Z^{0}$
masses and the electromagnetic coupling constant $e$ which are used
as physical input parameters in the OS scheme, isolating these three
parameters means that their renormalisation will proceed exactly as
within the OS renormalisation of the SM, see \cite{grace-1loop} for
details.

\noi In the SM, the fundamental parameters at the Lagrangian level for
the gauge sector are $g$ and $g'$. The Higgs potential with the
Higgs doublet ${\cal H}$

\begin{eqnarray}
V({\cal H})&=&-\mu^{2} {\cal H}^{\dag} {\cal H} + \lambda({\cal
H}^{\dag} {\cal H})^2 \quad {\rm with} \nonumber \\
\left|\left< 0|{\cal H}|0 \right> \right|^2&=&\frac{v^2}{2} \neq 0
\, ,
\end{eqnarray}

\noi furnishes the following: $\mu^2$ (the ``Higgs mass''), $\lambda$ (the Higgs self coupling) and $v$ (the value of the vacuum expectation value). We thus have at
Lagrangian level, $5$ parameters between the Higgs sector and the
gauge sector. $\mu^2, \lambda,v$ are not all independent. $v$, the
vacuum expectation value (vev), is defined as the minimum of the
potential, this is equivalent to requiring no tadpoles. The no
tadpole requirement amounts to no terms linear in the scalar Higgs.
With the tadpole defined as $T$, we have at tree-level
\beqn
T=v (\mu^2-\lambda v^2) \rightarrow 0 \, .
\eeqn
\noi This requirement is to be carried to any loop level. Out of
this constraint, the $5$ physical parameters in the OS scheme are
$e, M_{W^{\pm}},M_{Z^0},M_{H^0},T$. At all orders one defines,
$c_W=M_{W^{\pm}}/M_{Z^0}$. The latter is not an independent physical
parameter. Therefore in the SM a one-to-one mapping between the
physical set $e, M_{W^{\pm}},M_{Z^0},M_{H^0},T$ and the Lagrangian
parameters $g,g^\prime,v,\mu,\lambda$ is made.

\noi In the MSSM, the Higgs sector furnishes $m_1^2,m_2^2,m_{12}^2$ the
Higgs doublets soft masses and $v_1,v_2$ the vev of the Higgs
doublets . The gauge sector is still governed by the $U(1)_Y$ and
$SU(2)_{W}$ gauge couplings $g,g'$. The requirement of no tadpoles
from both Higgs doublets, and hence any linear combination of them,
is also a strong constraint. From these seven parameters in all, the
physical parameters are usually split between the SM physical
On-Shell parameters
\beqn
e, M_{W^\pm},M_{Z^0} \, ,
\eeqn
which are a trade-off for $g,g',v^2=v_1^2+v_2^2$ and the MSSM Higgs
parameters
\beqn
M_{A^0},T_{\phi_1^{0}}, T_{\phi_2^{0}}; ``t_\beta" \,
,\label{input-higgs}
\eeqn
which are a trade-off for $m_1^2,m_2^2,m_{12}^2,v_2/v_1$. At
tree-level we can set $\tb=v_2/v_1$ but this is, as yet, not
directly related to an observable. While $v$ can directly be
expressed as a physical gauge boson mass, the ratio $v_2/v_1$ within
the Higgs sector does not have an immediate simple physical
interpretation. Hence the difficulty with this Lagrangian parameter.
One possibility is to trade it with the mass of one of the CP-even
neutral Higgs through Eq.~(\ref{det-mh}).

\section{Non-linear gauge fixing}
\def\db{\delta_{\rm BRS}}
In {\tt SloopS} we have generalised the usual 't Hooft linear
gauge condition to a more general non-linear gauge that involves,
thanks to the extra scalars in the Higgs sector, eight extra
parameters
$(\tilde{\alpha},\tilde{\beta},\tilde{\delta},\tilde{\omega},\tilde{\kappa},\tilde{\rho},
\tilde{\epsilon},\tilde{\gamma})$. Such gauges within the Standard
Model had proved useful and powerful \cite{chopin-nlg, grace-1loop}
to check the correctness of the calculation. We have also
exploited these gauges in the one-loop calculation of $\neuto
\neuto \ra \gamma \gamma, Z^{0}\gamma$ \cite{sloopsgg} and to
corrections to the relic density in \cite{baro07}. A
7-parameter non-linear gauge-fixing Lagrangian based on the one we
introduce here is used in \cite{Grace-susy1loop}. We can extend
this non-linear gauge fixing so that the gauge-fixing function
involves the sfermions also. We refrain, in this paper, from
working through this generalisation. \\
We will take these gauge fixing terms to be
\textit{renormalised}. In particular the gauge functions involve
the physical fields. Although this will not make all Green's
functions finite, it is enough to make all $S$-matrix elements
finite. The gauge-fixing writes
\begin{eqnarray}
\mathcal{L}^{GF}&=&-\frac{1}{\xi_{W}}F^{+}F^{-}-\frac{1}{2\xi_{Z}}|F^{Z}|^{2}-\frac{1}{2\xi_{\gamma}}|F^{\gamma}|^{2}
\, ,
\end{eqnarray}
where
\begin{eqnarray}
F^{+}&=&(\partial_{\mu}-ie\tilde{\alpha}\gamma_{\mu}-ie\frac{c_{W}}{s_{W}}\tilde{\beta}Z_{\mu})W^{\mu +}
+i\xi_{W}\frac{e}{2s_{W}}(v+\tilde{\delta}h^{0}+\tilde{\omega}H^{0}+i \tilde{\rho}A^{0}+i\tilde{\kappa}G^{0})G^{+} \, ,\nonumber\\
F^{Z}&=&\partial_{\mu}Z^{\mu}
+\xi_{Z}\frac{e}{s_{2W}}(v+\tilde{\epsilon}h^{0}+\tilde{\gamma}H^{0})G^{0} \, ,\nonumber \\
F^{\gamma}&=&\partial_{\mu}\gamma^{\mu} \, .
\end{eqnarray}
The ghost Lagrangian ${{\cal L}}^{Gh}$ is derived by requiring
that the full effective Lagrangian, ${\cal L}^Q$, be invariant
under the BRST transformation. As discussed in \cite{grace-1loop},
this is a much more appropriate procedure than the usual
Fadeev-Popov approach especially when dealing with the quantum
symmetries of the generalised non-linear gauges we are using. \noi
$\db {\cal L}^Q=0$ therefore implies $\db {\cal L}^{GF}= - \db
{\cal L}^{Gh}$.

\noi It is very useful to also introduce the auxiliary $B$-field
formulation of the gauge-fixing Lagrangian ${\cal L}^{GF}$,
especially from the perspective of deriving some Ward identities.
The gauge fixing can then be expressed as
\beqn
\label{lgfB} {\cal L}^{GF}=\xi_W B^+B^- + \frac{\xi_Z}{2}|B^Z|^2 +
 \frac{\xi_\gamma} {2}|B^\gamma|^2 + B^-F^+ + B^+F^- + B^Z F^Z + B^\gamma
 F^\gamma \, .
\eeqn
From the equations of motion for the $B$-fields we recover the usual
${\cal L}^{GF}$ together with the condition $B^i=-\frac{F^i}{\xi_i}$
($\xi_{i}=\{\xi_W,\xi_Z,\xi_\gamma\}$). The anti-ghost, $\bar c^i$, is
defined from the gauge fixing functions, we write
\beqn
\label{antigtransf} \db \bar c^i= B^i \, .
\eeqn
Then the ghost Lagrangian writes as
\begin{eqnarray}
\mathcal{L}^{Gh}=-(\overline{c}^{+}\delta_{\textrm{BRS}}F^{+}+\overline{c}^{-}\delta_{\textrm{BRS}}F^{-}
+\overline{c}^{Z}\delta_{\textrm{BRS}}F^{Z}+\overline{c}^{\gamma}\delta_{\textrm{BRS}}F^{\gamma})
\db \tilde{{\cal L}}^{Gh}\, .
\end{eqnarray}

\noi The Fadeev-Popov prescription is therefore readily recovered,
${\cal L}^{FP}$, but only up to an overall function, $\db
\tilde{{\cal L}}^{Gh}$, which is BRST invariant. The latter is set
to zero for one-loop calculations. Our code {\tt SloopS} implements
this prescription automatically leading to the automatic
generation
of the whole set of Feynman rules for the ghost sector. \\
For all applications we set the Feynman parameters $\xi_{W,Z,\gamma}$
to one. This allows one to use the minimum set of libraries for the
tensor reduction. Indeed, $\xi_{W,Z,\gamma} \neq 1 $ can generate high
rank tensor loop functions, that would take much time to reduce to
the set of scalar functions. \\
It is important to stress, once more, that since we do not seek to
have all Green's functions finite but only the S-matrix elements,
we take the gauge fixing Lagrangian as being renormalised.

\noi Judicious choices of the the non-linear gauge parameters
can lead to simplifications like the vanishing of certain
vertices. For example, with $\tilde{\alpha}=1$, the $W^{+\,
\mu}G^{-}\gamma_{\mu}$ vertex cancels. More examples can be found in
Appendix~A for the vanishing of some ghost couplings to Higgses.

\section{Renormalisation}
Our renormalisation procedure is within the spirit of the on-shell
scheme borrowing as much as possible from the programme carried
strictly within the Standard Model in \cite{grace-1loop}. For the
gauge sector and the fermion sector, beside the electromagnetic
coupling which we fix from the Thomson limit, we take therefore the
same set of physical input parameters, namely the masses of the
$W^{\pm}$ and $Z^{0}$ together with the masses of all the standard
model fermions. To define the Higgs sector parameters, the set of
Eq.~(\ref{input-higgs}) looks most appropriate were it not for the
ill defined $\tb$. Indeed, the mass of the pseudoscalar $M_{A^0}$ within
the MSSM with CP conservation is a physical parameter. As within the
Standard Model, we also take the tadpole. For $\tb$ the aim of this
paper is to review, propose and compare different schemes.
Renormalisation of these parameters would then lead to finite
S-matrix elements. For the mass eigenstates and thus a proper
identification of the physical particles that appear as external
legs in our processes, field renormalisation is needed. S-matrix
elements obtained from these rescaled Green's functions will lead to
external legs with unit residue and will avoid mixing. Therefore one
also needs wave function renormalisation of the fields. Especially
for the unphysical sector of the theory, the precise choice of the
fields redefinition is not essential if one is only interested in
S-matrix elements of physical processes. It has to be stressed that
one can do without this if one is willing to include loop
corrections on the external legs. In the MSSM and in the Higgs
sector in particular mixing effects, especially at one-loop, are a
nuisance that has introduced some confusion especially in defining
$\tb$ with the help of wave-function renormalisation constants or
equivalently from two-point function describing particle mixing. For
the Higgs sector one needs to be wary about mixing of the Goldstones
with CP-odd Higgs or almost equivalently between the $Z^{0}$ and the
CP-odd Higgs or the $W^\pm$ and the charged Higgs. These two-point
functions are related through gauge invariance and impose strong
constraints on the wave function renormalisation constants. We will
derive Ward-Slavnov-Taylor identities relating these two-point
functions, and hence their associated counterterms, before imposing
any ad-hoc condition.

\subsection{Shifts in mass parameters and gauge couplings} All
fields and parameters introduced so far are considered as bare
parameters with the exception of the gauge fixing Lagrangian which
we choose to write in terms of {\em renormalised fields}. Care
should then be exercised when we split the tree-level
contributions and the counterterms. Shifts are then introduced for
the Lagrangian parameters and the fields with the notation that a
bare quantity is labeled as $X_0$. It will split in terms of
renormalised quantities $X$ and counterterms $\delta X$
\begin{eqnarray}
g_0&=&g +\delta g \, ,\quad g'_0 \ra g'+ \delta g' \, ,
\\
m_{i\,0}^2&=&m_i^2 +\delta m_i^2 \quad {\rm for \;\; } i=1,2\, , \quad
m_{12\,0}^2=m_{12}^2 +\delta m_{12}^2 \, ,
\\
v_{i\,0}&=&v_i -\delta v_i \quad {\rm for \;\;} i=1,2 \quad {\rm
hence} \quad \frac{\delta \tb}{\tb}= \frac{\delta
v_{1}}{v_{1}}-\frac{\delta v_{2}}{v_{2}} \, .\label{deftanbeta}
%\delta Z_{\varphi^{0}_{i}}, \delta
%Z_{\varphi^{0}_{i}\varphi^{0}_{j}}\\
%\delta Z_{\phi^{c}_{i}}, \delta Z_{\phi^{c}_{i}\phi^{c}_{j}}\\
%\delta Z_{\phi^{0}_{i}}, \delta Z_{\phi^{0}_{i}\phi^{0}_{j}}
%\end{array}\right.
\end{eqnarray}
In our approach the angles defining the rotation matrices, $\beta$
 and $\alpha$ in Eq.~(\ref{diagonal_beta_alpha}) are defined as
\underline{renormalised} quantities. For example the relation
between the Goldstone boson/pseudoscalar Higgs boson and the fields
$\varphi_{1,2}^{0}$ is maintained at all orders. Indeed,
\beqn
\left(\begin{array}{c} G^{0}\\A^{0}\end{array}\right)_0
=U(\beta)\left(\begin{array}{c} \varphi_{1}^{0}\\
\varphi_{2}^{0}\end{array}\right)_0 \quad {\rm implies \;\; also }
\quad \left(\begin{array}{c} G^{0}\\A^{0}\end{array}\right)
=U(\beta)\left(\begin{array}{c} \varphi_{1}^{0}\\
\varphi_{2}^{0}\end{array}\right) \, .
\eeqn
Since in our approach we will always perform a field
renormalisation there is no need in inducing more shifts from
$U(\alpha,\beta)$. Therefore $U(\alpha,\beta)$ is taken as
renormalised. For example, if we perform a field renormalisation
in the $\varphi_{1,2}^{0}$ basis
\begin{eqnarray}
\left(\begin{array}{c} \varphi_{1}^{0}\\
\varphi_{2}^{0}\end{array}\right)_{0}&=& Z_{\varphi^0} \left(\begin{array}{c} \varphi_{1}^{0}\\
\varphi_{2}^{0}\end{array}\right)= \left(\begin{array}{cc}
Z_{\varphi_{1}^{0}}^{1/2} & Z_{\varphi_{1}^{0}\varphi_{2}^{0}}^{1/2} \\
Z_{\varphi_{2}^{0}\varphi_{1}^{0}}^{1/2} &
Z_{\varphi_{2}^{0}}^{1/2}\end{array}\right)
\left(\begin{array}{c} \varphi_{1}^{0}\\
\varphi_{2}^{0}\end{array}\right)\, ,
%%%%%%%%%%%%
\end{eqnarray}
\noi this will imply
\begin{eqnarray}
\left(\begin{array}{c} G^{0}\\
A^{0}\end{array}\right)_{0}=U(\beta) Z_{\varphi^0} U(-\beta) \left(\begin{array}{c} G^{0}\\
A^{0}\end{array}\right)= Z_{P} \left(\begin{array}{c} G^{0}\\
A^{0}\end{array}\right)=\left(\begin{array}{cc}
Z_{G^0G^0}^{1/2} & Z_{G^0A^0}^{1/2} \\
Z_{A^0G^0}^{1/2} & Z_{A^0A^0}^{1/2}\end{array}\right)
\left(\begin{array}{c} G^{0}\\
A^{0}\end{array}\right) \, .
\label{wfr-Zp}
%%%%%%%%%%%%
\end{eqnarray}
For the field renormalisation we can perform this either at the
level of the $\varphi_i^{0}$, {\it i.e.} before any rotation on the
field in the Lagrangian is made, through $Z_{\varphi^{0}}$ as is
done in \cite{DabelsteinHiggs,DCPR,Grace-susy1loop} or in a much
efficient way directly in the basis $G^0 A^0$ since the latter are
directly related to our renormalisation conditions on the physical
fields as we will see later. For instance, there is no need for
$Z_{G^{0}G^{0}}$ in our approach since we will not be dealing with
Goldstone bosons in the external legs.

\subsection{Tadpole terms}
We start with the terms linear in the Higgs fields which will lead
to renormalisation of the tadpoles. With the tree-level condition on
the tadpoles $T_{\phi_{1}^{0}}=T_{\phi_{2}^{0}}=0$, field
normalisation if it were performed does not contribute, we therefore
have
\begin{eqnarray}
\left.V_{linear}\right|_{0}&=&(\delta
T_{\phi_{1}^{0}}\phi_{1}^{0}+\delta T_{\phi_{2}^{0}}\phi_{2}^{0}) \, ,
\end{eqnarray}
with
\begin{eqnarray}
\frac{\delta
T_{\phi_{1}^{0}}}{v_{1}}&=&\frac{M_{Z^0}^{2}}{2}c_{2\beta}
\frac{\delta g^{2}+\delta
g^{'\,2}}{g^{2}+g^{'\,2}}+\delta m_{1}^{2}+\tb \delta m_{12}^{2}\nonumber\\&-&\biggl(m_{1}^{2}
+\frac{M_{Z^0}^{2}}{2}c_{2\beta}+M_{Z^0}^{2}c_{\beta}^{2}\biggr)\frac{ \delta v_{1}}{v_{1}}+\biggl(-m_{12}^{2}
+\frac{M_{Z^0}^{2}}{2}s_{2 \beta}\biggr)\tb \frac{ \delta v_{2}}{v_{2}}\, , \\
\frac{\delta T_{\phi_{2}^{0}}}{v_{2}}&=&\frac{\delta
T_{\phi_{1}^{0}}}{v_{1}}(v_1 \leftrightarrow v_2, m_1
\leftrightarrow m_2)\, .
\end{eqnarray}
The minimum condition requires the one-loop tadpole contribution
generated by one-loop diagrams, $T_{\phi_{i}^{0}}^{\textrm{loop}}$
 is cancelled by the tadpole counterterm. This imposes
\begin{eqnarray}
%T_{\phi_{i}^{0}}&=&0\label{tadeq1}\\
\delta T_{\phi_{i}^{0}}&=&-T_{\phi_{i}^{0}}^{\textrm{loop}} \,
.\label{tadeq2}
\end{eqnarray}
$T_{\phi_{i}^{0}}^{\textrm{loop}}$ is calculated from the one-loop
tadpole amplitude for $H^{0}$, $T_{H^{0}}^{\textrm{loop}}$ and
$h^{0}$, $T_{H^{0}}^{\textrm{loop}}$ by simply moving to the
physical basis
\begin{eqnarray}
\left(\begin{array}{c} T_{\phi_{1}^{0}}^{\textrm{loop}}\\
T_{\phi_{2}^{0}}^{\textrm{loop}}\end{array}\right)=\left(\begin{array}{cc}
c_{\alpha}& -s_{\alpha}\\ s_{\alpha}&
c_{\alpha}\end{array}\right)\left(\begin{array}{c}
T_{H^{0}}^{\textrm{loop}}\\
T_{h^{0}}^{\textrm{loop}}\end{array}\right)\, .\label{TphiTH}
\end{eqnarray}

\subsection{Mass counterterms in the Higgs sector}
We now move to the mass counterterms induced by shifts in the
Lagrangian parameters. We need to consider all terms bi-linear in
the fields. From the bare matrices $M_{\varphi^{0}}^{2}$,
$M_{\phi^{\pm}}^{2}$ and $M_{\phi^{0}}^{2}$
(Eqs.~(\ref{Vmass})~,~(\ref{HiggsMassMatrix1})~-~(\ref{HiggsMassMatrix3})),
we find the corresponding counterterms in matrix form in the basis
$\varphi_{1,2}^{0}$, $\phi_{1,2}^{0}$ and $\phi_{1,2}^\pm$
\begin{eqnarray}
\delta M^{2}_{\varphi^{0}}&=&\left(\begin{array}{cc} \delta
m_{1}^{2}+\frac{1}{2}c_{2\beta}\delta
M_{Z^0}^{2}-\frac{M_{Z^0}^{2}}{2}s_{2\beta}^{2}\frac{\delta t_{\beta}}{t_{\beta}}& \delta m_{12}^{2}\\
\delta m_{12}^{2} & \delta m_{2}^{2}-\frac{1}{2}c_{2\beta}\delta
M_{Z^0}^{2}+\frac{M_{Z^0}^{2}}{2}s_{2\beta}^{2}\frac{\delta t_{\beta}}{t_{\beta}}\end{array}\right)\nonumber\\
%%%%%%%%%%%%%%%%%%
\delta M^{2}_{\phi^{\pm}}&=&\left(\begin{array}{cc} \delta m_{1}^{2}
+\frac{1}{2}c_{2\beta}\delta M_{Z^0}^{2}+s_{\beta}^{2}\delta
M_{W^{\pm}}^{2}-\frac{M_{Z^0}^{2}}{2} s_{2\beta}^2 s_W^2\frac{\delta
t_{\beta}}{t_{\beta}}& \delta
m_{12}^{2}-\frac{1}{2}s_{2\beta}\delta M_{W^{\pm}}^{2}
-\frac{M_{W^{\pm}}^{2}}{4}s_{4\beta}\frac{\delta t_{\beta}}{t_{\beta}}\\
\delta m_{12}^{2}-\frac{1}{2}s_{2\beta}\delta M_{W^{\pm}}^{2}
-\frac{M_{W^{\pm}}^{2}}{4}s_{4\beta}\frac{\delta
t_{\beta}}{t_{\beta}}& \delta m_{2}^{2}
-\frac{1}{2}c_{2\beta}\delta M_{Z^0}^{2}+c_{\beta}^{2}\delta
M_{W^{\pm}}^{2}+\frac{M_{Z^0}^{2}}{2} s_{2\beta}^2 s_W^2\frac{\delta
t_{\beta}}{t_{\beta}}\end{array}\right)
\nonumber\\
%%%%%%%%%%%%%%%%%%
\delta M^{2}_{\phi^{0}}&=&\left(\begin{array}{cc} \delta
m_{1}^{2}+\frac{1}{2}(4c_{\beta}^{2}-1)\delta
M_{Z^0}^{2}-M_{Z^0}^{2}s_{2\beta}^{2}\frac{\delta
t_{\beta}}{t_{\beta}}& \delta
m_{12}^{2}-\frac{1}{2}s_{2\beta}\delta
M_{Z^0}^{2}-\frac{M_{Z^0}^{2}}{4}s_{4\beta}\frac{\delta t_{\beta}}{t_{\beta}}\\
\delta m_{12}^{2}-\frac{1}{2}s_{2\beta}\delta
M_{Z^0}^{2}-\frac{M_{Z^0}^{2}}{4}s_{4\beta}\frac{\delta
t_{\beta}}{t_{\beta}}& \delta
m_{2}^{2}+\frac{1}{2}(4s_{\beta}^{2}-1)\delta
M_{Z^0}^{2}+M_{Z^0}^{2}s_{2\beta}^{2}\frac{\delta
t_{\beta}}{t_{\beta}}\end{array}\right)\nonumber
\end{eqnarray}
It is then straightforward to move to the physical fields through
the rotation matrices $U(\alpha)$ and $U(\beta)$, to find the mass
counterterms $\delta M_{A^0}^{2},\delta M_{H^{\pm}}^{2},\delta
M_{h^{0}}^{2}, \delta M_{H^{0}}^{2}$ for, respectively, the
pseudoscalar Higgs, $A^0$, the charged Higgs $H^\pm$, and the two
CP-even Higgses $h^0$, $H^0$. A mass mixing between these two Higgses, $\delta
M_{H^{0}h^{0}}^{2}$ is also induced

\begin{eqnarray}
\delta M_{A^0}^{2}&=&s_{\beta}^{2}\delta
m_{1}^{2}+c_{\beta}^{2}\delta m_{2}^{2}-s_{2\beta}\delta
m_{12}^{2} -\frac{1}{2}c_{2\beta}^{2}\delta
M_{Z^0}^{2}+\frac{M_{Z^0}^{2}}{2}s_{2\beta}^{2}c_{2\beta}\frac{\delta t_{\beta}}{t_{\beta}} \, , \nonumber\\
%%%%%%%%%%%%%
\delta M_{H^{\pm}}^{2}&=&\delta M_{A^0}^{2}+\delta M_{W^{\pm}}^{2} \, , \nonumber\\
%%%%%%%%%%%%%
\delta M_{H^{0}}^{2}&=&c_{\alpha}^{2}\delta
m_{1}^{2}+s_{\alpha}^{2}\delta m_{2}^{2}+s_{2\alpha}\delta m_{12}^{2}\nonumber\\
&+&\frac{1}{2}\biggl(4(c_{\alpha}^{2}c_{\beta}^{2}+s_{\alpha}^{2}s_{\beta}^{2}-c_{\alpha}s_{\alpha}
c_{\beta}s_{\beta})-1\biggr)\delta
M_{Z^0}^{2}-\frac{M_{Z^0}^{2}}{2}s_{2\beta}\biggl(2c_{2\alpha}s_{2\beta}+s_{2\alpha}c_{2\beta}\biggr)\frac{\delta t_{\beta}}{t_{\beta}} \, , \nonumber\\
%%%%%%%%%%%%%
\delta M_{h^{0}}^{2}&=&s_{\alpha}^{2}\delta
m_{1}^{2}+c_{\alpha}^{2}\delta m_{2}^{2}-s_{2\alpha}\delta m_{12}^{2}\nonumber\\
&+&\frac{1}{2}\biggl(4(c_{\alpha}^{2}s_{\beta}^{2}+s_{\alpha}^{2}c_{\beta}^{2}+c_{\alpha}s_{\alpha}
c_{\beta}s_{\beta})-1\biggr)\delta
M_{Z^0}^{2}+\frac{M_{Z^0}^{2}}{2}s_{2\beta}\biggl(2c_{2\alpha}s_{2\beta}+s_{2\alpha}c_{2\beta}\biggr)\frac{\delta t_{\beta}}{t_{\beta}} \, , \nonumber\\
%%%%%%%%%%%%%
\delta M_{H^{0}h^{0}}^{2}&=&c_{2\alpha}\delta m_{12}^{2}+\frac{1}{2}s_{2\alpha}(\delta m_{2}^{2}-\delta m_{1}^{2})\nonumber\\
&-&\frac{1}{2}\biggl(2s_{2\alpha}c_{2\beta}+s_{2\beta}c_{2\alpha}\biggr)\delta
M_{Z^0}^{2}+\frac{M_{Z^0}^{2}}{2}s_{2\beta}\biggl(2s_{2\alpha}s_{2\beta}-c_{2\alpha}c_{2\beta}\biggr)\frac{\delta
t_{\beta}}{t_{\beta}} \, . \label{dma-m12}
\end{eqnarray}
A mass term seems to be induced for the Goldstone bosons as well as
a mixing between the Goldstones and the corresponding CP-odd Higgs
\begin{eqnarray}
\delta M_{G^{0}}^{2}&=&c_{\beta}^{2}\delta
m_{1}^{2}+s_{\beta}^{2}\delta m_{2}^{2} +s_{2\beta}\delta
m_{12}^{2}+\frac{1}{2}c_{2\beta}^{2}\delta
M_{Z^0}^{2}-\frac{1}{2}M_{Z^0}^{2}s_{2\beta}^{2}c_{2\beta}\frac{\delta t_{\beta}}{t_{\beta}} \, ,\\
%%%%%%%%%%%%%%%%
\delta M_{G^{\pm}}^{2}&=&\delta M_{G^{0}}^{2} \, ,\\
%%%%%%%%%%%%%%%%
\delta M_{G^{0}A^{0}}^{2}&=&c_{2\beta}\delta
m_{12}^{2}+c_{\beta}s_{\beta}(\delta m_{2}^{2}-\delta
m_{1}^{2})-\frac{1}{2}c_{2\beta}s_{2\beta}\delta
M_{Z^0}^{2}+M_{Z^0}^{2}s_{2\beta}^{2}c_{\beta}s_{\beta}\frac{\delta t_{\beta}}{t_{\beta}} \, ,\\
%%%%%%%%%%%%%%%%
\delta M_{G^{\pm}H^{\pm}}^{2}&=&\delta
M_{G^{0}A^{0}}^{2}-M_{W^{\pm}}^{2}c_{\beta}s_{\beta}\frac{\delta
t_{\beta}}{t_{\beta}} \, .\label{dmg-m12}
\end{eqnarray}
It is much more transparent to re-express these mass counterterms
by trading-off $\delta m_{1,2}$ and $\delta m_{12}$ with our input
parameters $\delta T_{\phi_{1,2}^{0}}, \delta M_{A^0}^2, \dtb$ through
\begin{eqnarray}
\begin{array}{l}
\delta m_{1}^{2}=c_{\beta}^{2}(s_{\beta}^{2}+1)\frac{\delta
T_{\phi_{1}^{0}}}{v_{1}}-c_{\beta}^{2}s_{\beta}^{2}\frac{\delta
T_{\phi_{2}^{0}}}{v_{2}}+s_{\beta}^{2}\delta
M_{A^0}^{2}-\frac{1}{2}c_{2\beta}\delta
M_{Z^0}^{2}+\frac{1}{2}s_{2\beta}^{2} (M_{A^0}^{2}+M_{Z^0}^{2})\frac{\delta t_{\beta}}{t_{\beta}} \, ,\\
\\%%%%%%%%%%%%%%%%%%%%%%
\delta
m_{12}^{2}=\frac{1}{2}s_{2\beta}\biggl(s_{\beta}^{2}\frac{\delta
T_{\phi_{1}^{0}}}{v_{1}}+c_{\beta}^{2}\frac{\delta
T_{\phi_{2}^{0}}}{v_{2}}-\delta
M_{A^0}^{2}-c_{2\beta}M_{A^0}^{2}\frac{\delta t_{\beta}}{t_{\beta}}\biggr) \, ,\\
\\%%%%%%%%%%%%%%%%%%%%%%
\delta m_{2}^{2}=-c_{\beta}^{2}s_{\beta}^{2}\frac{\delta
T_{\phi_{1}^{0}}}{v_{1}}+s_{\beta}^{2}(c_{\beta}^{2}+1)\frac{\delta
T_{\phi_{2}^{0}}}{v_{2}}+c_{\beta}^{2}\delta
M_{A^0}^{2}+\frac{1}{2}c_{2\beta}\delta
M_{Z^0}^{2}-\frac{1}{2}s_{2\beta}^{2}(M_{A^0}^{2}+M_{Z^0}^{2})\frac{\delta
t_{\beta}}{t_{\beta}} \, .
\end{array} \label{soldm2}
\end{eqnarray}
In terms of $\delta T_{\phi_{1,2}^0}, \delta M_{A^0}^2, \dtb$, the mass
counterterms of Eq.~(\ref{dma-m12}) and Eq.~(\ref{dmg-m12}) write
\begin{eqnarray}
\delta M_{G^{0}}^{2}&=&\delta
M_{G^{\pm}}^{2}=\frac{1}{v}(c_{\alpha-\beta}\delta
T_{H^{0}}-s_{\alpha-\beta}\delta T_{h^{0}})\, ,\nonumber\\
%%%%%%%%%%%%%%%%%%%%%%%%%%%
\delta M_{G^{0}A^{0}}^{2}&=&\frac{1}{v}(s_{\alpha-\beta}\delta
T_{H^{0}}+c_{\alpha-\beta}\delta
T_{h^{0}})-s_{2\beta}\frac{M_{A^0}^{2}}{2}\frac{\delta t_{\beta}}{t_{\beta}} \, ,\nonumber\\
%%%%%%%%%%%%%%%%%%%%%%%%%%%
\delta M_{G^{\pm}H^{\pm}}^{2}&=&\frac{1}{v}(s_{\alpha-\beta}\delta
T_{H^{0}}+c_{\alpha-\beta}\delta
T_{h^{0}})-s_{2\beta}\frac{M_{H^{\pm}}^{2}}{2}\frac{\delta t_{\beta}}{t_{\beta}} \, ,\nonumber\\
%%%%%%%%%%%%%%%%%%%%%%%%%%%
\delta M_{H^{\pm}}^{2}&=&\delta M_{A^0}^{2}+\delta M_{W^{\pm}}^{2} \, ,\nonumber\\
%%%%%%%%%%%%%%%%%%%%%%%%%%%
%%%%%%%%%%%%%%%%%%%%%%%%%%%
\delta
M_{h^{0}}^{2}&=&-\frac{1}{v}\biggl(c_{\alpha-\beta}s_{\alpha-\beta}^{2}\delta
T_{H^{0}}+s_{\alpha-\beta}(1+c_{\alpha-\beta}^{2})\delta
T_{h^{0}}\biggr)+c_{\alpha-\beta}^{2}\delta
M_{A^0}^{2}+s_{\alpha+\beta}^{2}\delta M_{Z^0}^{2}\nonumber
\\
& & \quad +s_{2\beta}s_{2(\alpha+\beta)}M_{Z^0}^{2}\frac{\delta
t_{\beta}}{t_{\beta}} \, , \nonumber \\
%%%%%%%%%%%%%%%%%%%%%%%%%%%%%%%%%%%%
\delta
M_{H^{0}}^{2}&=&\frac{1}{v}\biggl(c_{\alpha-\beta}(1+s_{\alpha-\beta}^{2})\delta
T_{H^{0}}+s_{\alpha-\beta}c_{\alpha-\beta}^{2}\delta
T_{h^{0}}\biggr)+s_{\alpha-\beta}^{2}\delta
M_{A^0}^{2}+c_{\alpha+\beta}^{2}\delta M_{Z^0}^{2} \nonumber
\\
& & \quad - s_{2\beta}s_{2(\alpha+\beta)}M_{Z^0}^{2}\frac{\delta t_{\beta}}{t_{\beta}} \, ,\nonumber\\
%%%%%%%%%%%%%%%%%%%%%%%%%%%
\delta M_{H^{0}h^{0}}&=&-\frac{1}{v}s_{\alpha-\beta}^{3}\delta
T_{H^{0}}+\frac{1}{v}c_{\alpha-\beta}^{3}\delta
T_{h^0}+\frac{1}{2}s_{2(\alpha-\beta)}\delta
M_{A^0}^{2}-\frac{1}{2}s_{2(\alpha+\beta)}\delta
M_{Z^0}^{2}-\frac{s_{2\beta}}{2}\bigg(M_{A^0}^{2}c_{2(\alpha-\beta)}\nonumber
\\
& & \quad +M_{Z^0}^{2}c_{2(\alpha+\beta)}\bigg)\frac{\delta
t_{\beta}}{t_{\beta}} \, .\label{deltaM_all}
\end{eqnarray}
It is very satisfying to see that $\delta M_{G^{0}}^{2}=\delta
M_{G^{\pm}}^{2}$ is accounted for totally by the tadpole
counterterms.

\subsection{Field renormalisation}
We can now introduce field renormalisation at the level of the
physical fields without the need to first go through field
renormalisation in the basis $\phi_{1,2}^{0}, \varphi_{1,2}^{0},
\phi_{1,2}^\pm$. In most generality we can write, as in
Eq.~(\ref{wfr-Zp})
\begin{eqnarray}
\left(\begin{array}{c} G^{0}\\
A^{0}\end{array}\right)_{0}&=&Z_P \; \left(\begin{array}{c} G^{0}\\
A^{0}\end{array}\right)\equiv \left(\begin{array}{cc}
Z_{G^{0}}^{1/2} & Z_{G^{0}A^{0}}^{1/2} \\
Z_{A^{0}G^{0}}^{1/2} & Z_{A^{0}}^{1/2}\end{array}\right)
\left(\begin{array}{c} G^{0}\\
A^{0}\end{array}\right) \, ,\nonumber\\
%%%%%%%%%%%%%%
\left(\begin{array}{c} G^{\pm}\\
H^{\pm}\end{array}\right)_{0}&=&Z_C \; \left(\begin{array}{c} G^{\pm}\\
H^{\pm}\end{array}\right) \equiv \left(\begin{array}{cc}
Z_{G^{\pm}}^{1/2} & Z_{G^{\pm}H^{\pm}}^{1/2} \\
Z_{H^{\pm}G^{\pm}}^{1/2} & Z_{H^{\pm}}^{1/2}\end{array}\right)
\left(\begin{array}{c} G^{\pm}\\
H^{\pm}\end{array}\right) \, ,\nonumber\\
%%%%%%%%%%%%%%
\left(\begin{array}{c} H^{0}\\
h^{0}\end{array}\right)_{0}&=&Z_S \; \left(\begin{array}{c}H^{0}\\
h^{0}\end{array}\right) \equiv \left(\begin{array}{cc}
Z_{H^{0}}^{1/2} & Z_{H^{0}h^{0}}^{1/2} \\
Z_{h^{0}H^{0}}^{1/2} & Z_{h^{0}}^{1/2}\end{array}\right)
\left(\begin{array}{c}H^{0}\\
h^{0}\end{array}\right) \, .\label{wfr-All}
\end{eqnarray}
It is always possible to move basis through Eq.~(\ref{wfr-Zp}). Field
renormalisation will help get rid of mixing between physical
fields when these are on-shell and set the residue to $1$.

\subsection{Self-energies in the Higgs sector}
Collecting the contribution of all the counterterms, including
wave function renormalisation, the renormalised self-energies
write as
\begin{eqnarray}
&&\left\{\begin{array}{l}
\hat{\Sigma}_{G^{0}G^{0}}(q^{2})=\Sigma_{G^{0}G^{0}}(q^{2})+\delta
M_{G^{0}}^{2}-q^{2}\delta Z_{G^{0}}\\
\hat{\Sigma}_{G^{0}A^{0}}(q^{2})=\Sigma_{G^{0}A^{0}}(q^{2})+\delta
M_{G^{0}A^{0}}^{2}-\frac{1}{2}q^{2}\delta Z_{G^{0}A^{0}}
-\frac{1}{2}(q^{2}-M_{A^{0}}^{2})\delta Z_{A^{0}G^{0}}\\
\hat{\Sigma}_{A^{0}A^{0}}(q^{2})=\Sigma_{A^{0}A^{0}}(q^{2})+\delta
M_{A^{0}}^{2}-(q^{2}-M_{A^{0}}^{2})\delta Z_{A^{0}}
\end{array}\right.\nonumber\\
%%%%%%%%%%%%%%%%%%%%%%%%
&&\left\{\begin{array}{l}
\hat{\Sigma}_{G^{\pm}G^{\pm}}(q^{2})=\Sigma_{G^{\pm}G^{\pm}}(q^{2})+\delta
M_{G^{\pm}}^{2}-q^{2}\delta Z_{G^{\pm}}\\
\hat{\Sigma}_{G^{\pm}H^{\pm}}(q^{2})=\Sigma_{G^{\pm}H^{\pm}}(q^{2})+\delta
M_{G^{\pm}H^{\pm}}^{2}-\frac{1}{2}q^{2}\delta Z_{G^{\pm}H^{\pm}}
-\frac{1}{2}(q^{2}-M_{H^{\pm}}^{2})\delta Z_{H^{\pm}G^{\pm}}\\
\hat{\Sigma}_{H^{\pm}H^{\pm}}(q^{2})=\Sigma_{H^{\pm}H^{\pm}}(q^{2})+\delta
M_{H^{\pm}}^{2}-(q^{2}-M_{H^{\pm}}^{2})\delta Z_{H^{\pm}}
\end{array}\right.\nonumber\\
%%%%%%%%%%%%%%%%%%%%%%%%
&&\left\{\begin{array}{l}
\hat{\Sigma}_{H^{0}H^{0}}(q^{2})=\Sigma_{H^{0}H^{0}}(q^{2})+\delta
M_{H^{0}}^{2}-(q^{2}-M_{H^{0}}^2)\delta Z_{H^{0}}\\
\hat{\Sigma}_{H^{0}h^{0}}(q^{2})=\Sigma_{H^{0}h^{0}}(q^{2})+\delta
M_{H^{0}h^{0}}^{2}-\frac{1}{2}(q^{2}-M_{H^{0}}^{2})\delta
Z_{H^{0}h^{0}}
-\frac{1}{2}(q^{2}-M_{h^{0}}^{2})\delta Z_{h^{0}H^{0}}\\
\hat{\Sigma}_{h^{0}h^{0}}(q^{2})=\Sigma_{h^{0}h^{0}}(q^{2})+\delta
M_{h^{0}}^{2}-(q^{2}-M_{h^{0}}^{2})\delta Z_{h^{0}}
\end{array}\right.\nonumber
\end{eqnarray}
Note that as we stressed all along, since we are only interested in
having finite $S$-matrix transitions and not finite Green's
functions there is no need trying to make all two-point functions
finite. For instance the diagonal Goldstone self-energies
$\hat{\Sigma}_{G^{0}G^{0}}(q^{2})$ and
$\hat{\Sigma}_{G^{\pm}G^{\pm}}(q^{2})$ do not need any field
renormalisation. Therefore we can set for example $\delta
Z_{G^{0}}=\delta Z_{G^{\pm}}=0$ for simplicity. $\delta Z_{A^0 G^0}$
is also not needed as it is only involved in the transition of
Golsdtone boson to the pseudo-scalar Higgs.

\subsection{$A^0 Z^{0}$ and $H^\pm W^\pm$ transitions}
The (massive) gauge bosons and the pseudo-scalar mix. This
originates from the same part of the gauge Lagrangian where the
gauge bosons, at tree-level, mix with the Goldstone bosons as in the
Standard Model, see for example \cite{grace-1loop}. The latter is
eliminated through the usual 't Hooft gauge fixing. To wit, from
\begin{eqnarray}
\mathcal{L}^{GV}_{0}&=&\frac{g}{2}i(v_{1}\partial^{\mu}\phi_{1}^{-}+v_{2}\partial^{\mu}\phi_{2}^{-})W_{\mu}^{+}+h.c.\nonumber\\
&-&\frac{g}{2c_{W}}(v_{1}\partial^{\mu}\varphi_{1}^{0}+v_{2}\partial^{\mu}\varphi_{2}^{0})Z^{0}_{\mu}|_{0}
\, ,
\end{eqnarray}
we end up with
\begin{eqnarray}
\mathcal{L}^{GV}_{0}=\mathcal{L}^{GV}&+&\frac{1}{2}\biggl(\delta
Z_{G^{\pm}}+\delta Z_{W^{\pm}} + \frac{\delta
M_{W^{\pm}}^{2}}{M_{W^{\pm}}^{2}}\biggr)(iM_{W^{\pm}}\partial^{\mu}G^{-}W_{\mu}^{+}+h.c.)\nonumber\\
&-&\frac{1}{2}\biggl(\delta Z_{G^{0}}+\delta Z_{Z^{0}Z^{0}} + \frac{\delta
M_{Z^0}^{2}}{M_{Z^0}^{2}}\biggr)M_{Z^0}\partial^{\mu}G^{0}Z^{0}_{\mu}\nonumber\\
&-&\frac{1}{2}\delta Z_{Z^{0}\gamma}M_{Z^0}\partial^{\mu}G^{0}\gamma_{\mu}\nonumber\\
&+&\frac{1}{2}\biggl(\delta
Z_{G^{\pm}H^{\pm}}+s_{2\beta}\frac{\delta t_{\beta}}{t_{\beta}}\biggr)(iM_{W^{\pm}}\partial^{\mu}H^{-}W_{\mu}^{+}+h.c.)\nonumber\\
&-&\frac{1}{2}\biggl(\delta Z_{G^{0}A^{0}}+s_{2\beta}\frac{\delta
t_{\beta}}{t_{\beta}}\biggr)M_{Z^0}\partial^{\mu}A^{0}Z^{0}_{\mu} \,
.\label{LGV}
\end{eqnarray}
For the sake of completeness, we have also kept in Eq.~(\ref{LGV})
the wave-function renormalisation constants of the gauge bosons,
namely $\delta Z_{W^{\pm}}, \delta Z_{Z^{0}Z^{0}}$ and $\delta Z_{Z^{0}\gamma}$ (for the
$Z^{0}\ra \gamma$ transition), see \cite{grace-1loop}. The conditions on
the latter are the same as
in the Standard Model, details are found in \cite{grace-1loop}.\\
The novelty however is that now we have $A^{0}-Z^{0}$ and
$H^{\pm}-W^{\pm}$ transitions whose self-energies write:
\begin{eqnarray}
\hat{\Sigma}_{A^{0}Z^{0}}(q^2)&=&\Sigma_{A^{0}Z^{0}}(q^2)+\frac{M_{Z^0}}{2}\biggl(\delta
Z_{G^{0}A^{0}}+s_{2\beta}\frac{\delta t_{\beta}}{t_{\beta}}\biggr) \, ,\\
\hat{\Sigma}_{H^{\pm}W^{\pm}}(q^2)&=&\Sigma_{H^{\pm}W^{\pm}}(q^2)+\frac{M_{W^{\pm}}}{2}\biggl(\delta
Z_{G^{\pm}H^{\pm}}+s_{2\beta}\frac{\delta
t_{\beta}}{t_{\beta}}\biggr) \, .\label{A0Z-trans}
\end{eqnarray}
Note that apart from $\dtb$ the same counterterm $\delta
Z_{G^{0}A^{0}}$ appears in the $G^0 A^0$ transition. In fact there
is a Ward identity relating these two transitions. Contrary to what
one might see in some
papers \cite{DabelsteinHdecay,ShanHiggs,logan02}, the relation is
much more complicated for $q^2 \neq M_{A^0}^2$ and gets more subtle
in the case of the non-linear gauge. This identity is very important
especially that in many approaches the transition has been used as a
{\em definition} for $\dtb$. The identity can be most easily derived
by considering the BRST transformation on the (``ghost'') operator
$\langle 0|\overline{c}^{Z}(x)A^{0}(y)|0\rangle =0$. Full details
are given in Appendix~\ref{app_ward_azag}. We have the constraint
\begin{eqnarray}
q^{2}\hat{\Sigma}_{A^{0}Z^{0}}(q^{2})+M_{Z^0}\hat{\Sigma}_{A^{0}G^{0}}(q^{2})&=&
(q^{2}-M_{Z^0}^{2})\frac{1}{(4\pi)^{2}}\frac{e^{2}M_{Z^0}}{s_{2W}^{2}}s_{2\beta}\mathcal{F}_{GA}^{\tilde{\epsilon},\tilde{\gamma}}(q^2) \label{constraint-az-ag2} \\
&+&\frac{M_{Z^0}}{2}(q^{2}-M_{A^0}^{2})\biggl(\frac{1}{(4\pi)^{2}}\frac{2e^{2}}{s_{2W}^{2}}\mathcal{F}_{cc}^{\tilde{\epsilon},\tilde{\gamma}}(q^2)+
s_{2\beta}\frac{ \delta t_{\beta}}{t_{\beta}}-\delta
Z_{A^{0}G^{0}}\biggr)\, . \nonumber
\end{eqnarray}
$\mathcal{F}_{GA}^{\tilde{\epsilon},\tilde{\gamma}}(q^2)$ and
$\mathcal{F}_{cc}^{\tilde{\epsilon},\tilde{\gamma}}(q^2)$ are
functions defined in Appendix~\ref{app_ward_azag}. They vanish in
the linear gauge with $\tilde{\epsilon}=\tilde{\gamma}=0$. The
constraint shows that even in the linear gauge
$q^{2}\hat{\Sigma}_{A^{0}Z^{0}}(q^{2})+M_{Z^0}\hat{\Sigma}_{A^{0}G^{0}}(q^{2})$ is zero only for $q^2=M_{A^0}^2$ and not for {\em any} $q^2$. We will get back to the exploitation of this constraint later. A similar constraint relates also $\hat{\Sigma}_{H^\pm W^\pm}(q^{2})$ and $\hat{\Sigma}_{G^\pm H^\pm}(q^{2})$
\begin{eqnarray}
q^{2}\hat{\Sigma}_{H^{\pm}W^{\pm}}(q^{2})+M_{W^{\pm}}\hat{\Sigma}_{H^{\pm}G^{\pm}}(q^{2})&=&
(q^{2}-M_{W^{\pm}}^{2})\frac{1}{(4\pi)^{2}}\frac{e^{2}M_{W^{\pm}}}{s_{2W}^{2}}\mathcal{G}_{HW}^{\tilde{\rho},\tilde{\omega},\tilde{\delta}}(q^2) \nonumber\\ &+&\frac{M_{W^{\pm}}}{2}(q^{2}-M_{H^{\pm}}^{2})\biggl(\frac{1}{(4\pi)^{2}}\frac{2e^{2}}{s_{2W}^{2}}\mathcal{G}_{cc}^{\tilde{\rho},\tilde{\omega},\tilde{\delta}}(q^2)+ s_{2\beta}\frac{ \delta t_{\beta}}{t_{\beta}}-\delta Z_{H^\pm G^\pm}\biggr)\, . \nonumber
\end{eqnarray}
$\mathcal{G}_{HW}^{\tilde{\rho},\tilde{\omega},\tilde{\delta}}(q^2)$ and $\mathcal{G}_{cc}^{\tilde{\rho},\tilde{\omega},\tilde{\delta}}(q^2)$ are defined in Eq.~(\ref{ward-gh}), see Appendix~\ref{app_ward_azag}.

\subsection{Renormalisation conditions}
\subsubsection{Pole masses, residues and mixing}
\label{th-ren-cdt} Masses are defined as pole masses from the
propagator. Moreover this propagator must have residue $1$ at the
pole mass. In the case of particle mixing, the mixing must vanish at
the pole mass of any physical particle, {\it i.e.} at
the pole mass. In general in the case of mixing this requires
solving a system of an inverse propagator matrix with solutions given
by the pole masses. For a 2-particle mixing one has to deal with the
determinant of a $2\times 2$ matrix which is a quadratic form in
the self-energies whose solutions are the corrected masses. The
equation reads
\beqn
\biggl[\biggl(q^2-M_{h^0,{\rm tree}}^2- \hat
\Sigma_{h^{0}h^{0}}(q^2)\biggr)\biggl(q^2-M_{H^0,{\rm tree}}^2- \hat
\Sigma_{H^{0}H^{0}}(q^2)\biggr)- \biggl(\hat
\Sigma_{h^{0}H^{0}}(q^2)\biggr)^2\biggr]=0\, .
\eeqn
$M_{h^0,{\rm tree}}$ refers to the tree-level mass. This equations
simplifies considerably at one-loop since one only has to keep the
linear term, or first order in the loop expansion, in the equation.
In principle the argument that appears in the self-energy two-point
functions is the pole mass which might get a correction from its
value at tree-level. To get the corrections one can proceed through
iteration, starting from the tree-level masses as argument of the
two-point function. Higher order terms in the expansion will appear
as higher orders in the loop expansion and we do not count them as
being part of the one-loop correction. A genuine one-loop results
for the pole mass, $M_{i,{\rm 1loop}}$, starting from a tree-level
mass $M_{i,{\rm tree}}$ with $\hat{\Sigma}_{ii}(q^{2})$ the diagonal
renormalised self-energy is therefore the solution of
\begin{eqnarray}
q^{2}-M_{i,{\rm tree}}^2-Re\hat{\Sigma}_{ii}(q^{2})=0 \quad {\rm
at} \quad q^2=M_{i,{\rm 1loop}}^2 \, ,
\end{eqnarray}
which in the one-loop approximation means
\begin{eqnarray}
M_{i,{\rm 1loop}}^2=M_{i,{\rm
tree}}^2+Re\hat{\Sigma}_{ii}(M_{i,{\rm tree}}^2)=M_{i,{\rm
tree}}^2+ \delta M_{ii}^{2}+Re{\Sigma}_{ii}(M_{i,{\rm tree}}^2) \, .
\label{m1loop-corr}
\eeqn
The latter condition will constrain the Lagrangian parameters with
$\delta M_{ii}^{2}$ a gauge invariant quantity. Likewise, at one-loop,
the requirement of a residue equal to one, for the diagonal
propagator and vanishing mixing when the physical particle is
 on-shell leads to
\beqn
Re\hat{\Sigma}_{ii}^{\prime}(M_{i,{\rm tree}}^2)=0 \quad {\rm
with} \quad \frac{\partial \hat{\Sigma}_{ii}(q^2)}{\partial
q^2}=\hat{\Sigma}_{ii}^{\prime}(q^2) \, ,\nonumber \\
Re\hat{\Sigma}_{ij}^{\prime}(M_{i,{\rm
tree}}^2)=Re\hat{\Sigma}_{ij}^{\prime}(M_{j,{\rm tree}}^2)=0 \,
\quad i \neq j. \label{mix-cdt}
\eeqn
In our renormalisation programme, Eqs.~(\ref{mix-cdt}) set the
field renormalisation constants and avoid having to include
corrections on the external legs. The field renormalisation constants are therefore
not necessarily gauge invariant nor gauge parameter independent.

\subsubsection{Renormalisation conditions and corrections on the mass parameters}
As we have explained earlier one needs to fix the counterterms for
$\delta M_{A^0}^2$ and $\dtb$
once tadpole renormalisation has been
carried through to arrive at finite and gauge invariant $S$-matrix
elements. Taking $M_{A^0}$ as an input parameter means that its mass
is fixed the same at all orders, we therefore set
\beqn
\delta M_{A^0}^{2}=-Re\Sigma_{A^{0}A^{0}}(M_{A^0}^{2}) \, .
\eeqn
Finding a condition to fix $\dtb$ is an arduous task that has been
debated for sometime. We will study many schemes for $\dtb$ in
Section~\ref{section_tb_scheme}. \\

\noi The charged Higgs mass is independent of $\tb$, it gets a
finite correction at one-loop once $M_{A^0}$ is used as an input
parameter
\begin{eqnarray}
M_{H^{\pm}, {\rm 1 loop}}^{2}=M_{H^{\pm}, {\rm
tree}}^{2}+Re\Sigma_{H^{\pm}H^{\pm}}(M_{H^{\pm}, {\rm
tree}}^{2})-Re\Sigma_{A^{0}A^{0}}(M_{A^0}^{2})-Re\Pi_{W^{\pm}}^T(M_{W^{\pm}}^{2}) \, ,
\end{eqnarray}
we have used $\delta M_{W^{\pm}}^2=Re\Pi_{W^{\pm}}^T(M_{W^{\pm}}^{2})$ where
$\Pi_{W^{\pm}}^T(q^{2})$ is the transverse 2-point function of the $W^{\pm}$
following the same implementation as performed in \cite{grace-1loop}. The finiteness of the corrected charged Higgs
mass is the first non trivial check on the
code as concerns the Higgs sector.\\
The sum rule involving the CP-even Higgs masses Eq.~(\ref{trace-mh})
is also independent of $\tb$. This sum rule gets corrected at
one-loop
\begin{eqnarray}
M_{h^{0}, {\rm 1 loop}}^2+ M_{H^{0}, {\rm 1 loop}}^2&=&M_{A^0}^2+M_{Z^0}^2
+Re\Sigma_{h^{0}h^{0}}(M_{h^{0}}^{2})+Re\Sigma_{H^{0}H^{0}}(M_{H^{0}}^{2})
\nonumber\\
&+&\frac{g}{2M_{W^{\pm}}}\biggl(c_{\alpha-\beta}\delta
T_{H^{0}}-s_{\alpha-\beta}\delta
T_{h^{0}}\biggr)-Re\Sigma_{A^{0}A^{0}}(M_{A^0}^{2})-Re\Pi_{Z^{0}Z^{0}}^T(M_{Z^0}^{2})\,
.
\nonumber \\
\label{sumrule_1loop_higgs}
\end{eqnarray}
Here also we have used $\delta M_{Z^0}^2=Re\Pi_{Z^{0}Z^{0}}^T(M_{Z^0}^{2})$
where $\Pi_{Z^{0}Z^{0}}^T(q^{2})$ is the transverse 2-point function of the
$Z^{0}$ boson, see \cite{grace-1loop}. Otherwise to predict
$M_{h^{0}, {\rm 1 loop}}^2$ or $M_{H^{0}, {\rm 1 loop}}^2$ one needs
a prescription on $\dtb$, see Eq.~(\ref{deltaM_all}). Obviously
fixing one of these masses, for instance $M_{H^0}$ in particular in
analogy with $M_{A^0}$, is a scheme for $\tb$. In this scheme
therefore $Re \hat\Sigma_{H^0 H^0}(M_{H^0}^2)=0$ which sets a gauge
invariant counterterm for $\tb$, see Eq.~(\ref{osmh-scheme}).

\subsection{Constraining the field renormalisation constants}
We have introduced through the field renormalisation matrices
$Z_P,Z_C,Z_S$ a total of $12$ such constants, see
Eq.~(\ref{wfr-All}). However as argued repeatedly, some of these
constants are only involved in the transition involving an external
Goldstone bosons, {\em i.e.} in situations that do not correspond to
a physical process. Therefore we can give the constants $\delta
Z_{G^0},\delta Z_{G^\pm}, \delta Z_{A^0 G^0}, \delta Z_{H^\pm
G^\pm}$ any value, $S$-matrix elements will not depend on these
constants. It is therefore easiest to set these $4$ constants to $0$
in actual calculations and give them arbitrary values in preliminary
tests of a calculation of a physical process.\\
For the transitions involving physical Higgs particles we just go
along the general lines described in Section~\ref{th-ren-cdt}, in
order to avoid loop corrections on the external legs. In the
following, in order to avoid too much clutter the masses that will
appear as argument are the tree-level masses (or the input mass for
$M_{A^0}$). The conditions read
\begin{eqnarray}
Re\hat{\Sigma}_{A^{0}A^{0}}^{'}(M_{A^0}^{2})&=&0 \, ,\\
Re\hat{\Sigma}_{H^{\pm}H^{\pm}}^{'}(M_{H^{\pm}}^{2})&=&0 \, ,\\
Re\hat{\Sigma}_{H^{0}H^{0}}^{'}(M_{H^{0}}^{2})&=&0 \, , \\
Re\hat{\Sigma}_{h^{0}h^{0}}^{'}(M_{h^{0}}^{2})&=&0 \, ,\\
Re\hat{\Sigma}_{H^{0}h^{0}}(M_{H^{0}}^{2})&=&0\, ,\quad
Re\hat{\Sigma}_{H^{0}h^{0}}(M_{h^{0}}^{2})=0 \, .\label{residue_cdts}
\end{eqnarray}
From these we immediately derive $6$ out of the $8$ field renormalisation constants in the Higgs sector
\begin{eqnarray}
\delta Z_{A^0}&=&Re{\Sigma}_{A^{0}A^{0}}^{'}(M_{A^0}^{2}) \, ,\label{dzA}\\
\delta Z_{H^\pm}&=&Re{\Sigma}_{H^{\pm}H^{\pm}}^{'}(M_{H^{\pm}}^{2}) \, , \\
\delta Z_{H^0}&=&Re{\Sigma}_{H^{0}H^{0}}^{'}(M_{H^{0}}^{2}) \, ,\\
\delta Z_{h^0}&=&Re{\Sigma}_{h^{0}h^{0}}^{'}(M_{h^{0}}^{2}) \, ,\\
\delta Z_{h^{0}H^{0}}&=&2
\frac{Re{\Sigma}_{H^{0}h^{0}}(M_{H^{0}}^{2})+\delta
M_{H^{0}h^{0}}^2}{M_{H^0}^2-M_{h^0}^2} \, ,\\
\delta Z_{H^{0}h^{0}}&=&2
\frac{Re{\Sigma}_{H^{0}h^{0}}(M_{h^{0}}^{2})+\delta
M_{H^{0}h^{0}}^2}{M_{h^0}^2-M_{H^0}^2} \, .
\end{eqnarray}
When considering a process with $A^0$ as an external
leg\footnote{The argument with the charged Higgs is exactly the
same, therefore we will not make explicit the detailed derivation of
the field renormalisation constant $\delta Z_{G^\pm H^\pm}$ but only
quote the result.}, in principle it involves the $A^0 \ra A^0$
transition but also the $A^0 \ra Z^0$ and the $A^0 \ra G^0$
transitions. The field renormalisation constant $\delta Z_{A^0}$,
see Eq.~(\ref{dzA}) allows to set the $A^0 \ra A^0$ transition to $0$ and moves
its effect to a vertex counterterm correction. One therefore would
be tempted by setting $\hat{\Sigma}_{A^{0}Z^0}(M_{A^0}^2)=0$ together
with $\hat{\Sigma}_{A^{0}G^0}(M_{A^0}^2)=0$ as is done almost
everywhere in the literature. In our case this would mean that the remaining constant
$\delta Z_{G^0 A^0}$ could be derived equivalently from one of these
conditions. However the Ward identity we derived in
Eq.~(\ref{constraint-az-ag2}) imposes a very important constraint.
It shows that in a general non-linear gauge we can not impose
\underline{both} $\hat{\Sigma}_{A^{0}Z^{0}}(M_{A^0}^2)=0$ and
$\hat{\Sigma}_{A^{0}G^0}(M_{A^0}^2)=0$. It looks at first sight that
this requires that one introduces loop corrections on the external
legs when considering for example processes with the pseudoscalar
Higgs as an external leg. In the linear gauge on the other hand this
is possible since
$\mathcal{F}_{GA}^{\tilde{\epsilon},\tilde{\gamma}}(q^2)=0$, we
could then adjust $\delta Z_{G^{0}A^{0}}$ \underline{and} $\delta
Z_{A^{0}G^{0}}$ to have $\hat{\Sigma}_{A^{0}Z^{0}}(M_{A^0}^2)=0$
\underline{and} $\hat{\Sigma}_{A^{0}G^0}(M_{A^0}^2)=0$. Note however
that contrary to what we encounter in some publications, see for
example \cite{DabelsteinHdecay,ShanHiggs},
$q^{2}\hat{\Sigma}_{A^{0}Z^{0}}(q^{2})+M_{Z^0}\hat{\Sigma}_{A^{0}G^{0}}(q^{2})$
does not vanish for any value of $q^2$ but only for
$q^2=M_{A^0}^2$\footnote{The charged counterpart of this identity is
also not valid for {\em any} $q^2$ as is assumed sometimes,
see \cite{logan02}.}.

\noi Let us show how despite the constraint in
Eq.~(\ref{constraint-az-ag2}) we can still avoid one-loop
corrections and counterterms in the external legs associated with an
external pseudoscalar $A^0$. Of concern to us are the transition
$A^0-Z^0$ and $A^0-G^0$. The idea is that although we can not make
both $\hat{\Sigma}_{A^{0}Z^{0}}(M_{A^0}^2)=0$ and
$\hat{\Sigma}_{A^{0}G^0}(M_{A^0}^2)=0$, we will try to make the
combined contribution to the external leg vanish. This combined
contribution is pictured in Fig.~\ref{fig-az-ag}.
\begin{figure}[htb]
\begin{center}
\hspace*{-3cm}
\includegraphics[width=0.75\textwidth]{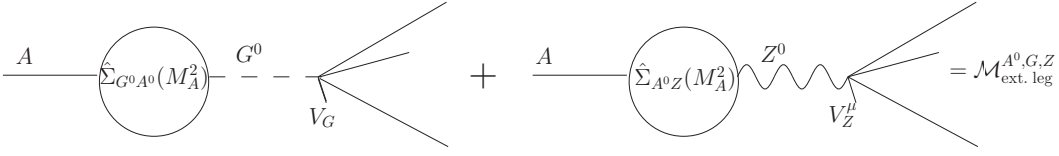}
\caption{\label{box_diag_3D_plots}{\em The combined contribution
of the $A^0-Z^0$ and $A^0-G^0$ transitions}} \label{fig-az-ag}
\end{center}
\end{figure}

\noi To the tree-level coupling of the $A^0$ to some vertex $V$, at
one-loop the transition $A^0-G^0$ involves the coupling of the
tree-level neutral Goldstone to this vertex, $V_G$ while the $Z^{0}$
transition involves the corresponding vertex $V_Z^{\mu}$. The total contribution of Fig.~\ref{fig-az-ag} for $A^0$ with momentum $q$ on-shell with $q^2=M_{A^0}^2$ writes
\beqn
{\cal M}^{A^{0},{G,Z}}_{{\rm
ext.\;leg}}&=&\frac{\hat{\Sigma}_{A^{0}G^0}(M_{A^0}^2) V_G \;+\; q.V_Z
\hat{\Sigma}_{A^{0}Z^{0}}(M_{A^0}^2)}{M_{A^0}^2-M_{Z^0}^2}\nonumber \\
&=&\frac{V_G}{M_{A^0}^2-M_{Z^0}^2} \left(\hat{\Sigma}_{A^{0}G^0}(M_{A^0}^2)
\;+\; M_{Z^0} \hat{\Sigma}_{A^{0}Z^{0}}(M_{A^0}^2) \right) \, .\label{mazg}
\eeqn

\noi In the second step of Eq.~(\ref{mazg}) we used another identity
that can be readily derived at tree-level from the invariance of
the Lagrangian under gauge transformations\footnote{Consider the
part of the Lagrangian with the $Z^{0}$ and the neutral Goldstone
$G^0$. Before gauge-fixing this Lagrangian is invariant under the
transformation $Z^{0}_\mu \ra Z^{0}_\mu + i\partial_\mu \omega$, $G^0 \ra
G^0+M_{Z^0}\omega$. If the $Z^{0}$ (vector) current is $V_Z^\alpha$ and the
Goldstone current $V_G$, that is we have the interaction $Z^0 .V_Z +
G^0 V_G$, invariance of the Lagrangian implies $-i
\partial_\alpha V_Z^\alpha +M_{Z^0} V_G=0$. In Eq.~(\ref{mazg}), this implies $q.V_Z=M_{Z^0}V_G$ where $q$ is the $Z^0$ momentum.}. Therefore in order not
to deal with any correction on the external pseudo-scalar leg we
require
\beqn
\hat{\Sigma}_{A^{0}G^{0}}(M_{A^0}^2) \;+\; M_{Z^0}
\hat{\Sigma}_{A^{0}Z^{0}}(M_{A^0}^2) = 0\, .\label{wfr-azg-cdt}
\eeqn

\noi For this requirement Eq.~(\ref{wfr-azg-cdt}), which is a
renormalisation condition, to be consistent with the Ward identity
in Eq.~(\ref{constraint-az-ag2}) leads to
\beqn
\hat{\Sigma}_{A^{0}Z^{0}}(M_{A^0}^2)=-\frac{1}{M_{Z^0}}
\hat{\Sigma}_{A^{0}G^0}(M_{A^0}^2)=\frac{1}{(4\pi)^{2}}\frac{e^{2}M_{Z^0}}{s_{2W}^{2}}s_{2\beta}\mathcal{F}_{GA}^{\tilde{\epsilon},\tilde{\gamma}}(M_{A^0}^2) \, .
\eeqn

\noi In particular with $\mathcal{F}_{GA}^{\tilde{\epsilon},\tilde{\gamma}}(M_{A^0}^2)=0$ in the linear
gauge, we can make $\hat{\Sigma}_{A^{0}Z^{0}}(M_{A^0}^2)=
\hat{\Sigma}_{A^{0}G^0}(M_{A^0}^2)=0$. This condition readily gives
\beqn
\delta
Z_{G^{0}A^{0}}=-s_{2\beta}\frac{\delta t_{\beta}}{t_{\beta}}
-2 \frac{\Sigma^{\textrm{tad}}_{A^{0}Z^{0}}(M_{A^0}^2)}{M_{Z^0}}+
\frac{2}{(4\pi)^{2}}\frac{e^{2}}{s_{2W}^{2}}s_{2\beta}\mathcal{F}_{GA}^{\tilde{\epsilon},\tilde{\gamma}}(M_{A^0}^2) \, .
\label{dzga}
\eeqn

\noi Since $\delta Z_{A^{0}G^{0}}$ only enters in off-shell
processes, $A^{0}$ off-shell or an external Goldstone boson, there
is no need to constrain it through some other renormalisation
condition. Our aim, as stressed repeatedly, is not to renormalise
all Green's functions, but only S-matrix elements without the need
for external leg corrections. The Ward identities that we derived in
this section were, numerically, checked extensively in our code
for various values of $q^2$ including $q^2=M_{A^0}^2$ and $q^2=M_{Z^0}^2$
and for different values of the non-linear gauge parameters.
Moreover it is thanks to the
$\mathcal{F}_{GA}^{\tilde{\epsilon},\tilde{\gamma}}(M_{A^0}^2)$
contribution in $\delta Z_{G^{0}A^{0}}$ that we are able to obtain
finite and gauge invariant results for processes involving $A^0$ as
an external particle. For $\delta Z_{G^{\pm}H^{\pm}}$ a similar
derivation gives
\beqn
\delta Z_{G^{\pm}H^{\pm}}=-s_{2\beta}\frac{\delta t_{\beta}}{t_{\beta}}
-2 \frac{\Sigma^{\textrm{tad}}_{H^{\pm}W^{\pm}}(M_{H^{\pm}}^2)}{M_{W^{\pm}}}+
\frac{2}{(4\pi)^{2}}\frac{e^{2}}{s_{2W}^{2}}\mathcal{G}_{HW}^{\tilde{\rho},\tilde{\omega},
\tilde{\delta}}(M_{H^{\pm}}^2)
\, .
\eeqn
With $\delta Z_{G^{0}A^{0}}$ (and $\delta Z_{G^{\pm}H^{\pm}}$) all
our field renormalisation constants are set and defined.

\section{Definitions of $\tb$ and the $\tb$ schemes}
\label{section_tb_scheme}
\subsection{Dabelstein-Chankowski-Pokorski-Rosiek Scheme (DCPR)}
This scheme, which we will refer to as the DCPR scheme, has been
quite popular and is based on an OS renormalisation scheme in the
Higgs sector \cite{DabelsteinHiggs, DCPR} working in the usual
linear gauge. The definition of $\tb$ however is difficult to
reconcile with an On-Shell quantity that represents a direct
interpretation in terms of a physical observable. One first
introduces a wave function renormalisation constant, $\delta
Z_{H_{i}}$, for each Higgs doublet $H_i$, {\it i.e.} before rotation
\begin{eqnarray}
H_{i}\rightarrow (1+\frac{1}{2}\delta Z_{H_{i}})H_{i} \quad
{i=1,2} \, .
\end{eqnarray}
To make contact with our approach and parameters, as concerns wave
function renormalisation, we refer to
Appendix~\ref{appendix-wfr-dcpr}. The vacuum expectation values are
also shifted such that the counterterm for each $v_i$ writes
\begin{eqnarray}
v_{i}\rightarrow v_{i}\left(1-\frac{\tilde{\delta}
v_{i}}{v_{i}}+\frac{1}{2}\delta Z_{H_{i}} \right) \, ,
\end{eqnarray}
giving
\begin{eqnarray}
\label{dbdcpro} \frac{\delta
t_{\beta}}{t_{\beta}}=\frac{\tilde{\delta}
v_{1}}{v_{1}}-\frac{\tilde{\delta}
v_{2}}{v_{2}}-\frac{1}{2}(\delta Z_{H_{1}}-\delta Z_{H_{2}}) \, .
\end{eqnarray}
The DCPR scheme takes $\frac{\tilde{\delta}
v_{1}}{v_{1}}=\frac{\tilde{\delta} v_{2}}{v_{2}}$ such that in
effect
\begin{eqnarray}
\label{dbdcprt} \frac{\delta
t_{\beta}}{t_{\beta}}=\frac{1}{2}(\delta Z_{H_{2}}-\delta
Z_{H_{1}}) \, .
\end{eqnarray}
$\tb$ is defined by requiring that the (renormalised) $A^{0}Z^{0}$
transition vanish at $q^2=M_{A^0}^2$, therefore from
\begin{eqnarray}
Re\hat{\Sigma}_{A^{0}Z^{0}}(M_{A^0}^{2})=0 \, ,
\end{eqnarray}
with
\begin{eqnarray}
\hat{\Sigma}_{A^{0}Z^{0}}(q^{2})=\Sigma_{A^{0}Z^{0}}(q^{2})+\frac{M_{Z^0}}{4}s_{2\beta}(\delta
Z_{H_{2}}-\delta Z_{H_{1}}+2\frac{\delta t_{\beta}}{t_{\beta}}) \, ,
\end{eqnarray}
one obtains that
\begin{eqnarray}
\frac{\delta
t_{\beta}}{t_{\beta}}^{\textrm{DCPR}}=-\frac{1}{M_{Z^0}s_{2\beta}}Re\Sigma_{A^{0}Z^{0}}(M_{A^0}^{2}) \, .
\end{eqnarray}
This definition is clearly not directly related to an observable.
Moreover $\delta t_{\beta}$ is expressed in terms of wave function
renormalisation constants, see Eq.~(\ref{dbdcprt}).

\subsection{$\overline{\textrm{DR}}$ Scheme
($\overline{\textrm{DR}}$)} In this scheme the counterterm for $\tb$
is taken to be a pure divergence proportional to the ultraviolet
(UV) factor in dimensional reduction, $\cuv$
\beqn
\cuv=2/(4-n)-\gamma_{E}+\ln(4\pi) \, ,
\eeqn
where $n$ is the dimensionality of space-time. In this scheme the
finite part of the counterterm is therefore set zero:
\begin{eqnarray}
\frac{\delta
t_{\beta}^{\textrm{fin}}}{t_{\beta}}^{\overline{\textrm{DR}}}=0 \, .
\end{eqnarray}
The divergent part can be related to a few quantities not
necessarily directly related to an observable. In the vein of the
DCPR approach within the linear gauge where $\delta \tb$ is defined
in Eq.~(\ref{dbdcprt}), solving for $\delta Z_{H_{2}}-\delta
Z_{H_{1}}$ leads to the HHW prescription of Hollik, Heinemeyer and
Weiglein \cite{HHWtb}, see also Eq.~(\ref{Holliktb}),
\begin{eqnarray}
\frac{\delta
t_{\beta}}{t_{\beta}}^{\overline{\textrm{DR}}-\textrm{HHW}}=\frac{1}{2c_{2\alpha}}(Re\Sigma_{h^{0}h^{0}}^{'}(M_{h^0}^{2})
-Re\Sigma_{H^{0}H^{0}}^{'}(M_{H^{0}}^{2}))^{\infty} \, .
\end{eqnarray}
The superscript $^{\infty}$ means that only the infinite $C_{UV}$
part in dimensional reduction is taken into account. A more
satisfactory $\overline{\textrm{DR}}$ scheme can be based on a
physical observable. Pierce and Papadopoulos \cite{RCHiggstoZZ} have
defined $\delta t_{\beta}$ by relating it to the {\em divergent} part of
$M_{H^0}^2-M_{h^0}^2$. Note that the sum $M_{H^0}^2+M_{h^0}^2$ does not depend on $t_{\beta}$ 
as can be seen from the tree-level sum rule in Eq.~(\ref{trace-mh}). Hence,
see also Eq.~(\ref{sumrule_1loop_higgs}),
\beqn
\frac{\delta
t_{\beta}}{t_{\beta}}^{\overline{\textrm{DR}}-\textrm{PP}}&=&\frac{1}{2s_{2\beta}s_{2(\alpha+\beta)}M_{Z^0}^{2}}
\biggl(\frac{1}{v}(c_{\alpha-\beta}(1+2s_{\alpha-\beta}^{2})\delta T_{H^{0}}+s_{\alpha-\beta}(1+2c_{\alpha-\beta}^{2})\delta T_{h^{0}}) \nonumber \\
&+&Re\Sigma_{H^{0}H^{0}}(M_{H^{0}}^{2})-Re\Sigma_{h^{0}h^{0}}(M_{h^{0}}^{2})+c_{2(\alpha+\beta)}Re\Sigma_{A^{0}A^{0}}(M_{A^{0}}^{2})
-c_{2(\alpha+\beta)}Re\Pi_{Z^{0}Z^{0}}^T(M_{Z^0}^{2})\biggr)^{\infty} \, .\nonumber \\
\eeqn

\subsection{An On-Shell Scheme (${\rm OS}_{M_H}$) with $M_{H^0}$ as an input}
In this scheme one takes $M_{H^0}$, the largest of the two scalar
Higgs masses, as an input parameter. This trade-off is operative
in the Higgs sector independently of any process. Therefore $M_{H^0}$
is no longer a prediction but is extracted from a measurement
together with $M_{A^0}$. As such it does not receive a correction at
any loop order, $\dtb$
is defined from the constraint
\begin{eqnarray}
Re\hat{\Sigma}_{H^{0}H^{0}}(M_{H^{0}}^{2})=0 \, ,
\end{eqnarray}
which leads to
\beqn
\label{osmh-scheme}
 \frac{\delta t_{\beta}}{t_{\beta}}^{\textrm{${\rm
OS}_{M_{H}}$}}&=&\frac{1}{s_{2\beta}s_{2(\alpha-\beta)}M_{A^0}^{2}}\biggl((c_{\alpha}^{2}-s_{\beta}^{2}s_{\alpha-\beta}^{2})\frac{\delta
T_{\phi_{1}^{0}}}{v_{1}}+(s_{\alpha}^{2}-c_{\beta}^{2}s_{\alpha-\beta}^{2})\frac{\delta
T_{\phi_{2}^{0}}}{v_{2}} \nonumber \\
&+&Re\Sigma_{H^{0}H^{0}}(M_{H^{0}}^{2})-s_{\alpha-\beta}^{2}Re\Sigma_{A^{0}A^{0}}(M_{A^{0}}^{2})
-c_{\alpha+\beta}^{2}Re\Pi_{Z^{0}Z^{0}}^T(M_{Z^{0}}^{2})\biggr) \, .
\eeqn
This scheme has been advocated in \cite{ShanHiggs,Grace-susy1loop}
and is one of the scheme implemented in {\tt SloopS}. At tree-level,
$\tb$ is extracted from the relation defined in Eq.~(\ref{det-mh})
\begin{eqnarray}
\label{c2bHiggsdef}
c_{2\beta}^{2}=\frac{(M_{A^{0}}^{2}+M_{Z^{0}}^{2}-M_{H^{0}}^{2})M_{H^{0}}^{2}}{M_{A^0}^{2}M_{Z^0}^{2}} \, .
\label{c2bmh}
\end{eqnarray}
In our numerical examples the input parameters are such that the
requirement $c_{2\beta}^{2} \leq 1$ is always met. In fact given a
set $M_{A^0},M_{Z^0}$ we {\em generate} $M_{H^0}$ through a given value of
$\tb$. The value $M_{H^0}$ is taken as the physical mass at all loop
orders, in particular at one-loop it does not receive a correction.
As pointed out in Section~2, in general with a set $M_{H^0},M_{A^0},M_{Z^0}$
$c_{2\beta}^{2} \leq 1$ is not guaranteed. With this important
proviso, we extract $\tgb$ (with $\tgb >1$) as
\begin{eqnarray}
t_{\beta}=\sqrt{\frac{M_{A^0}M_{Z^0}+M_{H^0}\sqrt{M_{A^0}^2+M_{Z^0}^2-M_{H^0}^2}}{M_{A^0}M_{Z^0}-M_{H^0}\sqrt{M_{A^0}^2+M_{Z^0}^2-M_{H^0}^2}}}
\, .
\end{eqnarray}
That this choice might lead to large corrections and large
uncertainty can already be guessed by considering the uncertainty
on $\tgb$ given an uncertainty on $M_{H^0},M_{A^0},M_{Z^0}$ with respectively
$\delta M_{H^0},\delta M_{A^0},\delta M_{Z^0}$. For clarity let us take
$\delta M_{Z^0}=0$ as would be fit from an experimental point of
view since $M_{Z^0}$ is known with an excellent precision from the LEP measurements. We find
\begin{eqnarray}
\frac{\delta t_{\beta}}{t_{\beta}}=\frac{M_{A^0}^2}{M_{H^0}^2-M_{A^0}^2}
\frac{M_{H^0}^2}{M_{H^0}^2-M_{Z^0}^2}
\biggl(-\frac{M_{H^0}^{2}-M_{Z^0}^{2}}{M_{A^0}^2}\frac{\delta
M_{A^0}^{2}}{M_{A^0}^{2}} +
\frac{M_{H^0}^2}{M_{A^0}^2}\frac{2M_{H^0}^{2}-M_{A^0}^{2}-M_{Z^0}^{2}}{M_{H^0}^2}\frac{\delta
M_{H^0}^{2}}{M_{H^0}^{2}} \biggr) \, .
\end{eqnarray}
With typical input parameters in the decoupling limit $M_{A^0} \gg
M_{Z^0}$ with $M_{A^0}/M_{H^0} \sim 1$ a large uncertainty ensues, to wit
\begin{eqnarray}
\frac{\delta t_{\beta}}{t_{\beta}} \simeq \frac{1}{M_{H^0}^2/M_{A^0}^2-1}
\biggl(-\frac{\delta M_{A^0}^{2}}{M_{A^0}^{2}} + \frac{\delta
M_{H^0}^{2}}{M_{H^0}^{2}} \biggr) \, .
\end{eqnarray}
Therefore although $\delta \tb$ is manifestly gauge invariant one
should expect large uncertainty from loop corrections. This scheme
is similar to the one considered in \cite{stockinger02}
based on Eq.~(\ref{c2bHiggsdef}).

\subsection{$A_{\tau \tau}$ as an input parameter (${\rm OS}_{A_{\tau \tau}}$)}
\label{subsec-att} $\beta$ which appears in the Higgs sector relies
on the assumption of a basis, only quantities which are basis
independent are physical quantities \cite{DavidsonHaber,me-3h-2hdm}. The Higgs potential of the MSSM appears as a general two-Higgs
doublet model if one restricts oneself solely to the Higgs sector.
The degeneracy is lifted when defining the Yukawa Higgs coupling to
fermions. This picks up a specific direction. One should therefore
define $\tgb$ from the Higgs couplings to fermions. Since $M_{A^0}$ is
used as an input parameter assuming one has had access to the
pseudoscalar Higgs, it looks natural to take a coupling $A^{0} f\bar
f$. Since couplings to quarks are subject to large QCD radiative
corrections the best choice is to consider the $A_{\tau \tau}$
coupling which is the largest coupling to leptons,
\beqn
{\cal {L}}_{A_{\tau \tau}}^0= i \frac{m_\tau}{v_1} \xb \; \bar
\tau \gamma_5 \tau \; A^0 =i \frac{g m_\tau}{2 M_{W^{\pm}}} \; \tb \; \bar
\tau \gamma_5 \tau \; A^0 \quad {\rm with} \quad v_1= v \cb \, .
\eeqn
This coupling can be extracted from the measurement of the width
$\Gamma_{A_{\tau \tau}}$ with $m_{\tau}$ the mass of the $\tau$.
Note also that $\delta \Gamma_{A_{\tau \tau}}=2\dtb/\tb $ so that
contrary to the On-Shell scheme based on $M_{H^0}$, ${\rm OS}_{M_H}$,
this scheme should therefore not introduce additional large
uncertainties assuming of course that this decay can be large and be
measured precisely. This scheme appears therefore very natural,
however it has not been used in practice because one has considered
it as being a {\em process dependent} definition set outside the
purely Higgs sector which moreover implies that fixing the
counterterm involves a three-point function. This last argument is
unjustified, take for example the $G_\mu$ scheme in the SM where muon decay is
used as a trade-off for $M_{W^{\pm}}$ taking advantage of the fact that
$G_\mu$ has been for a long time so much better measured than $M_{W^{\pm}}$.
The $G_\mu$ scheme involves four-point functions. We find that
technically this scheme is not more difficult to implement than a
scheme based on two-point functions. The full counterterm to
$A_{\tau \tau}$ involves the $G^{0} \ra A^{0}$ shift, the $A^{0}$
and $\tau^{\pm}$ wave function renormalisation constants among other
things, we get
\beqn
\delta {\cal L}_{A_{\tau \tau}}&=&{\cal L}_{A_{\tau \tau}}^0
\biggl(
\delta_{\rm CT}^{A_{\tau \tau}} + \frac{\dtb}{\tb} \biggr) \quad {\rm with} \nonumber \\
\delta_{\rm CT}^{A_{\tau \tau}}&=&\biggl( \frac{\delta
m_\tau}{m_\tau} + \frac{\delta e}{e} +\frac{c_W^2}{2 s_W^2}
\frac{\delta M_{W^{\pm}}^2}{M_{W^{\pm}}^2}-\frac{1}{2 s_W^2} \frac{\delta
M_{Z^0}^2}{M_{Z^0}^2} + \frac{1}{2} \delta
Z_{A^{0}A^{0}}-\frac{1}{2\tb} \delta \tilde{Z}_{G^{0}A^{0}} \nonumber \\
& & \;\;\;\;\;\;\;\;\; + \frac{1}{2}(\delta Z_L^\tau +\delta
Z_R^\tau) \biggr) \, ,\nonumber \\
-\frac{1}{2 \tb}\delta\tilde{Z}_{G^{0}A^{0}}&=&\frac{1}{\tb}
\frac{\Sigma_{A^{0}Z^{0}}(M_{A^0}^2)}{M_{Z^0}}-\frac{1}{1+\tb^2}
\frac{\alpha}{2\pi} M_{Z^0} {\cal F}_{GA}^{\tilde{\epsilon},\tilde{\gamma}} (M_{Z^0}^2)\, .\label{CT-att}
\eeqn
$\delta m_{\tau}$, the $\tau$ mass counterterm, $\delta e$ the
electromagnetic coupling counterterm, $\delta M_{W^{\pm},Z^{0}}$ the gauge
bosons mass counterterms and the $\tau$ wave function
renormalisation constant $\delta Z_{L,R}^\tau$ counterterms are
defined on-shell exactly as in the SM \cite{grace-1loop}. The full
one-loop virtual corrections consist of the vertex corrections,
$\delta_{\rm V}^{A_{\tau \tau}}$ which contributes a one-loop vertex
correction to the decay rate as:
\beqn
\delta \Gamma_{1}^{{\rm Vertex}}=2 \Gamma_0 \; \delta_{\rm
V}^{A_{\tau \tau}} \, .
\eeqn
The latter are made UV-finite by the addition of the counterterm in
Eq.~(\ref{CT-att}). These virtual QED corrections, both vertex and
counterterm (from $\delta m_\tau$ and $\delta Z_{L,R}^\tau$) include
genuine QED corrections through photon exchange which are infrared
divergent. In our case the infrared divergence can be trivially
regularised through the introduction of a small fictitious mass,
$\lambda$, for the photon. As known, the fictitious mass dependence
is cancelled when photon bremmstrahlung is added. Taking into
account the latter may depend on the experimental set-up that often
requires cuts on the additional photon kinematical variables.
Therefore it is much more appropriate to take as an observable a
quantity devoid of such cuts, knowing that hard/soft radiation can
be easily added. Fortunately for a {\em neutral} decay such as this
one which is of an Abelian nature, the virtual QED correction
constitutes a gauge-invariant subset that can be trivially
calculated separately. The virtual QED corrections to the decay
width $A^{0} \ra \tau^+ \tau^-$ are known \cite{Drees-qcd-aqq}, they
contribute a one-loop correction
\begin{eqnarray}
\delta \Gamma_{1}^{QED}&=&2\Gamma_{0} \; \delta_v^{QED} \quad {\rm
with} \nonumber \\
 \delta_v^{QED}&=&
\frac{\alpha}{2\pi}\Biggl( -\left(\frac{1+\beta^2}{2\beta}\ln\frac{1+\beta}{1-\beta}-1\right)\ln \frac{m_\tau^2}{\lambda^2}-1 \\
&+& \frac{1+\beta^2}{\beta}\left[\textrm{Li}_{2}\left(\frac{1-\beta}{1+\beta}\right)+\ln\frac{1+\beta}{2\beta}\ln\frac{1+\beta}{1-\beta}-\frac{1}{4}\ln^{2}\frac{1+\beta}{1-\beta}+\frac{\pi^{2}}{3}\right]\Biggr) \, ,\nonumber\\
\beta&=&\sqrt{1-\frac{4m_\tau^2}{M_{A^0}^{2}}} \, ,\\
\textrm{Li}_{2}(x)&=&-\int_{0}^{x}\frac{dt}{t}\ln(1-t) \, .
\end{eqnarray}
This QED correction only depends on $M_{A^0},e,m_\tau$ as it should
and does not involve any other (MSSM) parameter. Subtracting this
QED correction from the full one-loop virtual correction in
Eq.~(\ref{CT-att}) will give the genuine SUSY non QED contribution
that does not depend on any fictitious photon mass nor any
experimental cut. Our scheme is to require that $\dtb$ is such that
this contribution vanishes and that therefore $A^{0} \ra \tau^{+} \tau^{-}$
is only subject to QED corrections. This gives
\begin{eqnarray}
\frac{\delta t_{\beta}}{t_{\beta}}^{{\rm OS}_{A_{\tau \tau}}}=
-\biggl( \delta_{\rm V}^{A_{\tau \tau}}+\delta_{\rm CT}^{A_{\tau
\tau}} -\delta_v^{QED} \biggr) \, .
\end{eqnarray}
This definition is independent of the fictitious mass of the photon $\lambda$ used as a regulator. We have checked this explicitly within {\tt SloopS}.

\section{Set-up of the automatic calculation of the cross sections}
All the steps necessary for the renormalisation of the Higgs sector
as presented here together with a complete definition of the MSSM
have been implemented in {\tt SloopS}. As we will discuss in a
forthcoming publication \cite{Sloops-allren} the other sectors have
also been implemented and results relying on the complete
renormalisation of the MSSM have been given in \cite{baro07}. Since
even the calculation of a single two-point function in the MSSM
requires the calculation of a hundred of diagrams, some
automatisation is unavoidable. Even in the \smp, one-loop
calculations of $2\ra 2$ processes involve hundreds of diagrams and
a hand calculation is practically impracticable. Efficient automatic
codes for any generic $2\ra 2$ process, that have now been exploited
for many $2\ra
 3$ \cite{grace2to3,other2to3} and even some $2\ra
 4$ \cite{grace2to4,Dennereeto4f} processes, are almost unavoidable
for such calculations. For the electroweak theory these are the
{\tt GRACE-loop} \cite{grace-1loop} code and the bundle of packages
based on {\tt FeynArts} \cite{feynarts}, {\tt
FormCalc} \cite{formcalc} and {\tt LoopTools} \cite{looptools}, that we will refer to as {\tt FFL} for short. \\
\noindent With its much larger particle content, far greater
number of parameters and more complex structure, the need for an
automatic code at one-loop for the Minimal Supersymmetric Standard
Model is even more of a must. A few parts that are needed for such
a code have been developed based on an extension of \cite{FeynArtsusy} but, as far as we know, no complete code exists
or is, at least publicly, available. {\tt Grace-susy} \cite{grace-susy} is now also being developed at one-loop and many
results exist \cite{Grace-susy1loop}. One of the main difficulties
that has to be tackled is the implementation of the model file,
since this requires that one enters the thousands of vertices that
define the Feynman rules. On the theory side a proper
renormalisation scheme needs to be set up, which then means
extending many of these rules to include counterterms. When this
is done one can just use, or hope to use, the machinery developed
for the \smp, in particular the symbolic manipulation part and
most importantly the loop integral routines including tensor
reduction algorithms or any other efficient set of basis integrals.\\

\noi {\tt SloopS} combines {\tt LANHEP} \cite{lanhep} (originally part of
the package {\tt COMPHEP} \cite{comphep}) with the {\tt FFL} bundle
but with an extended and adapted {\tt LoopTools} \cite{sloopsgg}.
{\tt LANHEP} is a very powerful routine that {\em automatically}
generates all the sets of Feynman rules of a given model, the
latter being defined in a simple and compact format very similar to
the canonical coordinate representation. Use of multiplets and the
superpotential is built-in to minimize human error. The ghost
Lagrangian is derived directly from the BRST transformations. The
{\tt LANHEP} module also allows to shift fields and parameters and
thus generates counterterms most efficiently. Understandably the
{\tt LANHEP} output file must be in the format of the model file of
the code it is interfaced with. In the case of {\tt FeynArts} both
the {\it generic} (Lorentz structure) and {\it classes} (particle
content) files had to be given. Moreover, because we use a
non-linear gauge fixing condition \cite{grace-1loop}, see below, the
{\tt FeynArts} default {\it generic} file had to be extended.

\section{$\tb$ scheme dependence of physical observables, gauge invariance: A numerical investigation}
In this first investigation we will restrict ourselves to Higgs
observables. Other observables involving other supersymmetric
particles require that we first expose and detail our
renormalisation procedure of the chargino/neutralino and the
sfermion sector. This will be presented in \cite{Sloops-allren}. We
have however presented some results on the $\tgb$ scheme dependence
of a few cross sections that are needed for the calculation of the
relic density in the MSSM \cite{baro07}.

\subsection{Parameters}
To make contact with the analysis of \cite{stockinger02} and also
allow comparisons we will consider the $3$ sets of benchmarks points
for the Higgs based on \cite{BenchmarkHiggstb}. The $3$ sets of
parameters called \textit{mhmax}, \textit{large $\mu$} and
\textit{nomix} are as in \cite{BenchmarkHiggstb} except that we set
a common tri-linear $A_f$ to all sfermions for convenience. For each
set there are two values of $\tb$, $\tb=3,50$.
\\
\begin{table}[h]
\begin{center}
\begin{tabular}{|c|c|}
\hline Parameter&Value\\
\hline
$s_{W}$&0.48076\\
$e$&0.31345\\
$g_{s}$&1.238\\
$M_{Z^{0}}$&91.1884\\
$m_{e}$&0.000511\\
\hline
\end{tabular}
\begin{tabular}{|c|c|}
\hline Parameter&Value\\
\hline
$m_{\mu}$&0.1057\\
$m_{\tau}$&1.777\\
$m_{u}$&0.046\\
$m_{d}$&0.046\\
$m_{c}$&1.42\\
\hline
\end{tabular}
\begin{tabular}{|c|c|}
\hline Constant&Value\\
\hline
$m_{s}$&0.2\\
$m_{t}$&174.3\\
$m_{b}$&3\\
$M_{A^0}$&500\\
$t_{\beta}$&3;50\\
\hline
\end{tabular}
\end{center}
\begin{center}
\begin{tabular}{|c|c|}
\hline \textit{mhmax} &Value\\
\hline
$\mu$&-200\\
$M_{2}$&200\\
$M_{3}$&800\\
$M_{\tilde{F}_L}$&1000\\
$M_{\tilde{f}_R}$&1000\\
$A_{f}$&2000+$\mu/t_{\beta}$\\
\hline
\end{tabular}
\begin{tabular}{|c|c|}
\hline \textit{nomix} &Value\\
\hline
$\mu$&-200\\
$M_{2}$&200\\
$M_{3}$&800\\
$M_{\tilde{F}_L}$&1000\\
$M_{\tilde{f}_R}$&1000\\
$A_{f}$&$\mu/t_{\beta}$\\
\hline
\end{tabular}
\begin{tabular}{|c|c|}
\hline \textit{large $\mu$} &Value\\
\hline
$\mu$&1000\\
$M_{2}$&400\\
$M_{3}$&200\\
$M_{\tilde{F}_L}$&400\\
$M_{\tilde{f}_R}$&400\\
$A_{f}$&-300+$\mu/t_{\beta}$\\
\hline
\end{tabular}
%\end{tiny}
\caption{{\em The set of SM and MSSM parameters for the benchmark
points. All mass parameters are in GeV. We take $M_1$ according
to the so-called gaugino mass unification with
$M_{1}=\frac{5s_{W}^{2}M_{2}}{3c_{W}^{2}}$. }
 \label{tab-param-higgs} }
\end{center}
\end{table}

\subsection{Gauge independence and the finite part of $\tb$}
If $\tb$ is defined as a physical parameter then $\dtb$
must be gauge invariant and gauge parameter independent. Our non-linear
gauge fixing allows us to check the gauge parameter independence
of $\dtb$ and hence $\tb$. Even when two schemes are gauge
parameter independent the values of $\dtb$
are not expected to be
the same. It is therefore also interesting to inquire how much two
schemes differ from each other. Naturally since $\dtb$ is not ultraviolet finite we split this contribution into a finite part
and infinite part, the latter being regularised in dimensional
reduction, such that
\begin{eqnarray}
\delta t_{\beta}=\delta t_{\beta}^{\textrm{fin}}+\delta
t_{\beta}^{\infty}C_{UV}\, \, .
\end{eqnarray}
The $\overline{\textrm{DR}}$ schemes have by definition $\delta
t_{\beta}^{\textrm{fin}}=0$. When calculating observables in this
scheme we will also need to specify a scale $\bar \mu$ which we
associate with the scale introduced by dimensional reduction. For
the latter our default value is $\bar \mu=M_{A^0}$. Our set of
non-linear gauge parameters is
defined as nlgs = ($\tilde{\alpha}$, $\tilde{\beta}$, $\tilde{\delta}$, $\tilde{\omega}$, $\tilde{\rho}$, $\tilde{\kappa}$, $\tilde{\epsilon}$, $\tilde{\gamma}$).\\
The usual linear gauge, nlgs = 0, corresponds to all these parameters
set to $0$. For the gauge parameter independence we will compare the
results of the linear gauge to a non-linear gauge where all the
non-linear gauge parameters have been set to $10$, referring to this
as nlgs = 10.

\noi To make the point about the gauge parameter dependence it is
enough to consider only one of the benchmarks points.

\begin{table}[h]
\begin{center}
\begin{tabular}{|c|c|c|}
\hline $\delta t_{\beta}^{\infty}$ & nlgs = 0 & nlgs = 10 \\
\hline
DCPR& -3.19$\times 10^{-2}$ & -1.04$\times 10^{-1}$ \\
${\rm OS}_{M_{H}}$& -3.19$\times 10^{-2}$ & -3.19$\times 10^{-2}$\\
${\rm OS}_{A_{\tau \tau}}$& -3.19$\times 10^{-2}$ & -3.19$\times 10^{-2}$\\
$\overline{\textrm{DR}}$-HHW & -3.19$\times 10^{-2}$ & +5.32$\times 10^{-2}$\\
$\overline{\textrm{DR}}$-PP & -3.19$\times 10^{-2}$ & -3.19$\times 10^{-2}$\\
\hline
\end{tabular}
\begin{tabular}{|c|c|c|}
\hline $\delta t_{\beta}^{\textrm{fin}}$ & nlgs = 0 & nlgs = 10 \\
\hline
DCPR& -0.10 & -0.27 \\
${\rm OS}_{M_{H}}$& +0.92 & +0.92\\
${\rm OS}_{A_{\tau \tau}}$& -0.10 & -0.10\\
$\overline{\textrm{DR}}$-HHW & 0 & 0\\
$\overline{\textrm{DR}}$-PP & 0 & 0\\
\hline
\end{tabular}\\
\caption{{\em Gauge dependence of $\delta t_{\beta}$ at the scale
$\overline{\mu}=M_{A^{0}}$ for the set \textit{mhmax} at
$t_{\beta}=3$.}\label{dtbgd}}
\end{center}
\end{table}
\noi As expected we see from Table~\ref{dtbgd} that only the schemes
based on a physical definition of $\tb$ are gauge parameter
independent. Therefore neither DCPR nor a $\overline{\textrm{DR}}$
manifestation of it based on \cite{HHWtb} are gauge independent.
Within the physical definitions note that although the divergent
part is, as expected, the same for all the schemes in all gauges,
the finite parts are quite different from each other, in particular
the ${\rm OS}_{M_{H}}$ scheme introduces a ``correction" of about
$30\%$ to $\tb$. This is just an indication that this scheme might
induce large corrections on observables. However one needs to be
cautious, in the same way that the $C_{UV}$ part cancels in
observables, a large finite correction could, in principle, also be
absorbed when we consider a physical process. Our rather extensive
analysis will show that this is, after all, not the case. Schemes
where the finite part of $\dtb$ is large do, generally, induce large
corrections. It is important to note that for the linear gauge all
schemes give the same $C_{UV}$ part. Having made the point about the
gauge parameter dependence, we will now work purely in the linear
gauge since some of the schemes introduced in the literature are
{\em acceptable} only within the linear gauge. Therefore in this
case the results for $\overline{\textrm{DR}}$-HHW and
$\overline{\textrm{DR}}$-PP are the same and will be denoted as
$\overline{\textrm{DR}}$ in what follows.

\subsection{$\delta t_{\beta}^{fin}$ }
\begin{table}[h]
\begin{center}
\begin{tabular}{|c|c|c|c|}
\hline $t_{\beta}=3$ & \textit{mhmax} & \textit{large $\mu$} & \textit{nomix}\\
\hline
DCPR& -0.10 & -0.06 & -0.08 \\
${\rm OS}_{M_H}$& +0.92 & -1.31 & +0.64\\
${\rm OS}_{A_{\tau \tau}}$& -0.10 & -0.06 & -0.08 \\
$\overline{\textrm{DR}}$ & 0 & 0 & 0\\
\hline
\end{tabular}
\begin{tabular}{|c|c|c|c|}
\hline $t_{\beta}=50$ & \textit{mhmax} & \textit{large $\mu$} & \textit{nomix}\\
\hline
DCPR& +3.42 & +14.57 & +0.48 \\
${\rm OS}_{M_H}$& -385.53 & -2010.84 & -290.18\\
${\rm OS}_{A_{\tau \tau}}$& +0.12 & -4.72 & +0.16 \\
$\overline{\textrm{DR}}$ & 0 & 0 & 0\\
\hline
\end{tabular}\\
\caption{{\em $\delta t_{\beta}^{fin}$ for the Higgs benchmark
points.}\label{dtbfin}}
\end{center}
\end{table}

\noi First of all let us mention that our numerical results concerning
the DCPR and $\overline{\textrm{DR}}$ schemes
 agree quite well with those of \cite{stockinger02}
concerning the shifts in $\tb$ and the lightest CP-even Higgs mass.
Our results for ${\rm OS}_{M_H}$ follow sensibly the same trend as
the scheme defined as the Higgs mass scheme in \cite{stockinger02}.
We see that for small $\tb$ DCPR and ${\rm OS}_{A_{\tau \tau}}$ give
sensibly the same result with a finite relative shift of a few
percent. For larger $\tb$ the difference is much larger, we notice
that ${\rm OS}_{A_{\tau \tau}}$ gives much smaller shifts. On the
other hand the ${\rm OS}_{M_H}$ gives huge corrections for $\tb=50$
well above $100\%$. As we will see this will have an impact on the
radiative corrections on some observables based on this scheme.

\subsection{Higgs masses and their scheme dependence}
\begin{table}[h]
\begin{center}
\begin{tabular}{|c|c|c|c|}
\hline $t_{\beta}=3$ & \textit{mhmax} & \textit{large $\mu$} & \textit{nomix}\\
\hline
$M_{h^0}^{TL}=72.51$ & & & \\
\hline
DCPR& 134.28 & 97.57 & 112.26 \\
${\rm OS}_{M_H}$& 140.25 & 86.68 & 117.37\\
${\rm OS}_{A_{\tau \tau}}$& 134.25 & 97.59 & 112.27 \\
$\overline{\textrm{DR}}$ $\overline{\mu}=M_{A^{0}}$ & 134.87 & 98.10 & 112.86\\
$\overline{\textrm{DR}}$ $\overline{\mu}=M_{t}$ & 134.47 & 97.55 & 112.38\\
\hline
\end{tabular}
\begin{tabular}{|c|c|c|c|}
\hline $t_{\beta}=50$ & \textit{mhmax} & \textit{large $\mu$} & \textit{nomix}\\
\hline
$M_{h^0}^{TL}=91.11$ & & & \\
\hline
DCPR& 144.50 & 35.88 & 124.80 \\
${\rm OS}_{M_H}$& 143.76 & 13.21 & 124.16\\
${\rm OS}_{A_{\tau \tau}}$& 144.50 & 35.73 & 124.80 \\
$\overline{\textrm{DR}}$ $\overline{\mu}=M_{A^{0}}$ & 144.50 & 35.77 & 124.80\\
$\overline{\textrm{DR}}$ $\overline{\mu}=M_{t}$ & 144.50 & 35.77 & 124.80\\
\hline
\end{tabular}
\\
\caption{{\em Mass of the lightest CP-even Higgs at one loop in
different schemes All masses are in GeV.}\label{mh-tbdep}}
\end{center}
\end{table}
\noi We start with the one-loop correction to the lightest CP-even Higgs.
Of course, this has now been calculated beyond one-loop as the
one-loop correction is large, however a study of the scheme
dependence is important. Moreover this study represents a direct
application of the code that can be compared to results in the
literature. We note that all schemes apart from ${\rm OS}_{M_H}$ are
in very good agreement with each other for both values of $\tb$.
Leaving aside the case of $\tb=50$ in the large $\mu$ scenario,
despite the very large shifts we observed in $\delta
t_{\beta}^{fin}$ for the ${\rm OS}_{M_H}$ scheme, the $\tb$
dependence is much suppressed such that the ${\rm OS}_{M_H}$ scheme
compares favourably with the other schemes. In the case of the
correction to the heaviest CP-even Higgs at one loop, by definition
there is no correction in the ${\rm OS}_{M_H}$ scheme, the other
schemes agree with each other at a very high level of precision.
Moreover especially at high $\tb$ the correction is very small.

\begin{table}[h]
\begin{center}
\begin{tabular}{|c|c|c|c|}
\hline $t_{\beta}=3$ & \textit{mhmax} & \textit{large $\mu$} & \textit{nomix}\\
\hline
$M_{H^0}^{TL}=503.05$ & & & \\
\hline
DCPR& 504.68 & 501.05 & 504.21 \\
${\rm OS}_{M_H}$& 503.05 & 503.05 & 503.05\\
${\rm OS}_{A_{\tau \tau}}$& 504.68 & 501.05 & 504.21 \\
$\overline{\textrm{DR}}$ $\overline{\mu}=M_{A^{0}}$ & 504.52 & 500.95 & 504.08\\
$\overline{\textrm{DR}}$ $\overline{\mu}=M_{t}$ & 504.63 & 501.05 & 504.19\\
\hline
\end{tabular}
\begin{tabular}{|c|c|c|c|}
\hline $t_{\beta}=50$ & \textit{mhmax} & \textit{large $\mu$} & \textit{nomix}\\
\hline
$M_{H^0}^{TL}=500.01$ & & & \\
\hline
DCPR& 499.80 & 498.90 & 499.85 \\
${\rm OS}_{M_H}$& 500.01 & 500.01 & 500.01\\
${\rm OS}_{A_{\tau \tau}}$& 499.80 & 498.91 & 499.85 \\
$\overline{\textrm{DR}}$ $\overline{\mu}=M_{A^{0}}$ & 499.80 & 498.91 & 500.01\\
$\overline{\textrm{DR}}$ $\overline{\mu}=M_{t}$ & 499.80 & 498.91 & 499.85\\
\hline
\end{tabular}\\
\caption{{\em Mass of the heaviest CP-even Higgs at one loop in
different schemes. All masses are in GeV. The one-loop result is
based on the relation $M_{h^{0}}^2=M_{h^{0},{\rm
tree}}^2+Re\hat{\Sigma}(M_{h^{0},{\rm tree}}^2)$. }\label{mH-tbdep}}
\end{center}
\end{table}
\noi The mass of the charged Higgs does not depend on $\dtb$, therefore
the correction is scheme independent, with the counterterm $\delta
M_{H^{\pm}}^2=\delta M_{W^{\pm}}^2+\delta M_{A^0}^2$.

\subsection{Higgs decays to SM particles and their scheme
dependence}
\subsubsection{$A^{0}\rightarrow \tau^{+}\tau^{-}$, the non QED one-loop corrections}
\begin{table}[h]
\begin{center}
\begin{tabular}{|c|c|c|c|}
\hline $t_{\beta}=3$ & \textit{mhmax} & \textit{large $\mu$} & \textit{nomix}\\
\hline
$\Gamma^{TL}=9.40\times 10^{-3}$ & & & \\
\hline
DCPR& +3.56$\times 10^{-5}$ & -8.71$\times 10^{-6}$ & -7.37$\times 10^{-6}$ \\
${\rm OS}_{M_H}$& +6.41$\times 10^{-3}$ & -7.82$\times 10^{-3}$ & +4.56$\times 10^{-3}$\\
${\rm OS}_{A_{\tau \tau}}$& 0 & 0 & 0 \\
$\overline{\textrm{DR}}$ $\overline{\mu}=M_{A^{0}}$ & +6.51$\times 10^{-4}$ & +3.94$\times 10^{-4}$ & +5.18$\times 10^{-4}$\\
$\overline{\textrm{DR}}$ $\overline{\mu}=M_{t}$ & +2.30$\times 10^{-4}$ & -2.66$\times 10^{-5}$ & +9.67$\times 10^{-5}$\\
\hline
\end{tabular}
\begin{tabular}{|c|c|c|c|}
\hline $t_{\beta}=50$ & \textit{mhmax} & \textit{large $\mu$} & \textit{nomix}\\
\hline
$\Gamma^{TL}=2.61\times 10^{0}$ & & & \\
\hline
DCPR& +3.45$\times 10^{-1}$ & +2.01$\times 10^{0}$ & +3.35$\times 10^{-2}$ \\
${\rm OS}_{M_H}$& -4.03$\times 10^{1}$ & -2.09$\times 10^{2}$ & -3.03$\times 10^{1}$\\
${\rm OS}_{A_{\tau \tau}}$& 0 & 0 & 0 \\
$\overline{\textrm{DR}}$ $\overline{\mu}=M_{A^{0}}$ & -1.21$\times 10^{-2}$ & +4.92$\times 10^{-1}$ & -1.66$\times 10^{-2}$\\
$\overline{\textrm{DR}}$ $\overline{\mu}=M_{t}$ & -3.00$\times 10^{-2}$ & +4.75$\times 10^{-1}$ & -3.44$\times 10^{-2}$\\
\hline
\end{tabular}\\
\caption{{\em Corrections to the decay $A^{0}\rightarrow
\tau^{+}\tau^{-}$ at one loop without the universal QED correction. All
widths in GeV.}\label{table-att-tbdep}}
\end{center}
\end{table}

\noi We now study the non QED corrections to the decay width
$A^{0}\rightarrow \tau^{+}\tau^{-}$, see Section~\ref{subsec-att} for our
benchmark points. By definition there is no correction in the ${\rm
OS}_{A_{\tau \tau}}$ scheme. Many interesting and important
conclusions can be drawn from Table~\ref{table-att-tbdep}. First of
all we note that the scheme dependence is quite large here. After
all this is an observable which is directly proportional to $\dtb$. In fact the difference between schemes can be accounted for by $2
\dtb$ read off from Table~\ref{dtbfin}. For this decay, the ${\rm
OS}_{M_H}$ scheme is totally unsuitable, for $\tb=3$ the correction
are of order $100\%$, whereas for $\tb=50$ the one-loop correction
is an order of magnitude, at least, larger than the tree-level.
Especially for $\tb=3$ in $\overline{\textrm{DR}}$ the scale
dependence is not negligible. For example with $\bar \mu=m_t$ in
$\overline{\textrm{DR}}$ the correction is of order $\sim 1\%$ and
$5\%$ for $\bar \mu=M_{A^{0}}$. The corrections are much smaller in DCPR
being at the per-mil level. The scale dependence is much smaller for
$\tb=30$ and the corrections in $\overline{\textrm{DR}}$ are now
smaller than in DCPR. Note also that in the large $\mu$ scenario the
corrections are large.

\subsubsection{$H^{0}\rightarrow \tau^{+}\tau^{-}$, the non QED one-loop corrections}
\begin{table}[h]
\begin{center}
\begin{tabular}{|c|c|c|c|}
\hline $t_{\beta}=3$ & \textit{mhmax} & \textit{large $\mu$} & \textit{nomix}\\
\hline
$\Gamma^{TL}=9.35\times 10^{-3}$ & & & \\
\hline
DCPR& -1.09$\times 10^{-4}$ & -7.96$\times 10^{-5}$ & -1.09$\times 10^{-4}$ \\
${\rm OS}_{M_H}$& +6.28$\times 10^{-3}$ & -7.91$\times 10^{-3}$ & +4.47$\times 10^{-3}$\\
${\rm OS}_{A_{\tau \tau}}$& -1.45$\times 10^{-4}$ & -7.09$\times 10^{-5}$ & -1.01$\times 10^{-4}$ \\
$\overline{\textrm{DR}}$ $\overline{\mu}=M_{A^{0}}$ & +5.08$\times 10^{-4}$ & +3.24$\times 10^{-4}$ & +4.17$\times 10^{-4}$\\
$\overline{\textrm{DR}}$ $\overline{\mu}=M_{t}$ & +8.57$\times 10^{-5}$ & -9.75$\times 10^{-5}$ & -4.52$\times 10^{-6}$\\
\hline
\end{tabular}
\begin{tabular}{|c|c|c|c|}
\hline $t_{\beta}=50$ & \textit{mhmax} & \textit{large $\mu$} & \textit{nomix}\\
\hline
$\Gamma^{TL}=2.61\times 10^{0}$ & & & \\
\hline
DCPR& +3.54$\times 10^{-1}$ & +2.02$\times 10^{0}$ & +4.31$\times 10^{-2}$ \\
${\rm OS}_{M_H}$& -4.03$\times 10^{1}$ & -2.09$\times 10^{2}$ & -3.03$\times 10^{1}$\\
${\rm OS}_{A_{\tau \tau}}$& +9.52$\times 10^{-3}$ & +1.94$\times 10^{-3}$ & +9.55$\times 10^{-3}$ \\
$\overline{\textrm{DR}}$ $\overline{\mu}=M_{A^{0}}$ & -2.59$\times 10^{-3}$ & +4.94$\times 10^{-1}$ & -7.00$\times 10^{-3}$\\
$\overline{\textrm{DR}}$ $\overline{\mu}=M_{t}$ & -2.04$\times 10^{-2}$ & +4.76$\times 10^{-1}$ & -2.49$\times 10^{-2}$ \\
\hline
\end{tabular}\\
\caption{{\em Corrections to the decay $H^{0}\rightarrow
\tau^{+}\tau^{-}$ at one loop without the universal QED correction.
All widths in GeV.}\label{table-Htt-tbdep}}
\end{center}
\end{table}

\noi Similar conclusions can be drawn from the study of the non-QED
corrections to $H^{0}\rightarrow \tau^{+}\tau^{-}$, see
Table~\ref{table-Htt-tbdep}. The QED corrections for this decay can
be implemented as in \cite{Drees-qcd-aqq}. The only difference is
that now there is a correction also in the case of the ${\rm
OS}_{A_{\tau \tau}}$ scheme. But as expected this correction is very
small for both values of $\tb$. Note that for $\tb=50$ the DCPR
scheme gives very large corrections in the large $\mu$ scenario. For
this process we have not taken into account the one-loop correction
to $M_{H^0}$, since as we have seen this correction is very small for
all schemes and also because one is much too far from the $\tau
\tau$ threshold, $M_{H^0}\sim 500$ GeV $\gg 2 m_\tau$, where this effect can
play a role.

\subsubsection{$H^{0}\rightarrow Z^{0}Z^{0}$ and $A^{0}\rightarrow
Z^{0}h^{0}$}
\begin{table}[h]
\begin{center}
\begin{tabular}{|c|c|c|c|}
\hline $t_{\beta}=3$ & \textit{mhmax} & \textit{large $\mu$} & \textit{nomix}\\
\hline
$\Gamma^{TL}=8.97\times 10^{-3}$ & & & \\
\hline
DCPR& +1.59$\times 10^{-2}$ & -6.32$\times 10^{-3}$ & +8.47$\times 10^{-3}$ \\
${\rm OS}_{M_H}$& +1.40$\times 10^{-2}$ & -4.00$\times 10^{-3}$ & +7.12$\times 10^{-3}$\\
${\rm OS}_{A_{\tau \tau}}$& +1.59$\times 10^{-2}$ & -6.32$\times 10^{-3}$ & +8.47$\times 10^{-3}$ \\
$\overline{\textrm{DR}}$ $\overline{\mu}=M_{A^{0}}$ & +1.57$\times 10^{-2}$ & -6.44$\times 10^{-3}$ & +8.32$\times 10^{-3}$\\
$\overline{\textrm{DR}}$ $\overline{\mu}=M_{t}$ & +1.58$\times 10^{-2}$ & -6.32$\times 10^{-3}$ & +8.44$\times 10^{-3}$ \\
\hline
\end{tabular}
\begin{tabular}{|c|c|c|c|}
\hline $t_{\beta}=50$ & \textit{mhmax} & \textit{large $\mu$} & \textit{nomix}\\
\hline
$\Gamma^{TL}=6.40\times 10^{-5}$ & & & \\
\hline
DCPR& +2.18$\times 10^{-5}$ & -5.14$\times 10^{-4}$ & +3.89$\times 10^{-5}$ \\
${\rm OS}_{M_H}$& +1.01$\times 10^{-2}$ & +4.66$\times 10^{-3}$ & +7.81$\times 10^{-4}$\\
${\rm OS}_{A_{\tau \tau}}$& +3.02$\times 10^{-5}$ & -4.65$\times 10^{-4}$ & +3.97$\times 10^{-5}$ \\
$\overline{\textrm{DR}}$ $\overline{\mu}=M_{A^{0}}$ & +3.05$\times 10^{-5}$ & -4.77$\times 10^{-4}$ & +4.01$\times 10^{-5}$\\
$\overline{\textrm{DR}}$ $\overline{\mu}=M_{t}$ & +3.09$\times 10^{-5}$ & -4.76$\times 10^{-4}$ & +4.05$\times 10^{-5}$ \\
\hline
\end{tabular}\\
\caption{{\em Corrections to the decay $H^{0}\rightarrow
\tau^{+}\tau^{-}$ at one loop. All widths in
GeV.}\label{table-HZZ-tbdep}}
\end{center}
\end{table}
\noi $H^{0}\rightarrow Z^{0}Z^{0}$ was studied by \cite{RCHiggstoZZ}
where a large correction was found. We confirm here, see
Table~\ref{table-HZZ-tbdep}, that a large correction is indeed
induced with the one-loop result of the same order if not
exceeding both at $\tb=3$ and $\tb=50$ the tree-level result. This
larger correction is not due to the scheme dependence since in this process the latter is very small whereas one sees a large
correction with all the schemes. The correction is large because
the benchmark points with $M_{A^0}=500$ GeV are in the decoupling
regime where $H^{0}\ra Z^{0}Z^{0}$ practically vanishes at tree-level. The
$H^0Z^{0}Z^{0}$ is proportional to $\cbma \sim M_{Z^0}/M_{A^0}$, the coupling is
therefore almost induced at one loop without the $1/M_{A^0}$
suppression. Here again because $M_{H^0} \gg 2 M_{Z^0}$ the one-loop
correction on $M_{H^{0}}$ is negligible. Very similar results and
conclusions can be drawn for the process $A^{0}\rightarrow
Z^{0}h^{0}$, see Table~\ref{table-AZh-tbdep}.

\begin{table}[h]
\begin{center}
\begin{tabular}{|c|c|c|c|}
\hline $t_{\beta}=3$ & \textit{mhmax} & \textit{large $\mu$} & \textit{nomix}\\
\hline
$\Gamma^{TL}=9.03\times 10^{-3}$ & & & \\
\hline
DCPR& +2.42$\times 10^{-2}$ & +3.86$\times 10^{-3}$ & +1.68$\times 10^{-2}$ \\
${\rm OS}_{M_H}$& +2.23$\times 10^{-2}$ & +6.20$\times 10^{-3}$ & +1.55$\times 10^{-2}$\\
${\rm OS}_{A_{\tau \tau}}$& +2.50$\times 10^{-2}$ & +3.86$\times 10^{-3}$ & +1.64$\times 10^{-2}$ \\
$\overline{\textrm{DR}}$ $\overline{\mu}=M_{A^{0}}$ & +2.48$\times 10^{-2}$ & +3.74$\times 10^{-3}$ & +1.67$\times 10^{-2}$\\
$\overline{\textrm{DR}}$ $\overline{\mu}=M_{t}$ & +2.41$\times 10^{-2}$ & 3.87$\times 10^{-3}$ & +1.68$\times 10^{-2}$\\
\hline
\end{tabular}
\begin{tabular}{|c|c|c|c|}
\hline $t_{\beta}=50$ & \textit{mhmax} & \textit{large $\mu$} & \textit{nomix}\\
\hline
$\Gamma^{TL}=6.30\times 10^{-5}$ & & & \\
\hline
DCPR& +2.39$\times 10^{-5}$ & +8.75$\times 10^{-4}$ & +4.31$\times 10^{-5}$ \\
${\rm OS}_{M_H}$& +1.00$\times 10^{-3}$ & +5.97$\times 10^{-3}$ & +7.74$\times 10^{-4}$\\
${\rm OS}_{A_{\tau \tau}}$& +3.48$\times 10^{-5}$ & +9.26$\times 10^{-4}$ & +4.39$\times 10^{-5}$ \\
$\overline{\textrm{DR}}$ $\overline{\mu}=M_{A^{0}}$ & +3.51$\times 10^{-5}$ & +9.12$\times 10^{-4}$ & +4.43$\times 10^{-5}$\\
$\overline{\textrm{DR}}$ $\overline{\mu}=M_{t}$ & +3.30$\times 10^{-5}$ & +9.12$\times 10^{-4}$ & +4.47$\times 10^{-5}$\\
\hline
\end{tabular}\\
\caption{{\em Corrections to the decay $A^{0}\rightarrow
Z^{0}h^{0}$ at one loop. All widths in
GeV.}\label{table-AZh-tbdep}}
\end{center}
\end{table}

\section{Conclusions}
The use of the non-linear gauge has allowed us, for the first time,
to quantitatively and qualitatively, study different proposals for
the ubiquitous parameter $\tgb$ and its effect on the Higgs
observables, both the physical Higgs masses as well as their decays.
Our first preliminary conclusion is that the scheme based on the
extraction and definition of $\tgb$ from a decay such as $A^0 \ra
\tau^+ \tau^-$ is by far the most satisfactory. Not only is this
definition directly related to a physical observable and therefore
is gauge independent, the functional dependence of the physical
width in $\tgb$ is linear and is the same independently of the value
of the pseudoscalar Higgs mass. Moreover the definition is clean
once we subtract the universal gauge invariant QED correction. The
scheme is also most pleasing and satisfactory since it is the one
where the observables we have studied show the least corrections,
leading therefore to a stable prediction. On this last count the
$\overline{\textrm{DR}}$ scheme performs almost just as well.
However the widely used $\overline{\textrm{DR}}$ scheme extracted
from the $A^{0}Z^{0}$ transition is not gauge invariant and
therefore terribly unsatisfactory from a theoretical point of view.
In the non-linear gauge with a general gauge-fixing set of
parameters, the parameter gauge dependence shows up already at
one-loop, whereas it has been known that the scheme fails even in
the linear gauge but at two-loop \cite{Yamada-2looptb}. A gauge
independent $\overline{\textrm{DR}}$ scheme such as the one proposed
in \cite{RCHiggstoZZ} is most satisfactory. A scheme based on the
usage of $M_{H^0}$ as an independent parameter from the Higgs sector
leads to too large corrections in most of the observables we
considered so far. We therefore propose that the decay $A^0 \ra
\tau^{+}\tau^{-}$ be used as a definition of $\tgb$. This choice
assumes that this decay will one day be measured with high enough
precision but this depends much on the spectrum of the MSSM. Were it
not for the unambiguous extraction of the full QED corrections, the
decay of the charged Higgs to $\tau \nu$ may also qualify as a
suitable input parameter, see \cite{LHC-chargedHiggs} for prospects
on the measurement of this decay. Apart from the discussion on gauge
invariance and the issue of the scheme dependence for $\tgb$, we
have shown how a complete one-loop renormalisation of
the MSSM can be automatised and have given results and details as
concerns the Higgs sector which is the first step in a successful
implementation of this programme.

\vspace{1cm}
\noi {\bf \large Acknowledgments} \\
We would first like to thank David Temes whose help was invaluable
in the first stages of the project. We also owe much to our friends
of the Minami-Tateya group and the developers of the {\tt
Grace-SUSY} code, in particular we learned much from Masaaki Kuroda.
It is also a pleasure to acknowledge the fruitful discussions with
Ben Allanach. This work is supported in part by GDRI-ACPP of the
CNRS (France). The work of A.S. is supported by grants of the
Russian Federal Agency of Science NS-1685.2003.2 and RFBR
04-02-17448. This work is also part of the French ANR project, {\tt
ToolsDMColl}.

\newpage

\renewcommand{\thesection}{\Alph{section}}
\setcounter{section}{0}
\renewcommand{\theequation}{\thesection.\arabic{equation}}
\setcounter{equation}{0} \noi {\Large {\bf Appendices}}

\section{The Ward-Slavnov-Taylor identity for the transitions $A^{0}Z^{0}$ and $A^{0}G^{0}$}
\label{app_ward_azag} There is an identity relating the $A^{0}Z^{0}$
and $A^{0}G^{0}$ transition. This is most useful for $q^2=M_{A^0}^2$.
Contrary to what one might see in some papers, the relation is much
more complicated for $q^2 \neq M_{A^0}^2$ and gets more subtle in the
case of the non-linear gauge.

\noi The identity can be most easily derived by considering the BRST
transformation on the (``ghost") operator $\langle
0|\overline{c}^{Z}(x)A^{0}(y)|0 \rangle=0$. We find
\begin{eqnarray}
\delta_{\textrm{BRS}}\langle 0|\overline{c}^{Z}(x)A^{0}(y)|0
\rangle=\langle
0|(\delta_{\textrm{BRS}}\overline{c}^{Z}(x))A^{0}(y)|0\rangle
-\langle
0|\overline{c}^{Z}(x)(\delta_{\textrm{BRS}}A^{0}(y))|0\rangle =0
\, ,
\end{eqnarray}
with
\begin{eqnarray}
\delta_{\textrm{BRS}}A^{0}&=&-\frac{g}{2}(c^{+}H^{-}+c^{-}H^{+})+\frac{e}{s_{2W}}c^{Z}(c_{\alpha-\beta}h^{0}
+s_{\alpha-\beta}H^{0}) \, ,\\
\textrm{and}\quad \delta_{\textrm{BRS}}\overline{c}^{Z}&=&B^{Z} \,
.
\end{eqnarray}
Therefore,
\begin{eqnarray}
& &\langle 0|B^{Z}(x)A^{0}(y)|0\rangle+\frac{g}{2}\biggl(\langle 0|\overline{c}^{Z}(x)c^{+}(y)H^{-}(y)|0\rangle + \langle 0|\overline{c}^{Z}(x)c^{-}(y)H^{+}(y)|0\rangle\biggr)\nonumber\\
& &-\frac{e}{s_{2W}}\biggl(c_{\alpha-\beta} \langle
0|\overline{c}^{Z}(x)c^{Z}(y)h^{0}(y)|0
\rangle+s_{\alpha-\beta}\langle
0|\overline{c}^{Z}(x)c^{Z}(y)H^{0}(y)|0\rangle \biggr)=0 \, .
\end{eqnarray}
At tree-level, there is no vertex involving
$\overline{c}^{Z}c^{\pm}H^{\pm}$. Using the equation of motion of
the B field, we obtain a relation for the following Green's
functions (external legs are not amputated):
\begin{eqnarray}
&&\partial_{x}\langle 0|Z^{0}(x)A^{0}(y)|0\rangle +M_{Z^0}\langle 0|G^{0}(x)A^{0}(y)|0 \rangle\nonumber\\
&+&\frac{e}{s_{2W}}\biggl(\tilde{\epsilon}\langle
0|h^{0}(x)G^{0}(x)A^{0}(y)|0\rangle
+\tilde{\gamma}\langle 0|H^{0}(x)G^{0}(x)A^{0}(y)|0\rangle\biggr)\nonumber\\
&+&\frac{e}{s_{2W}}\biggl(c_{\alpha-\beta}\langle
0|\overline{c}^{Z}(x)c^{Z}(y)h^{0}(y)|0\rangle
+s_{\alpha-\beta}\langle
0|\overline{c}^{Z}(x)c^{Z}(y)H^{0}(y)|0\rangle\biggr)=0 \, .
\end{eqnarray}
In a diagrammatic form, we have
\begin{eqnarray}
&&\frac{1}{q^{2}-M_{Z^0}^{2}}\frac{1}{q^{2}-M_{A^0}^{2}}\biggl(iq_{\mu}
\times Z^{\mu}\dashrightarrow\circlearrowleft \dashrightarrow
A^{0}+M_{Z^0}\times G^{0}\dashrightarrow\circlearrowleft
\dashrightarrow A^{0}\biggr)\nonumber\\&=&
-\frac{i}{q^{2}-M_{A^0}^{2}}\frac{e}{s_{2W}}\biggl(\tilde{\epsilon}\times\circlearrowleft^{G^{0}}_{h^{0}}\dashrightarrow
A^{0}+\tilde{\gamma}\times\circlearrowleft^{G^{0}}_{H^{0}}\dashrightarrow
A^{0}\biggr)\nonumber\\
&+&\frac{i}{q^{2}-M_{Z^0}^{2}}\frac{e}{s_{2W}}\biggl(c_{\alpha-\beta}
\times
\overline{c}^{Z}\dashrightarrow\circlearrowleft^{c^{Z}}_{h^{0}}+
s_{\alpha-\beta}\times
\overline{c}^{Z}\dashrightarrow\circlearrowleft^{c^{Z}}_{H^{0}}\biggr)
\, ,
\end{eqnarray}
and obtain the relation
\begin{eqnarray}
&&q^{2}\Sigma_{A^{0}Z^{0}}(q^{2})+M_{Z^0}\Sigma_{A^{0}G^{0}}(q^{2})=-(q^{2}-M_{Z^0}^{2})\frac{ie}{s_{2W}}
\biggl(\tilde{\epsilon}\times\circlearrowleft^{G^{0}}_{h^{0}}\dashrightarrow
A^{0}+\tilde{\gamma}\times\circlearrowleft^{G^{0}}_{H^{0}}\dashrightarrow
A^{0}\biggr)\nonumber\\
&+&(q^{2}-M_{A^0}^{2})\frac{ie}{s_{2W}}\biggl(c_{\alpha-\beta}
\times
\overline{c}^{Z}\dashrightarrow\circlearrowleft^{c^{Z}}_{h^{0}}+
s_{\alpha-\beta}\times
\overline{c}^{Z}\dashrightarrow\circlearrowleft^{c^{Z}}_{H^{0}}\biggr)
\, .
\end{eqnarray}
With the following vertices
\begin{eqnarray}
\mathcal{L}&\supset&-\frac{eM_{Z^0}}{s_{2W}}s_{2\beta}\biggl(s_{\alpha+\beta}h^{0}-c_{\alpha+\beta}H^{0}\biggr)A^{0}G^{0}\, ,\\
\mathcal{L}^{Gh}&\supset&\frac{e
M_{Z^0}}{s_{2W}}\biggl((s_{\alpha-\beta}-\tilde{\epsilon})h^{0}
-(c_{\alpha-\beta}+\tilde{\gamma})H^{0}\biggr)\bar{c}^{Z}c^{Z} \,
,
\end{eqnarray}
we calculate all the ``lollipops"
\begin{eqnarray}
\circlearrowleft^{G^{0}}_{h^{0}}\dashrightarrow
A^{0}&=&-i\frac{eM_{Z^0}}{s_{2W}}s_{2\beta}s_{\alpha+\beta} B_{0}(q^2,M_{h^{0}}^{2},M_{Z^{0}}^{2}) \, ,\\
\circlearrowleft^{G^{0}}_{H^{0}}\dashrightarrow
A^{0}&=&i\frac{eM_{Z^0}}{s_{2W}}s_{2\beta}c_{\alpha+\beta} B_{0}(q^2,M_{H^{0}}^{2},M_{Z^{0}}^{2}) \, , \\
\overline{c}^{Z}\dashrightarrow\circlearrowleft^{c^{Z}}_{h^{0}}&=&i\frac{eM_{Z^0}}{s_{2W}}(s_{\alpha-\beta}-\tilde{\epsilon})
B_{0}(q^2,M_{h^{0}}^{2},M_{Z^{0}}^{2}) \, , \\
\overline{c}^{Z}\dashrightarrow\circlearrowleft^{c^{Z}}_{H^{0}}&=&-i\frac{eM_{Z^0}}{s_{2W}}(c_{\alpha-\beta}+\tilde{\gamma})
B_{0}(q^2,M_{H^{0}}^{2},M_{Z^{0}}^{2}) \, ,
\end{eqnarray}
with
\begin{eqnarray}
B_{0}(q^2, M_{1}^{2}, M_{2}^{2})&=& C_{UV}-\int_{0}^{1} {\rm d}x \ln( \Delta(q^2, M_{1}^{2}, M_{2}^{2}) ) \, ,\\
\Delta (q^2, M_{1}^{2}, M_{2}^{2})&=&q^{2}x^{2}-
(q^{2}+M_{2}^{2}-M_{1}^{2})x + M_{2}^{2} \, .
\end{eqnarray}
We finally obtain the identity
\begin{eqnarray}
%&&
q^{2}\Sigma_{A^{0}Z^{0}}(q^{2})+M_{Z^0}\Sigma_{A^{0}G^{0}}(q^{2})=\frac{1}{(4\pi)^{2}}\frac{e^{2}M_{Z^0}}{s_{2W}^{2}}\biggl(
(q^{2}-M_{Z^0}^{2})s_{2\beta}\mathcal{F}_{GA}^{\tilde{\epsilon},\tilde{\gamma}}(q^2)+(q^{2}-M_{A^0}^{2})\mathcal{F}_{cc}^{\tilde{\epsilon},\tilde{\gamma}}(q^2)\biggr)\, ,& & \nonumber \\
%&&\\
%&&
\textrm{with}\quad
\mathcal{F}_{GA}^{\tilde{\epsilon},\tilde{\gamma}}(q^2)=\tilde{\gamma}c_{\alpha+\beta}B_{0}(q^2,M_{H^{0}}^{2},M_{Z^{0}}^{2})
-\tilde{\epsilon}s_{\alpha+\beta}B_{0}(q^2,M_{h^{0}}^{2},M_{Z^{0}}^{2})
\, , \quad\quad\quad\quad\quad\quad\quad\quad\quad\quad\quad\,\,& &
\nonumber\\
%&&
%\,\,\,\,\,\,\,\,\,\,\,\quad
\mathcal{F}_{cc}^{\tilde{\epsilon},\tilde{\gamma}}(q^2)=\tilde{\epsilon}c_{\alpha-\beta}
B_{0}(q^2,M_{h^{0}}^{2},M_{Z^{0}}^{2}) +
\tilde{\gamma}s_{\alpha-\beta}
B_{0}(q^2,M_{H^{0}}^{2},M_{Z^{0}}^{2})
\quad\quad\quad\quad\quad\quad\quad\quad\quad\quad\quad\,\,\,\,\,& &
\nonumber\\
+\frac{1}{2}s_{2(\alpha-\beta)} \left(
B_{0}(q^2,M_{H^{0}}^{2},M_{Z^{0}}^{2})-
B_{0}(q^2,M_{h^{0}}^{2},M_{Z^{0}}^{2}) \right) \, .
\quad\quad\quad\quad\quad\quad\quad\quad\quad\quad\quad\,\,\,\,\,& &
\end{eqnarray}
To implement this formula into {\tt SloopS} and check it
numerically, we need to introduce the tadpole part in {\tt FormCalc}
and we define $\Sigma^{\textrm{tad}}$ the self-energy without
tadpole:
\begin{eqnarray}
q^{2}\Sigma^{\textrm{tad}}_{A^{0}Z^{0}}(q^{2})+M_{Z^0}\Sigma^{\textrm{tad}}_{A^{0}G^{0}}(q^{2})+M_{Z^0} \delta T=\frac{1}{(4\pi)^{2}}\frac{e^{2}M_{Z^0}}{s_{2W}^{2}}(
(q^{2}-M_{Z^0}^{2})s_{2\beta}\mathcal{F}_{GA}^{\tilde{\epsilon},\tilde{\gamma}}+(q^{2}-M_{A^0}^{2})\mathcal{F}_{cc}^{\tilde{\epsilon},\tilde{\gamma}})\, , & & \nonumber\\
\textrm{where}\quad \delta T=\frac{e}{s_{2W}M_{Z^0}}(s_{\alpha-\beta}\delta
T_{H^{0}}+c_{\alpha-\beta}\delta T_{h^{0}}) \, .
\quad\quad\quad\quad\quad\quad\quad\quad\quad\quad\quad\quad\quad\quad\quad\quad\quad\quad\quad\quad\quad& &
\end{eqnarray}
We remark on some simplifications in the functions $\mathcal{F}$ for
specific choices of the non linear gauge parameters
\begin{eqnarray}
\mathcal{F}_{GA}^{\tilde{\epsilon},\tilde{\gamma}}(\tilde{\epsilon}=0,\tilde{\gamma}=0)&=&0 \, ,\\
\mathcal{F}_{GA}^{\tilde{\epsilon},\tilde{\gamma}}(\tilde{\epsilon}=c_{\alpha+\beta},\tilde{\gamma}=s_{\alpha+\beta})&=&\frac{1}{2}s_{2(\alpha+\beta)}
\int_{0}^{1}dx\ln\left(\frac{\Delta(q^{2},M_{h^{0}}^{2}, M_{Z^{0}}^{2})}{\Delta(q^{2},M_{H^{0}}^{2}, M_{Z^{0}}^{2})}\right) \, , \\
\mathcal{F}_{cc}^{\tilde{\epsilon},\tilde{\gamma}}(\tilde{\epsilon}=s_{\alpha-\beta},\tilde{\gamma}=-c_{\alpha-\beta})&=&0\, , \\
\mathcal{F}_{cc}^{\tilde{\epsilon},\tilde{\gamma}}(\tilde{\epsilon}=0,\tilde{\gamma}=0)&=&\frac{1}{2}s_{2(\alpha-\beta)}
\int_{0}^{1}dx\ln\left(\frac{\Delta(q^{2},M_{h^{0}}^{2},
M_{Z^{0}}^{2})}{\Delta(q^{2},M_{H^{0}}^{2}, M_{Z^{0}}^{2})}\right)
\, .
\end{eqnarray}
In terms of renormalised self energies,
\begin{eqnarray}
\hat{\Sigma}_{A^{0}Z^{0}}(q^2)&=&\Sigma^{\textrm{tad}}_{A^{0}Z}(q^2)+\frac{M_{Z^0}}{2}(\delta
Z_{G^{0}A^{0}}+s_{2\beta}\frac{ \delta t_{\beta}}{t_{\beta}}) \, ,\\
\hat{\Sigma}_{A^{0}G^{0}}(q^{2})&=&\Sigma^{\textrm{tad}}_{A^{0}G^{0}}(q^{2})+\delta
M_{A^{0}G^{0}}^{2}-\frac{1}{2}q^{2}\delta Z_{G^{0}A^{0}}
-\frac{1}{2}(q^{2}-M_{A^0}^{2})\delta Z_{A^{0}G^{0}} \, ,
\label{ren-az-ag-nonren}
\end{eqnarray}
with (Eq.~\ref{deltaM_all})
\begin{eqnarray}
\delta M_{A^{0}G^{0}}^{2}&=&\delta
T-\frac{1}{2}s_{2\beta}M_{A^0}^{2}\frac{\delta t_{\beta}}{t_{\beta}}\,
,
\end{eqnarray}
we obtain the following constraint on the {\em renormalised}
two-point functions
\begin{eqnarray}
q^{2}\hat{\Sigma}_{A^{0}Z^{0}}(q^{2})+M_{Z^0}\hat{\Sigma}_{A^{0}G^{0}}(q^{2})&=&
(q^{2}-M_{Z^0}^{2})\frac{1}{(4\pi)^{2}}\frac{e^{2}M_{Z^0}}{s_{2W}^{2}}s_{2\beta}\mathcal{F}_{GA}^{\tilde{\epsilon},\tilde{\gamma}}(q^2)\\
&+&\frac{M_{Z^0}}{2}(q^{2}-M_{A^0}^{2})(\frac{1}{(4\pi)^{2}}\frac{2e^{2}}{s_{2W}^{2}}\mathcal{F}_{cc}^{\tilde{\epsilon},\tilde{\gamma}}(q^2)+
s_{2\beta}\frac{ \delta t_{\beta}}{t_{\beta}}-\delta
Z_{AG})\nonumber \, .\label{constraint-az-ag}
\end{eqnarray}
Note that in this identity $\delta T$ and more importantly $\delta Z_{G^{0}A^{0}}$ drop out.\\

\noi The derivation of the identity for the charged Higgses follows along
the same steps. We only quote the result
\begin{eqnarray}
\label{ward-gh}
q^{2}\hat{\Sigma}_{H^{+}W^{+}}(q^{2})+M_{W^{\pm}}\hat{\Sigma}_{H^{+}W^{+}}(q^{2})&=&
(q^{2}-M_{W^{\pm}}^{2})\frac{1}{(4\pi)^{2}}\frac{e^{2}M_{W^{\pm}}}{s_{2W}^{2}}\mathcal{G}_{HW}^{\tilde{\rho},\tilde{\omega},\tilde{\delta}}(q^2) \nonumber \\
&+&\frac{M_{W^{\pm}}}{2}(q^{2}-M_{H^{\pm}}^{2})\biggl(\frac{1}{(4\pi)^{2}}\frac{2e^{2}}{s_{2W}^{2}}\mathcal{G}_{cc}^{\tilde{\rho},\tilde{\omega},\tilde{\delta}}(q^2)+
s_{2\beta}\frac{ \delta t_{\beta}}{t_{\beta}}-\delta Z_{H^\pm
G\pm}\biggr)\, . \nonumber
\end{eqnarray}
with the functions
$\mathcal{G}_{HW}^{\tilde{\rho},\tilde{\omega},\tilde{\delta}}(q^2)$
and
$\mathcal{G}_{cc}^{\tilde{\rho},\tilde{\omega},\tilde{\delta}}(q^2)$
defined as
\begin{eqnarray}
\mathcal{G}_{HW}^{\tilde{\rho},\tilde{\omega},\tilde{\delta}}(q^2)&=&
\tilde{\delta}(s_{2\beta}s_{\alpha+\beta}-c_{W}^{2}c_{\alpha-\beta})B_{0}(q^{2},M_{W^{\pm}}^{2},M_{h^{0}}^{2})
- \tilde{\omega}(c_{2\beta}s_{\alpha+\beta}-s_{W}^{2}s_{\alpha-\beta})B_{0}(q^{2},M_{W^{\pm}}^{2},M_{H^0}^{2})\nonumber\\
& &+\tilde{\rho}c_{W}^{2}B_{0}(q^{2},M_{W^{\pm}}^{2},M_{A^{0}}^{2}) \, ,\nonumber\\
%%%%%%%%%%%%%%%%%%%%%%%%%%%%%%%%%%
\mathcal{G}_{cc}^{\tilde{\rho},\tilde{\omega},\tilde{\delta}}(q^2)&=&
 c_{\alpha-\beta}(s_{\alpha-\beta}-\tilde{\delta})c_{W}^{2}B_{0}(q^{2},M_{W^{\pm}}^{2},M_{h^{0}}^{2})
-s_{\alpha-\beta}(c_{\alpha-\beta}+\tilde{\omega})c_{W}^{2}B_{0}(q^{2},M_{W^{\pm}}^{2},M_{H^{0}}^{2}) \nonumber\\
& & - \tilde{\rho}c_{W}^{2}B_{0}(q^{2},M_{W^{\pm}}^{2},M_{A^{0}}^{2}) \, .
\end{eqnarray}

\section{Wave function renormalisation constants before rotation }
\label{appendix-wfr-dcpr}
In our approach field renormalisation
was performed on the physical fields, or better said, after
rotation to the $h^0,H^0,A^0,G^0,H^\pm,G^\pm$ basis. We could have
applied field renormalisation on the components of the doublets
$H_1,H_2$, Eq.~(\ref{doublet_H1H2}). To make contact with some of
the early papers \cite{DabelsteinHiggs,DCPR,HHWtb} on the
renormalisation of the Higgs sector we therefore introduce the
most general field renormalisation on the components of $H_1,H_2$.
We define
\begin{eqnarray}
\left(\begin{array}{c} \varphi_{1}^{0}\\
\varphi_{2}^{0}\end{array}\right)_{0}&=&\left(\begin{array}{cc}
Z_{\varphi_{1}^{0}}^{1/2} & Z_{\varphi_{1}^{0}\varphi_{2}^{0}}^{1/2} \\
Z_{\varphi_{2}^{0}\varphi_{1}^{0}}^{1/2} &
Z_{\varphi_{2}^{0}}^{1/2}\end{array}\right)
\left(\begin{array}{c} \varphi_{1}^{0}\\
\varphi_{2}^{0}\end{array}\right) \, ,
\label{transf-phu1}
%%%%%%%%%%%%
\end{eqnarray}
\begin{eqnarray}
\left(\begin{array}{c} \phi_{1}^{\pm}\\
\phi_{2}^{\pm}\end{array}\right)_{0}&=&\left(\begin{array}{cc}
Z_{\phi_{1}^{\pm}}^{1/2} & Z_{\phi_{1}^{\pm}\phi_{2}^{\pm}}^{1/2} \\
Z_{\phi_{2}^{\pm}\phi_{1}^{\pm}}^{1/2} &
Z_{\phi_{2}^{\pm}}^{1/2}\end{array}\right)
\left(\begin{array}{c} \phi_{1}^{\pm}\\
\phi_{2}^{\pm}\end{array}\right) \, ,
\label{transf-phu2}
%%%%%%%%%%%%
\end{eqnarray}
\begin{eqnarray}
\label{transf-phu3}
\left(\begin{array}{c} \phi_{1}^{0}\\
\phi_{2}^{0}\end{array}\right)_{0}&=&\left(\begin{array}{cc}
Z_{\phi_{1}^{0}}^{1/2} & Z_{\phi_{1}^{0}\phi_{2}^{0}}^{1/2} \\
Z_{\phi_{2}^{0}\phi_{1}^{0}}^{1/2} &
Z_{\phi_{2}^{0}}^{1/2}\end{array}\right)
\left(\begin{array}{c} \phi_{1}^{0}\\
\phi_{2}^{0}\end{array}\right) \, .
%\label{wfr-phis}
\end{eqnarray}
As explained in the text, these constants are immediately
transformed into the set of matrices $Z_P,Z_C,Z_S$. Or we can go
from the set $Z_P,Z_C,Z_S$ to the set defined by
Eqs.~(\ref{transf-phu1}-\ref{transf-phu3}). For example,
\begin{eqnarray}
&\,&\left\{\begin{array}{l} \delta Z_{G^{0}}=c_{\beta}^{2}\delta
Z_{\varphi_{1}^{0}}+s_{\beta}^{2}\delta
Z_{\varphi_{2}^{0}}+c_{\beta}s_{\beta}(\delta
Z_{\varphi_{1}^{0}\varphi_{2}^{0}}+\delta Z_{\varphi_{2}^{0}\varphi_{1}^{0}})\\
%%%%%%%%%%%%%%
\delta Z_{G^{0}A^{0}}=c_{\beta}s_{\beta}(\delta
Z_{\varphi_{2}^{0}}-\delta
Z_{\varphi_{1}^{0}})+c_{\beta}^{2}\delta
Z_{\varphi_{1}^{0}\varphi_{2}^{0}}-s_{\beta}^{2}\delta
Z_{\varphi_{2}^{0}\varphi_{1}^{0}}\\
%%%%%%%%%%%%%%
\delta Z_{A^{0}G^{0}}=c_{\beta}s_{\beta}(\delta
Z_{\varphi_{2}^{0}}-\delta
Z_{\varphi_{1}^{0}})+c_{\beta}^{2}\delta
Z_{\varphi_{2}^{0}\varphi_{1}^{0}}-s_{\beta}^{2}\delta
Z_{\varphi_{1}^{0}\varphi_{2}^{0}}\\
%%%%%%%%%%%%%%
\delta Z_{A^{0}}=s_{\beta}^{2}\delta
Z_{\varphi_{1}^{0}}+c_{\beta}^{2}\delta
Z_{\varphi_{2}^{0}}-c_{\beta}s_{\beta}(\delta
Z_{\varphi_{1}^{0}\varphi_{2}^{0}}+\delta
Z_{\varphi_{2}^{0}\varphi_{1}^{0}})
\end{array}\right. \label{defdelZGA}\\
%%%%%%%%%%%%%%
%%%%%%%%%%%%%%
&\,&\left\{\begin{array}{l} \delta Z_{G^{\pm}}=c_{\beta}^{2}\delta
Z_{\phi_{1}^{\pm}}+s_{\beta}^{2}\delta
Z_{\phi_{2}^{\pm}}+c_{\beta}s_{\beta}(\delta
Z_{\phi_{1}^{\pm}\phi_{2}^{\pm}}+\delta Z_{\phi_{2}^{\pm}\phi_{1}^{\pm}})\\
%%%%%%%%%%%%%%
\delta Z_{G^{\pm}H^{\pm}}=c_{\beta}s_{\beta}(\delta
Z_{\phi_{2}^{\pm}}-\delta Z_{\phi_{1}^{\pm}})+c_{\beta}^{2}\delta
Z_{\phi_{1}^{\pm}\phi_{2}^{\pm}}-s_{\beta}^{2}\delta
Z_{\phi_{2}^{\pm}\phi_{1}^{\pm}}\\
%%%%%%%%%%%%%%
\delta Z_{H^{\pm}G^{\pm}}=c_{\beta}s_{\beta}(\delta
Z_{\phi_{2}^{\pm}}-\delta Z_{\phi_{1}^{\pm}})+c_{\beta}^{2}\delta
Z_{\phi_{2}^{\pm}\phi_{1}^{\pm}}-s_{\beta}^{2}\delta
Z_{\phi_{1}^{\pm}\phi_{2}^{\pm}}\\
%%%%%%%%%%%%%%
\delta Z_{H^{\pm}}=s_{\beta}^{2}\delta
Z_{\phi_{1}^{\pm}}+c_{\beta}^{2}\delta
Z_{\phi_{2}^{\pm}}-c_{\beta}s_{\beta}(\delta
Z_{\phi_{1}^{\pm}\phi_{2}^{\pm}}+\delta
Z_{\phi_{2}^{\pm}\phi_{1}^{\pm}})
\end{array}\right. \label{defdelZGCHC}\\
%%%%%%%%%%%%%%
%%%%%%%%%%%%%%
&\,&\left\{\begin{array}{l} \delta Z_{H^{0}}=c_{\alpha}^{2}\delta
Z_{\phi_{1}^{0}}+s_{\alpha}^{2}\delta
Z_{\phi_{2}^{0}}+c_{\alpha}s_{\alpha}(\delta
Z_{\phi_{1}^{0}\phi_{2}^{0}}+\delta Z_{\phi_{2}^{0}\phi_{1}^{0}})\\
%%%%%%%%%%%%%%
\delta Z_{H^{0}h^{0}}=c_{\alpha}s_{\alpha}(\delta
Z_{\phi_{2}^{0}}-\delta Z_{\phi_{1}^{0}})+c_{\alpha}^{2}\delta
Z_{\phi_{1}^{0}\phi_{2}^{0}}-s_{\alpha}^{2}\delta
Z_{\phi_{2}^{0}\phi_{1}^{0}}\\
%%%%%%%%%%%%%%
\delta Z_{h^{0}H^{0}}=c_{\alpha}s_{\alpha}(\delta
Z_{\phi_{2}^{0}}-\delta Z_{\phi_{1}^{0}})+c_{\alpha}^{2}\delta
Z_{\phi_{2}^{0}\phi_{1}^{0}}-s_{\alpha}^{2}\delta
Z_{\phi_{1}^{0}\phi_{2}^{0}}\\
%%%%%%%%%%%%%%
\delta Z_{h^{0}}=s_{\alpha}^{2}\delta
Z_{\phi_{1}^{0}}+c_{\alpha}^{2}\delta
Z_{\phi_{2}^{0}}-c_{\alpha}s_{\alpha}(\delta
Z_{\phi_{1}^{0}\phi_{2}^{0}}+\delta Z_{\phi_{2}^{0}\phi_{1}^{0}})
\end{array}\right. \label{defdelZH0h0}\\
%%%%%%%%%%%%%%
%%%%%%%%%%%%%%
&\,&\left\{\begin{array}{l} \delta Z_{H^{0}h^{0}}+\delta
Z_{h^{0}H^{0}}= (c_{\alpha}^{2}-s_{\alpha}^{2})(\delta
Z_{\phi_{1}^{0}\phi_{2}^{0}}+\delta Z_{\phi_{2}^{0}\phi_{1}^{0}})+
2c_{\alpha}s_{\alpha}(\delta Z_{\phi_{2}^{0}}-\delta Z_{\phi_{1}^{0}})\\
\delta Z_{H^{0}h^{0}}-\delta Z_{h^{0}H^{0}}=\delta
Z_{\phi_{1}^{0}\phi_{2}^{0}}-\delta Z_{\phi_{2}^{0}\phi_{1}^{0}}
%%%%%%%%%%%%%%
\end{array}\right.
\end{eqnarray}
%=============================================================================
Our renormalisation conditions in Eq.~(\ref{residue_cdts}) on
$\hat \Sigma_{ii}$ will turn into
\begin{eqnarray}
Re\Sigma^{'}_{A^{0}A^{0}}(M_{A^0}^{2})&=&s_{\beta}^{2}\delta
Z_{\varphi_{1}^{0}}+c_{\beta}^{2}\delta
Z_{\varphi_{2}^{0}}-c_{\beta}s_{\beta}(\delta Z_{\varphi_{1}^{0}\varphi_{2}^{0}}+\delta Z_{\varphi_{2}^{0}\varphi_{1}^{0}}) \, , \label{resA0}\\
%%%%%%%%%%%
Re\Sigma^{'}_{H^{\pm}H^{\pm}}(M^{2}_{H^{\pm}})&=&s_{\beta}^{2}\delta
Z_{\phi_{1}^{\pm}}+c_{\beta}^{2}\delta
Z_{\phi_{2}^{\pm}}-c_{\beta}s_{\beta}(\delta
Z_{\phi_{1}^{\pm}\phi_{2}^{\pm}}+Z_{\phi_{2}^{\pm}\phi_{1}^{\pm}}) \, ,\label{resHc}\\
%%%%%%%%%%%
Re\Sigma^{'}_{H^{0}H^{0}}(M^{2}_{H^{0}})&=&c_{\alpha}^{2}\delta
Z_{\phi_{1}^{0}}+s_{\alpha}^{2}\delta
Z_{\phi_{2}^{0}}+c_{\alpha}s_{\alpha}(\delta Z_{\phi_{1}^{0}\phi_{2}^{0}}+\delta Z_{\phi_{2}^{0}\phi_{1}^{0}}) \, ,\label{resH0}\\
%%%%%%%%%%%
Re\Sigma^{'}_{h^{0}h^{0}}(M^{2}_{h^{0}})&=&s_{\alpha}^{2}\delta
Z_{\phi_{1}^{0}}+c_{\alpha}^{2}\delta
Z_{\phi_{2}^{0}}-c_{\alpha}s_{\alpha}(\delta
Z_{\phi_{1}^{0}\phi_{2}^{0}}+\delta
Z_{\phi_{2}^{0}\phi_{1}^{0}}) \, .\label{resh0}
\end{eqnarray}
In fact in \cite{DabelsteinHiggs,DCPR,HHWtb} only two
renormalisation constants are introduced, one for each doublet
through
\begin{eqnarray}
H_{i}\rightarrow (1+\frac{1}{2}\delta Z_{H_{i}})H_{i} \quad
{i=1,2} \, .\label{DZHi}
\end{eqnarray}
This means that
\beqn
\delta Z_{\phi_i^0}&=&\delta Z_{\varphi_i^0}=\delta
Z_{\phi_i^\pm}=\delta Z_{H_i} \, ,\nonumber \\
\delta Z_{\phi_{i}^{0}\phi_{j}^{0}}&=&\delta
Z_{\varphi_{i}^{0}\varphi_{j}^{0}}=\delta
Z_{\phi_{i}^{\pm}\phi_{j}^{\pm}}=0 \quad i \neq j \, .
\label{dzp1p2_0}
\eeqn
Since wave function renormalisation is applied on the doublets it
also contributes a shift to $v_i$. Another shift on this parameter
is also applied, $v_i \ra v_i -\tilde{\delta} v_i$ as to all other
Lagrangian parameters. Compared to our shift $\delta v_i$, we have
\beqn
\delta v_i=\tilde{\delta} v_i -\frac{1}{2} \delta Z_{H_i} v_i \, .
\eeqn
Note that with only $\delta Z_{H_1}$ and $\delta Z_{H_2}$, in view
of Eqs.~(\ref{resH0})-(\ref{resh0}) and Eq.~(\ref{dzp1p2_0}) we have
\beqn
\delta Z_{H_1}-\delta Z_{H_2}=-\frac{1}{2 c_{2 \alpha}}
\biggl(Re\Sigma^{'}_{h^{0}h^{0}}(M^{2}_{h^{0}})-
Re\Sigma^{'}_{H^{0}H^{0}}(M^{2}_{H^{0}})\biggr) \, . \label{Holliktb}
\eeqn

\renewcommand{\theequation}{\thesection.\arabic{equation}}
\setcounter{equation}{0}
\end{document}